\pdfoutput=1
\documentclass[useAMS,usenatbib,a4paper]{mn2e}
\usepackage{fixltx2e}
\usepackage{graphicx}
\usepackage{mathrsfs}
\usepackage{times}
\usepackage{txfonts}
\usepackage{color}
\usepackage{hyperref}
\usepackage{breakurl}
\newcommand{\fras}[2]{\mbox{\normalsize $\frac{#1}{#2}$}}
\newcommand{\spar}[2]{\frac{\upartial#1}{\upartial#2}}
\newcommand{\dpar}[2]{\frac{\upartial#1^{2}}{\upartial#2^{2}}}

\newcommand{\ie}{{i.e.}~}
\newcommand{\eg}{{e.g.}~}
\newcommand{\cf}{{cf.}~}

\DeclareSymbolFont{lettersC}{OML}{txmi}{bx}{it}
\SetSymbolFont{lettersC}{bold}{OML}{txmi}{bx}{it}
\title[Neutron stars with a toroidal
  magnetic field]{Equilibrium models of relativistic stars with a
  toroidal magnetic field}

\author[J. Frieben and L. Rezzolla]{J. Frieben$^{1}$ and L.
Rezzolla$^{1,2}$\\
$^{1}$Max-Planck-Institut f{\"u}r Gravitationsphysik, Albert-Einstein-Institut, Am M{\"u}hlenberg 1, D-14476 Golm, Germany \\
$^{2}$Department of Physics, Louisiana State University,
   Baton Rouge, LA 70803, USA}

\begin{document}
\date{}
\pagerange{\pageref{firstpage}--\pageref{lastpage}} \pubyear{2012}
\maketitle
\label{firstpage}
\begin{abstract}
We have computed models of rotating relativistic stars with a toroidal
magnetic field and investigated the combined effects of magnetic field
and rotation on the apparent shape (\ie the surface deformation),
which could be relevant for the electromagnetic emission, and on the
internal matter distribution (\ie the quadrupole distortion), which
could be relevant for the emission of gravitational waves. Using a
sample of eight different cold nuclear physics equations of state, we
have computed models of maximum field strength, as well as the
distortion coefficients for the surface and the quadrupolar
deformations. Surprisingly, we find that non-rotating models admit
arbitrary levels of magnetization, accompanied by a growth of size and
quadrupole distortion to which we could not find a limit. Rotating
models, on the other hand, are subject to a mass-shedding limit at
frequencies well below the corresponding ones for unmagnetized
stars. Overall, the space of solutions can be split into three
distinct classes for which the surface deformation and the quadrupole
distortion are either: prolate and prolate, oblate and prolate, or
oblate and oblate, respectively. We also derive a simple formula
expressing the relativistic distortion coefficients, which allows
one to compute the surface deformation and the quadrupole distortion
up to significant levels of rotation and magnetization, essentially
covering all known magnetars. Such a formula replaces Newtonian
equivalent expressions that overestimate the magnetic quadrupole
distortion by about a factor of 6 and are inadequate for
strongly relativistic objects like neutron stars.
\end{abstract}

\section{Introduction}
\label{se:intro}

In addition to rapid rotation, also strong magnetic fields can
introduce significant deformations in neutron stars, as shown, for
instance, by~\citet{Bocquet1995} and~\citet*{Cardall2001}, who have
computed fully non-linear models of relativistic stars with a poloidal
magnetic field. At the end of the collapse of the core of a massive
star, differential rotation could create strong toroidal magnetic
fields of the order of $10^{16}$--$10^{17} \, \rmn{G}$
inside the hot proto-neutron star~(\citealt*{Bonanno:2003uw};
 \citealt{Naso2008}; Bonazzola \& Haensel, unpublished).
As a result, realistic models of magnetized relativistic
stars require the simultaneous
presence of both poloidal and toroidal field components. As pointed
out in~\citet{Gourgoulhon1993}, however, this requires a formalism
capable of dealing with non-circular space--times (\ie with
convective currents in the meridional planes) like the one presented
here, but which has so far been implemented only in a perturbative
scheme~\citep{Ioka2003,Ioka2004}. Triggered also by the increasing
interest in strongly magnetized neutron stars due to their relation
with soft-gamma repeaters and anomalous X-ray pulsars
\citep{Duncan1992, Thompson1996}, a growing number of studies
adopting perturbative techniques have appeared investigating either
the field geometry and neglecting the influence of the magnetic field
on the matter distribution~\citep{Ciolfi2009}, or solving the coupled
Einstein--Maxwell--Euler system, from which the magnetic deformation
can be calculated~(\citealt{Ioka2004}; \citealt{Colaiuda2008};
\citealt*{Ciolfi2010}; \citealt*{Gualtieri2011}; \citealt*{Yoshida2012}).
Until recently, however, fully non-linear models of
relativistic magnetized stars were restricted to purely poloidal
magnetic fields \citep{Bocquet1995,Cardall2001}, for which the
generated space--time is circular like in the unmagnetized
case~\citep{Carter1973}. Following the recent insight that also a
magnetic field with only a toroidal component is compatible with the
circularity of space--time~\citep{Oron2002}, studies of relativistic
models of rotating stars with a toroidal magnetic field have emerged
(\citealt{Frieben2007}; \citealt{Kiuchi2008}; \citealt*{Kiuchi2009};
\citealt*{Yasutake2010a}; \citealt*{Yasutake2011}).
These new studies have complemented earlier Newtonian
investigations~\citep{Sinha1968,Sood1972,Miketinac1973}, which being
simpler, allowed for the investigation of more complex field
geometries, in particular of mixed poloidal and toroidal magnetic
fields (\citealt{Tomimura2005}; \citealt*{Yoshida2006};
\citealt{Haskell2008}; \citealt{Lander2009}; \citealt*{Fujisawa2012}).

In this work, we are mainly concerned with the deformation of neutron
stars endowed with a toroidal magnetic field because they might be an
important source of gravitational radiation due to the prolate
deformation induced by the magnetic field and provided that the axes
of symmetry and of rotation are different
\citep{Bonazzola1996}. Furthermore, \citet{Jones1975} has pointed out
that viscous processes can trigger a secular instability, which drives
the axis of symmetry of the prolate star into the plane perpendicular
to the angular momentum vector transforming it into a bar-shaped
rotating source of gravitational
radiation~\citep{Cutler2002,Stella2005}. From the astrophysical point
of view, both the deformation of the stellar surface as well as the
distortion of the matter distribution are relevant and can be measured
by appropriate quantities, namely the surface deformation (or apparent
oblateness) and the quadrupole distortion. Previous studies of
relativistic models of stars with a toroidal magnetic field have
provided a broad survey of non-rotating and rotating models for varying
masses, radii and magnetic fluxes. The focus of this work is a
complementary one to earlier studies: in order to obtain a
comprehensive picture of the impact of a toroidal magnetic field on
relativistic stars, we exclusively study models of fixed baryon mass
corresponding to a gravitational mass of $M=1.400 \, \mathrm{M}_{\sun}$ in the
unmagnetized and non-rotating case. Although we expect neutrons stars
to come in a narrow but non-zero range of masses, restricting to a
single value of the gravitational mass has the important advantage
that we can explore with unprecedented precision both the deformations
introduced by magnetic fields and those introduced by rapid
rotation. As we will comment later on, our increased accuracy has
allowed us also to discover novel results and equilibrium
configurations.

Our neutron stars are modelled assuming the matter to be a perfect
fluid at zero temperature well described by a single-parameter equation
of state (EOS), and as having infinite conductivity, as required by the
ideal magnetohydrodynamics (MHD) limit. We do not consider multifluid
models, which would allow for the stratification of neutron star
matter, nor do we treat the protons in the interior as a superconducting
fluid, which would greatly alter the magnetic properties of associated
equilibrium models. Nevertheless, important results obtained for
extremely magnetized models with field strengths of the order of
$10^{17} \, \mathrm{G}$ can be readily extended to configurations of
much lower and more realistic field strengths of the order of $10^{13}
\, \mathrm{G}$. More specifically, in addition to a standard $\gamma =
2$ polytropic EOS, we have considered a sample of seven realistic EOSs
resulting from calculations of cold catalysed dense matter, namely the
APR EOS~\citep*{Akmal1998}, the BBB2 EOS~\citep*{Baldo1997}, the BN1H1
EOS~\citep{Balberg1997}, the BPAL12 EOS~\citep{Bombaci1996}, the FPS
EOS~\citep{Pandharipande1989}, the GNH3 EOS ~\citep{Glendenning1985}
and the SLy4 EOS~\citep{Douchin2001}. Altogether, this set of EOSs
spans a wide range of physical properties and should cover any
realistic description of neutron stars. For the Pol2 EOS, we
have explored systematically the space of solutions and computed the
corresponding surface deformation and quadrupole distortion. In
addition, we have also computed the distortion coefficients, which
allow one to compute the deformation of neutron stars up to large
magnetizations and rotation rates through a simple algebraic
expression, following the procedure devised in~\citet{Cutler2002}.

The plan of the paper is the following. In Section~\ref{se:overview},
we give an overview of the novel results obtained in this study, in
Section~\ref{se:theoretical}, we discuss the theoretical framework on
which our approach, whose numerical implementation and testing is
discussed in Section~\ref{se:numerical}, is based. Results for static
magnetized models are presented in Section~\ref{se:statmag} and for
rotating magnetized models in Section~\ref{se:rotmag}. In
Section~\ref{se:distcoef}, we deal with the case of moderate magnetic
field and rotation and derive empirical distortion coefficients before
presenting our conclusions in Section~\ref{se:conclusions}.
%

\section{General Overview}
\label{se:overview}

Given the complexity of the numerous results found and the risk that
the most important ones may be lost in the details, in the following
we briefly summarize what we believe are the most salient properties
of (relativistic) stars with purely toroidal magnetic fields. We
recall that we measure with $\epsilon_\rmn{s}$ the deformation of the
surface shape, while we measure with $\epsilon$ the (quadrupolar)
deviation of the matter distribution from a spherically symmetric
one. Overall, equilibrium models of relativistic stars with a toroidal
magnetic field exhibit the following properties:

\begin{enumerate}

\item \textit{Non-rotating} and \textit{magnetized} models exhibit a
  prolate surface deformation and a prolate quadrupolar deformation,
  \ie $\epsilon_\rmn{s} < 0$, $\epsilon < 0$, both of which decrease
  as the magnetization parameter $\lambda_0$ is increased.

\item \textit{Rotating} and \textit{unmagnetized} models exhibit an
  oblate surface deformation and an oblate quadrupolar deformation,
  \ie $\epsilon_\rmn{s} > 0$, $\epsilon > 0$, both of which increase
  as the rotation frequency is increased.

\item Between these limiting cases, neutral lines $\epsilon_\rmn{s} =
  0$ and $\epsilon = 0$ divide models into having prolate/oblate
  surface deformations and prolate/oblate internal deformations,
  respectively.

\item For \textit{non-rotating} models no upper limit was found
  to the magnetization parameter $\lambda_{\rm 0}$, with stellar
  models that become increasingly prolate, but also increasingly
  extended as the magnetization is increased (see Figs~\ref{f:pol2xxl}
  and \ref{f:aprxxl}).

\item The magnetic pressure associated with the toroidal magnetic
  field also causes an expansion in the outer layers of the star, in
  particular a growth of its equatorial radius. This effect is present
  also for non-rotating models, for which, however, the polar radius
  grows more rapidly than the equatorial one, yielding a prolate
  surface deformation, \ie $\epsilon_\rmn{s} < 0$. However, as the
  rotation is increased and the magnetization decreased, models can be
  found which have a prolate interior deformation, \ie $\epsilon < 0$,
  and an oblate surface deformation, \ie $\epsilon_\rmn{s} > 0$.

\item For any given angular velocity, the model of maximum
  magnetization coincides with the mass-shedding model, \ie the model
  for which centrifugal and gravitational forces are equal at the
  equatorial radius. This point is characterized by the appearance of
  a cusp at the equator.

\item For \textit{rotating} models the magnetic pressure at the
  equator adds to the centrifugal force, favouring the loss of
  mass. As a result of the increase in the equatorial size, the
  mass-shedding frequency is systematically smaller than the
  corresponding one for unmagnetized models.

\item In the space of parameters considered, the average toroidal
  magnetic field strength $\langle B^{2}\rangle^{1/2}$ is not a
  monotonic function of the intrinsic magnetization parameter
  $\lambda_\rmn{0}$, and after reaching a maximum value, it gradually
  decreases. This implies the existence of double solutions in certain
  parts of the space of parameters $(\mathit{\Omega}^2,\langle B^{2}\rangle)$.

\item Although the space of parameters $(\mathit{\Omega}^2,\langle
  B^{2}\rangle)$ is potentially degenerate, the corresponding space of
  parameters $(\mathit{\Omega}^2,\lambda^2_0)$ is not. As a result, any stellar
  model considered is characterized uniquely by the values of the
  angular velocity $\mathit{\Omega}$ and of the magnetization parameter
  $\lambda_\rmn{0}$.

\end{enumerate}

\begin{figure}
\begin{center}
\includegraphics[width=7.5cm]{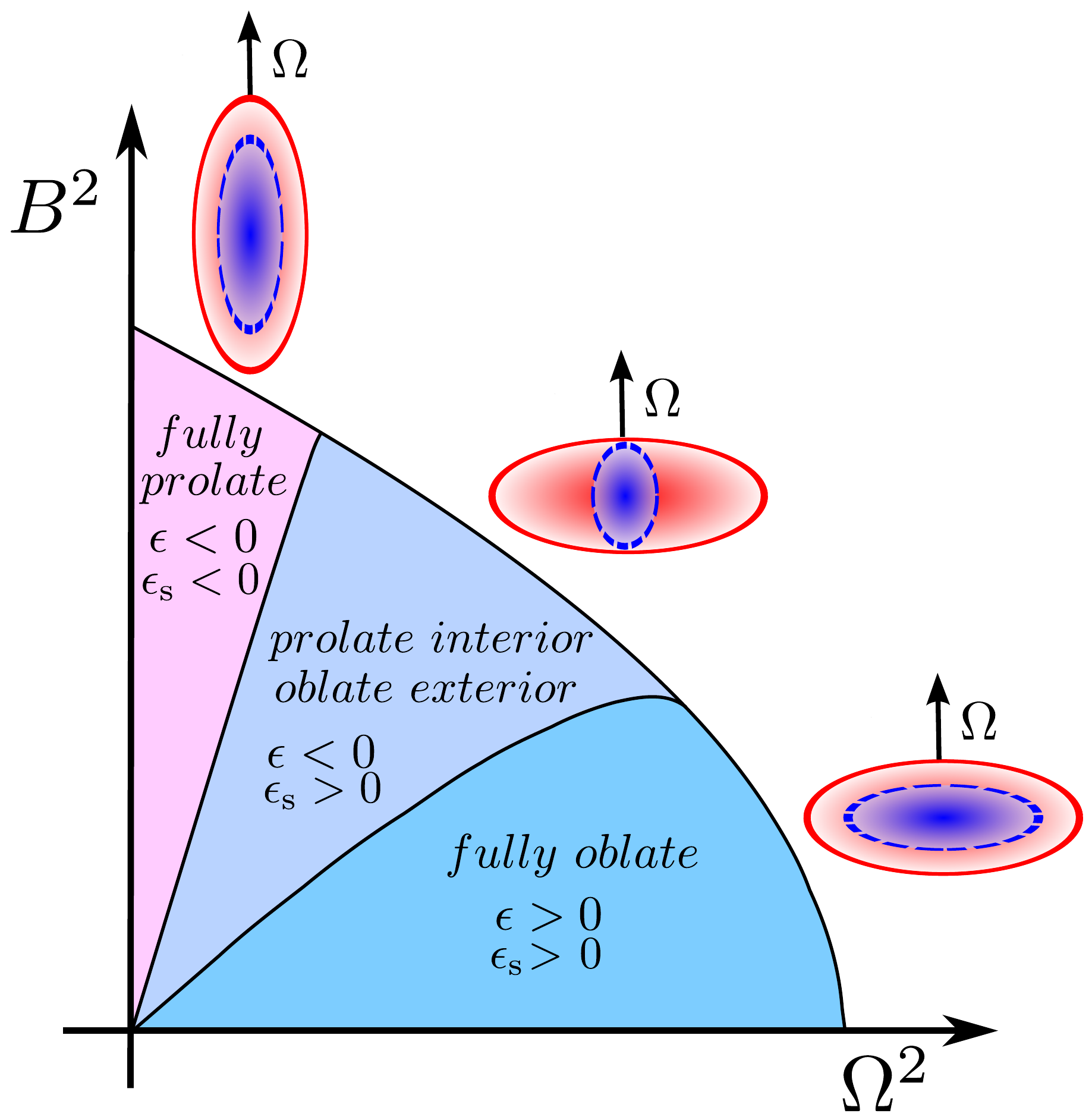}
\caption{Schematic diagram showing the lower part of the solution
  space of equilibrium models in the $(\mathit{\Omega}^2,\langle B^{2}\rangle)$
  plane from the unmagnetized limit up to the maximum field strength
  limit.
  According to the relative strength of the magnetic field and
  of the rotation rate, different combinations of the surface
  deformation, $\epsilon_{\rm s}$, and of the quadrupole deformation,
  $\epsilon$, are possible.}
\end{center}
\label{f:cartoon}
\end{figure}

Given these results, it is natural to divide our models into three
classes: (1) models labelled \texttt{PP} for prolate--prolate, for
which both apparent shape and matter distribution are prolate, \ie
$\epsilon_\rmn{s} < 0$, $\epsilon < 0$; (2) models labelled
\texttt{PO} for prolate--oblate, whose shape is oblate whereas their
matter distribution is prolate, \ie $\epsilon_\rmn{s} > 0$, $\epsilon
< 0$; (3) models labelled \texttt{OO} for oblate--oblate, which appear
oblate and also exhibit an oblate matter distribution, \ie
$\epsilon_\rmn{s} > 0$, $\epsilon > 0$. The latter class of models had
not been found by~\citet{Kiuchi2008}. A schematic picture
illustrating the three different classes for models below the
maximum field strength limit is shown in Fig.~\ref{f:cartoon}.
Finally, while different EOSs with their different stiffness introduce
quantitative differences in the behaviour described above, they all
follow the same qualitative behaviour.

\section{Mathematical Setup}
\label{se:theoretical}

\subsection{Basic assumptions}
\label{ss:basic}

We assume that the space--time generated by the rotating star is {\em
  stationary\/} and {\em axisymmetric\/}, with Killing vector fields
$\bmath{e}_{0}$ and $\bmath{e}_{3}$ associated with these
symmetries. If, in addition, the space--time is asymptotically flat and
there exists an axis where $\bmath{e}_{3}$ vanishes, then
$\bmath{e}_{0}$ and $\bmath{e}_{3}$ commute~\citep{Carter1970}. This
enables us to choose coordinates $(x^{\alpha}) = (t,r,\theta,\phi)$
with vector fields $\bmath{e}_{0}=\upartial/\upartial t$ and
$\bmath{e}_{3}=\upartial/\upartial \phi$. If furthermore the total
stress--energy tensor $\bmath{T}$ satisfies the {\em circularity\/}
conditions
\begin{equation}
\label{EQ:CIR1}
  \bmath{T} \cdot \bmath{e}_{0} = \alpha \bmath{e}_{0} +
    \beta \bmath{e}_{3} \,,
\end{equation}

\begin{equation}
     \label{EQ:CIR2}
  \bmath{T} \cdot \bmath{e}_{3} = \lambda \bmath{e}_{0} +
    \mu \bmath{e}_{3}\,,
\end{equation}
convective currents in the meridional planes of constant $(t,\phi)$
are absent by construction. In this case, quasi-isotropic (QI)
coordinates can be adopted and the line element reads
\begin{equation}
\label{EQ:GAB}
\begin{array}{ll}
  \mathrm{d}s^2 = g_{\alpha\beta} \, \rmn{d}x^{\alpha}
  \rmn{d}x^{\beta} = & \!\!\!\! - N^{2} \rmn{d}t^{2} +
  \mathit{\Phi}^{2} r^{2} \sin^{2} \theta^{2} ( \rmn{d}\phi -
  N^{\phi} \rmn{d}t )^{2} \\
  & \!\!\!\! + \mathit{\Psi}^{2} ( \rmn{d}r^{2} + r^{2}
  \rmn{d}\theta^{2} ) \,,
\end{array}
\end{equation}
where $N$, $N^{\phi}$, $\mathit{\Psi}$, and $\mathit{\Phi}$ are functions of
$(r,\theta)$. We further introduce the Eulerian observer
${\cal{O}}_{0}$ whose 4-velocity $\bmath{n}$ is the future-directed
unit vector normal to hypersurfaces of constant $t$. From
equation~(\ref{EQ:GAB}), we infer $n_{\alpha}=(-N,0,0,0)$.

The compatibility of the electromagnetic fields with the circularity
condition was established long ago~\citep{Carter1973} for the case in
which the electromagnetic field tensor $\bmath{F}$ is derived from a
potential 1-form $\bmath{A}$ with components $(A_{t}, 0, 0,
A_{\phi})$, whereas the contravariant components of the electric
current vector $\bmath{j}$ read $j^{\alpha}=(j^{t}, 0, 0,
j^{\phi})$. This fact has been exploited for computing fully
relativistic models of stars with a poloidal electromagnetic
field~\citep{Bocquet1995,Cardall2001}. However, it has been shown that
the case of a toroidal magnetic field satisfies the circularity
assumption too~\citep{Oron2002,Kiuchi2008}. In the following, we
adopt a vector potential which is orthogonal to the Killing vectors
$\bmath{e}_{0}$ and $\bmath{e}_{3}$, namely
$\bmath{A}\cdot\bmath{e}_{0}=0$ and
$\bmath{A}\cdot\bmath{e}_{3}=0$. The covariant components of
$\bmath{A}$ then become
\begin{equation} 
\label{EQ:BA1} 
A_{\alpha} = (0, A_{r}, A_{\theta},0) \,.
\end{equation}
This particular form of $\bmath{A}$ ensures, by construction, the
absence of any poloidal electric or magnetic field component. The only
non-vanishing component of the antisymmetric Faraday tensor
$F_{\alpha\beta} = A_{\beta,\alpha} - A_{\alpha,\beta}$ is then given
by
\begin{equation} 
\label{EQ:BA2} 
F_{r\theta} = \spar{A_\theta}{r}
   - \spar{A_r}{\theta} \,.
\end{equation}
For the {\em electric field\/} $\bmath{E}$ and the {\em magnetic field\/}
$\bmath{B}$ as measured by observer ${\cal{O}}_{0}$, we then obtain
\begin{equation} 
\label{EQ:EB1}  
   E_{\alpha} = F_{\alpha\beta} n^{\beta}  = (0, 0, 0, 0) \,,
\end{equation}
\begin{equation}
\label{EQ:EB2}
\begin{array}{ll}
    \!\!\! B_{\alpha} & \hspace*{-0.85em} = - \fras{1}{2}
    \eta_{\alpha\beta\gamma\delta}
    F^{\gamma\delta} n^{\beta} \\[0.85ex]
    & \hspace*{-0.85em} \displaystyle =
    \frac{\mathit{\Phi} \sin\theta}{\mathit{\Psi}^2}
    \left( N^{\phi} \left[\spar{A_r}{\theta} - \spar{A_\theta}{r} \right],
    0, 0, - \left[\spar{A_r}{\theta} - \spar{A_\theta}{r} \right]
    \right) \,,
\end{array}
\end{equation}
where $\eta_{\alpha\beta\gamma\delta}$ is the totally antisymmetric
tensor associated with the metric $\bmath{g}$. According to
equations~(\ref{EQ:EB1}) and (\ref{EQ:EB2}), the combination of a vanishing
electric field and a toroidal magnetic field holds for {\em any\/}
observer whose 4-velocity $\bmath{u}$ is a linear combination of the
two Killing vectors $\bmath{e}_{0}$ and $\bmath{e}_{3}$. The
electromagnetic contribution to the stress--energy tensor is given by
\begin{equation} 
\label{EQ:EB6} 
\bmath{\mathcal{T}}_{\alpha\beta} = \frac{1}{4\upi}
    \left(F_{\alpha\kappa} F^{\kappa}{}_{\beta} - \frac{1}{4}
    F_{\kappa\lambda} F^{\kappa\lambda} \, g_{\alpha\beta}\right) \,.
\end{equation}
Following the procedure adopted by \citet{Bonazzola1993},
$\bmath{\mathcal{T}}$ is split up into the total electromagnetic
energy density $\mathcal{E}$, the Poynting 3-vector
$\bmath{\mathcal{J}}$, and the electromagnetic stress 3-tensor
$\bmath{\mathcal{S}}$ as measured by the Eulerian observer
${\cal{O}}_{0}$. With the projection tensor $\bmath{h}=\bmath{g} +
\bmath{n}\otimes\bmath{n}$, these quantities are obtained as
projections of $\bmath{\mathcal{T}}$ on to and orthogonal to
$\bmath{n}$, namely $\mathcal{E} \equiv
\bmath{n}\cdot\bmath{\mathcal{T}}\cdot\bmath{n}$, $\bmath{\mathcal{J}}
\equiv -\bmath{h} \cdot \bmath{\mathcal{T}} \cdot \bmath{n}$, and
$\bmath{\mathcal{S}} \equiv \bmath{h} \cdot \bmath{\mathcal{T}} \cdot
\bmath{h}$. Specialized to the present case of no electric field and
no poloidal magnetic field components, the electromagnetic
contribution to the (3+1) matter variables [for an introduction to the
(3+1) formalism of general relativity,
\cf~\citealt{Smarr1978,Gourgoulhon2012a}], namely the total energy
density $E\equiv\bmath{n} \cdot \bmath{T} \cdot \bmath{n}$, the
momentum density 3-vector $\bmath{J}\equiv-\bmath{h} \cdot \bmath{T}
\cdot \bmath{n}$, and the stress 3-tensor $\bmath{S}\equiv\bmath{h}
\cdot \bmath{T} \cdot \bmath{h}$ reads
\begin{equation} \label{EQ:EB7A} \mathcal{E} =
    \frac{1}{8 \upi} \left(
    \frac{B_{\phi}}{\mathit{\Phi} \, r\sin\theta} \right)^{2} \,,
\end{equation}
\begin{equation}
    \label{EQ:EB7B}
    \mathcal{J}_{i} = 0 \,,
\end{equation}
\begin{equation}
    \label{EQ:EB7C}
    \mathcal{S}^{r}{}_{r} = \mathcal{E} \,, \quad
    \mathcal{S}^{\theta}{}_{\theta} =
    \mathcal{E} \,, \quad
    \mathcal{S}^{\phi}{}_{\phi} = -\mathcal{E} \,.
\end{equation}
In particular, all non-diagonal components of $\mathcal{S}^{i}{}_{j}$
are zero, and $\mathcal{S} = \mathcal{E}$. The circularity assumption
requires that for the Poynting vector $\bmath{\mathcal{J}}$
\begin{equation} 
\label{EQ:EB8} 
\mathcal{J}_{r} = 0\,, \qquad
    \mathcal{J}_{\theta} = 0 \,,
\end{equation}
and that the electromagnetic stress tensor $\bmath{\mathcal{S}}$
satisfies
\begin{equation} 
\label{EQ:EB9} \mathcal{S}_{r\phi} = 0\,, \qquad
    \mathcal{S}_{\theta\phi} = 0 \,.
\end{equation}

\subsection{Einstein equations}
\label{ss:graveq}
The Einstein equations for the metric tensor $\bmath{g}$ defined by
equation~(\ref{EQ:GAB}) and a general stress--energy tensor $\bmath{T}$
decomposed into the (3+1) quantities $E$, $\bmath{J}$ and $\bmath{S}$
become
\begin{equation} 
\label{EQ:EE1A}
  \Delta\nu = 4\upi \mathit{\Psi}^{2} (E + S) + \frac{\mathit{\Phi}^{2}r^{2}\sin^{2}\theta}{2N^{2}}
    (\upartial N^{\phi})^{2} - \upartial\nu \upartial(\nu + \beta) \,,
\end{equation}
\begin{equation}
  \label{EQ:EE1B}
  \tilde{\Delta}_{3} (N^{\phi}r\sin\theta) =
    - 16 \upi \frac{N \mathit{\Psi}^{2}}{\mathit{\Phi}^{2}} \frac{J_{\phi}}{r\sin\theta}
    - r\sin\theta \, \upartial N^{\phi} \upartial(3 \beta - \nu ) \,,
\end{equation}
\begin{equation}
  \label{EQ:EE1C}
  \Delta_{2}[(N\mathit{\Phi} - 1) \, r\sin\theta] = 8 \upi N \mathit{\Psi}^{2} \mathit{\Phi} \, (S^{r}{}_{r}
    + S^{\theta}{}_{\theta}) \, r\sin\theta \,,
\end{equation}
\begin{equation}
  \label{EQ:EE1D}
  \Delta_{2}\zeta = 8 \upi \mathit{\Psi}^{2} S^{\phi}{}_{\phi}
    + \frac{3 \mathit{\Phi}^{2}r^{2}\sin^{2}\theta}{4 N^{2}}
    (\upartial N^{\phi})^{2} - (\upartial\nu)^{2} \,, \quad
\end{equation}
where the following abbreviations have been used:
\begin{equation} 
\label{EQ:EE2} 
\nu \equiv \ln N \,, \quad \zeta \equiv \ln (N\mathit{\Psi}) \,, \quad
   \beta \equiv \ln \mathit{\Phi} \,,
\end{equation}
\begin{equation}
   \Delta_{2} \equiv \dpar{}{r} + \frac{1}{r} \spar{}{r} + \frac{1}{r^{2}}
   \dpar{}{\theta} \,,
\end{equation}
\begin{equation}
   \Delta_{3} \equiv \dpar{}{r} + \frac{2}{r} \spar{}{r} + \frac{1}{r^{2}}
   \dpar{}{\theta} + \frac{1}{r^{2}\tan\theta}\spar{}{\theta} \,,
\end{equation}
\begin{equation}
   \tilde{\Delta}_{3} \equiv \Delta_{3} -
      \frac{1}{r^{2}\sin^{2}\theta} \,,
\end{equation}
\begin{equation}
   \upartial a \, \upartial b \equiv \spar{a}{r} \spar{b}{r} + \frac{1}{r^{2}}
      \spar{a}{\theta} \spar{b}{\theta} \,.
\end{equation}
The resulting system of four non-linear elliptic equations for the
metric variables $N$, $N^{\phi}$, $\mathit{\Psi}$ and $\mathit{\Phi}$
can be solved iteratively once suitable boundary conditions of
asymptotic flatness have been adopted. Additional details can be
found in~\citet{Bonazzola1993}.

\subsection{Maxwell equations}
\label{ss:maxweq}
Since the electromagnetic field tensor $\bmath{F}$ is derived from
a potential 1-form $\bmath{A}$, the homogeneous Maxwell equations
\begin{equation} \label{EQ:EB10}
   F_{\alpha\beta;\gamma} + F_{\beta\gamma;\alpha} +
   F_{\gamma\alpha;\beta} = 0
\end{equation}
are satisfied by construction. The inhomogeneous Maxwell equations
$F^{\alpha\beta}{}_{;\beta} = 4\upi j^{\alpha}$ allow us to express
the electric current 4-vector $\bmath{j}$ in terms of the Faraday
tensor $\bmath{F}$, where the alternative expression
\begin{equation} \label{EQ:EB11} 4 \upi j^{\alpha} = \frac{1}{\sqrt{-g}}
   \left( \sqrt{-g} F^{\alpha\beta}\right)_{,\beta}
\end{equation}
with $\sqrt{-g} = \mathit{\Psi}^2 \mathit{\Phi} N r^2 \sin\theta$ is used. The
electromagnetic field tensor $\bmath{F}$ has only one non-vanishing
contra- and covariant component, which can now be expressed in terms
of the azimuthal component $B_{\phi}$ of the magnetic field
$\bmath{B}$,
\begin{equation} 
\label{EQ:EB12} 
   F_{r\theta} = \frac{\mathit{\Psi}^2}{\mathit{\Phi}\sin\theta}
   B_{\phi}\,,
\end{equation}
\begin{equation}
   F^{r\theta} = \frac{B_{\phi}}{\mathit{\Psi}^2 \mathit{\Phi} r^2\sin\theta} \,.
\end{equation}
Combining equations~(\ref{EQ:EB11}) and (\ref{EQ:EB12}), the poloidal
components of the electric current 4-vector $\bmath{j}$ can be written
as
\begin{equation} 
\label{EQ:EB13A} 
   4 \upi j^{r} = \frac{1}{\mathit{\Psi}^2 \mathit{\Phi} N r^2 \sin\theta}
   \spar{(N B_{\phi})}{\theta} \,,
\end{equation}
\begin{equation}
   \label{EQ:EB13B}
   4 \upi j^{\theta} = - \frac{1}{\mathit{\Psi}^2 \mathit{\Phi} N r^2 \sin\theta}
   \spar{(N B_{\phi})}{r} \,.
\end{equation}
The remaining components $j^{t}$ and $j^{\phi}$ are zero, as expected.
Note that the circularity condition, equations~(\ref{EQ:CIR1}) and
(\ref{EQ:CIR2}), forbids any meridional convective current
contributing to $\bmath{J}$, which is ensured by assuming the charge
carriers to be {\em massless\/}. By taking the divergence of
equation~(\ref{EQ:EB11}), it follows that $j^{\alpha}{}_{;\alpha}=0$ thanks
to the antisymmetry of $\bmath{F}$, and this continuity equation for
the electric current leads to a dependence between $j^{r}$ and
$j^{\theta}$, namely
\begin{equation} 
\label{EQ:EB14} 
\bmath{\nabla} \cdot \bmath{j} =
    \frac{1}{\sqrt{-g}} \left[ \spar{}{r} (\sqrt{-g} j^{r}) + 
    \spar{}{\theta} (\sqrt{-g} j^{\theta}) \right] = 0 \,,
\end{equation}
which however is trivially fulfilled by virtue of
equations~(\ref{EQ:EB13A}) and (\ref{EQ:EB13B}) which express $j^{r}$ and
$j^{\theta}$ as functions of a single quantity, namely $N B_{\phi}$,
without any restriction from equation~(\ref{EQ:EB14}). A useful consequence
of equations~(\ref{EQ:EB13A}) and (\ref{EQ:EB13B}) is that the flow lines
of the electric current coincide with the isocontours of $N B_{\phi}$,
which allows one to visualize the current distribution without
actually computing $\bmath{j}$.
\subsection{Equation of motion and condition on the Lorentz force}
\label{ss:moteq}

In the case of a perfect fluid the stress--energy tensor $\bmath{T}$
takes the form
\begin{equation} 
\label{EQ:EM00} 
\bmath{T} = (e + p) \, \bmath{u} \otimes
    \bmath{u} + p \, \bmath{g}\,,
\end{equation}
where $e$ is the energy density and $p$ the pressure as measured by
the {\em fluid comoving observer\/} $\mathcal{O}_{1}$ with 4-velocity
$\bmath{u}$. Since we assume the absence of meridional currents,
$\bmath{u}$ can be written as a linear combination of the Killing
vectors $\bmath{e}_{0}$ and $\bmath{e}_{3}$, namely
\begin{equation} 
\label{EQ:EM01} 
\bmath{u} = u^{t} \bmath{e}_{0} + u^{\phi}
   \bmath{e}_{3}\,.
\end{equation}
Introducing the Lorentz factor $\mathit{\Gamma} \equiv -\bmath{n}\cdot\bmath{u}
= N u^{t}$ linking obser\-vers $\mathcal{O}_{0}$ and $\mathcal{O}_{1}$
and the fluid coordinate angular velocity $\mathit{\Omega}\equiv
u^{\phi}/u^{t}$, the physical fluid velocity $U$ and the Lorentz
factor $\mathit{\Gamma}$ can be expressed, respectively, as
\begin{equation}
\label{EQ:EM02}
   U = \frac{\mathit{\Phi} r\sin\theta}{N}(\mathit{\Omega} - N^{\phi}) \,,
\end{equation}
\begin{equation}
   \mathit{\Gamma} = (1 - U^{2})^{1/2} \,.
\end{equation}
Taking into account the contribution of the magnetic field to the
(3+1) matter variables according to
equations~(\ref{EQ:EB7A})--(\ref{EQ:EB7C}), the following expressions are
obtained for a perfect fluid endowed with a toroidal magnetic field:
\begin{equation}
\label{EQ:EM03}
   E = \mathit{\Gamma}^{2}(e + p) - p + \frac{1}{8 \upi} \left(
    \frac{B_{\phi}}{\mathit{\Phi} r\sin\theta} \right)^{2} \,,
\end{equation}
\begin{equation}
  \label{EQ:EM03B}
   J_{\phi} = \mathit{\Gamma}^{2} (e + p) \mathit{\Phi} r\sin\theta \, U \,,
\end{equation}
\begin{equation}
  \label{EQ:EM03C}
   S^{r}{}_{r} = S^{\theta}{}_{\theta} = p + \frac{1}{8 \upi} \left(
    \frac{B_{\phi}}{\mathit{\Phi} r\sin\theta} \right)^{2} \,,
\end{equation}
\begin{equation}
  \label{EQ:EM03D}
   S^{\phi}{}_{\phi} = p + \mathit{\Gamma}^{2}(e + p) U^{2} - \frac{1}{8 \upi} \left(
    \frac{B_{\phi}}{\mathit{\Phi} r\sin\theta} \right)^{2} \,.
\end{equation}
All other components are zero, and $S = 3 p + \mathit{\Gamma}^{2}(e + p) U^{2}
+ 1/(8\upi) [B_{\phi}/(\mathit{\Phi} r\sin\theta)]^{2}$. The projection of
$\bmath{\nabla} \cdot \bmath{T} = \bmath{0}$ on to spatial hypersurfaces of
constant ${t}$ provides the equation of motion
\begin{equation} 
\label{EQ:EM04} 
\frac{1}{e + p} \spar{p}{x^i} + \spar{\nu}{x^i}
  - \spar{}{x^i}(\ln\mathit{\Gamma}) - \frac{1}{e + p} F_{i\beta} j^{\beta} =
  - F \spar{\mathit{\Omega}}{x^i} \,,
\end{equation}
where $F$ is defined as
\begin{equation} 
\label{EQ:EM05} 
F \equiv u_{\phi} u^{t} = \mathit{\Gamma}^{2} \frac{\mathit{\Phi}}{N}
  U r\sin\theta \,.
\end{equation}
Assuming a single-parameter EOS, $e=e(n)$ and $p=p(n)$, with $n$ the
baryon number density, and zero temperature, we introduce the fluid
log-enthalpy
\begin{equation}
\label{EQ:EM06}
H \equiv
   \ln\left(\frac{e + p}{n \, m_{\rmn{b}}}\right) \,,
\end{equation}
with a mean baryon rest mass of $m_{\rmn{b}} = 1.66 \times 10^{-27} \,
\rmn{kg}$. Using equations~(\ref{EQ:EB12})--(\ref{EQ:EB13B}),
(\ref{EQ:EM06}), and discarding the case of differential rotation,
equation~(\ref{EQ:EM04}) becomes
\begin{equation}
\label{EQ:EM1}
\begin{array}{l} \displaystyle
\spar{}{x^i}(H + \nu - \ln\mathit{\Gamma}) +
\left(\frac{1}{e + p}\right) 
\left(\frac{1}{8 \upi \mathit{\Phi}^2 N^2 r^2\sin^2\theta}\right) \\
    \displaystyle \quad
    \times \spar{(N B_{\phi})^2}{x^{i}} = 0 \,.
\end{array}
\end{equation}
Equation~(\ref{EQ:EM1}) can only be satisfied if the Lorentz force
term is derived from a scalar function $\tilde{M}(r,\theta)$ as
well, thus we require
\begin{equation}
\label{EQ:EM2}
\left(\frac{1}{e + p}\right)
\left(\frac{1}{8 \upi \mathit{\Phi}^2 N^2 r^2\sin^2\theta}\right) 
\spar{(N B_{\phi})^2}{x^{i}}
   = \spar{\tilde{M}}{x^{i}} \,.
\end{equation}
Using the Schwartz theorem, the integrability condition of 
equation~(\ref{EQ:EM2}) can be written as
\begin{equation}
\label{EQ:EM3}
\spar{G}{\theta} \spar{(N B_{\phi})}{r}
    - \spar{G}{r} \spar{(N B_{\phi})}{\theta} = 0 \,,
\end{equation}
where $G \equiv (e + p) \mathit{\Phi}^2 N^2 r^2 \sin^2\theta$. In other words,
equation~(\ref{EQ:EM3}) states that the Jacobian of the coordinate
transform $(r,\theta) \rightarrow (G,NB_{\phi})$ is zero, and hence
there exists a scalar function $\mathit{\Theta} : {\bmath{R}}^2 \rightarrow
{\bmath{R}}$ such that $\mathit{\Theta}(G,NB_{\phi})=0$. Two different cases
can be distinguished. (i) If $\upartial\mathit{\Theta}/\upartial (N
B_\phi)=0$, then $\mathit{\Theta}$ does not depend on $NB_\phi$, and the
relation $\mathit{\Theta}(G, N B_\phi)=0$ implies that $G$ is constant, which
is equivalent to the absence of any matter and $G = \rmn{0}$.
(ii) If $\upartial\mathit{\Theta}/\upartial (NB_\phi) \neq 0$, then the
implicit-function theorem enables us to write $NB_\phi =
NB_\phi(G)$. Discarding the case without matter, we retain possibility
(ii) and conclude that
\begin{equation} 
\label{EQ:EM4} 
N B_{\phi} =
    g((e + p) \mathit{\Phi}^2 N^2 r^2 \sin^2\theta) \,,
\end{equation}
with $g$ being an arbitrary scalar function. However, the regularity
properties of an axisymmetric vector field in the case of spatial
spherical coordinates $(r,\theta,\phi)$ impose some further
restrictions on $g$. For the covariant component $U_{\phi}$ of an
azimuthal vector field $\bmath{U} = U^{\phi} \bmath{e}_{3}$, it can be
shown~\citep{Bardeen1983} that
\begin{equation} 
\label{EQ:EM5} 
U_{\phi}(r,\theta) = r^2 \sin^2\theta \, m(r,\theta)\,,
\end{equation}
where $m$ is an arbitrary axisymmetric scalar function. Application of
equation~(\ref{EQ:EM5}) to $N B_{\phi}$ allows us to conclude that $g$
can be written as $g(x) \! = \! x f(x)$, with an arbitrary regular
scalar function $f$. The resulting expression for $N B_{\phi}$ after
application of equation~(\ref{EQ:EM5}) to equation~(\ref{EQ:EM4})
reads
\begin{equation} 
\label{EQ:EM6} 
N B_{\phi} = (e \! + \! p) \, \mathit{\Phi}^2 N^2 r^2
    \sin^2\theta f((e \! + \! p) \, \mathit{\Phi}^2 N^2 r^2
    \sin^2\theta) \,.
\end{equation}
By inserting equation~(\ref{EQ:EM6}) into equation~(\ref{EQ:EM2}),
the gradient of the magnetic potential $\tilde{M}$ reads
\begin{equation} 
\label{EQ:EM7} 
\spar{\tilde{M}}{x^{i}} = \frac{f}{4 \upi}
    \spar{}{x^{i}} [ (e \! + \! p) \, \mathit{\Phi}^2 N^2 r^2 \sin^2 
    \theta \, f] = R_{i} \,.
\end{equation}
In general, $\tilde{M}$ cannot be determined in closed form. However,
inspection of equation~(\ref{EQ:EM7}) reveals that for the sub-case
of a monomial function $f(x) \! = \! \lambda_{m} x^{m}$ (not to be
meant as a summation over repeated indices), a solution can be
written down immediately. In this case, equation~(\ref{EQ:EM6})
simplifies to
\begin{equation} 
\label{EQ:EM8} 
N B_{\phi} = \lambda_{m} \, ((e \! + \! p) \,
    \mathit{\Phi}^2 N^2 r^2 \sin^2\theta)^{m+1} \,,
\end{equation}
and the solution to equation~(\ref{EQ:EM7}) is given by an algebraic
expression, namely
\begin{equation} 
\label{EQ:EM9} 
\tilde{M} = \frac{\lambda^{2}_{m}}{4 \upi}
    \left(\frac{m + 1}{2m + 1}\right)
    ((e \! + \! p) \, \mathit{\Phi}^2 N^2 r^2 \sin^2\theta)^{2m + 1} \,.
\end{equation}
The simplest function $f$ is obtained for $m=0$ and corresponds to
a constant function of value $\lambda_{0}$. In this case,
equation~(\ref{EQ:EM6}) reduces to
\begin{equation} 
\label{EQ:EM10_0} 
    N B_{\phi} = \lambda_{0} \, (e \! + \! p) \,
    \mathit{\Phi}^2 N^2 r^2 \sin^2\theta \,, 
\end{equation}
and thus
\begin{equation} 
\label{EQ:EM10} 
    B_{\phi} = \lambda_{0} \, (e \! + \! p) \, \mathit{\Phi}^2 N r^2
    \sin^2 \theta \,,
\end{equation}
supplemented by equation~(\ref{EQ:EM9}), which yields the simplified
expression
\begin{equation} 
\label{EQ:EM12} 
\tilde{M} = \frac{\lambda^2_{0}}{4 \upi}
    (e \! + \! p) \, \mathit{\Phi}^2 N^2 r^2 \sin^2\theta \,.
\end{equation}
For all models computed in this study, the magnetic field component
$B_{\phi}$ is given by equation~(\ref{EQ:EM10}) and the magnetic potential
$\tilde{M}$ by equation~(\ref{EQ:EM12}). Because $\tilde{M}$ vanishes on
the axis of symmetry, $\tilde{M}_\mathrm{c} = 0$, and the integral of
equation~(\ref{EQ:EM1}) reads
\begin{equation} 
\label{EQ:EM13} 
H + \nu - \ln\mathit{\Gamma} + \tilde{M} = H_{\rmn{c}} +
    \nu_{\rmn{c}} \,,
\end{equation}
where $H_{\rmn{c}}$ and $\nu_{\rmn{c}}$ are the central values of $H$
and $\nu$, respectively. Note that a general prescription for the
determination of the magnetic field $B_{\phi}$ and the magnetic
potential $\tilde{M}$ can be found in \citet{Kiuchi2008}
and~\citet{Gourgoulhon2012}. 

\subsection{Perfect-conductor relation}
\label{ss:perfcon}

According to Ohm's law and assuming an infinite conductivity for the
neutron star matter, the electric field $\bmath{E}'$ as measured by
the fluid comoving observer ${\cal{O}}_{1}$ has to vanish, namely
\begin{equation} 
\label{EQ:EB4} 
 E_{\alpha}'  =  F_{\alpha\beta} u^{\beta}  =
    (0,0,0,0) \,.
\end{equation}
In the case of a purely poloidal magnetic field~\citep{Bocquet1995},
the perfect-conductor relation equation~(\ref{EQ:EB4}) is non-trivial
as it establishes a dependence between $A_t$ and $A_\phi$ and forces
a dependence of the angular velocity on $A_\phi$, \ie
$\mathit{\Omega}=\mathit{\Omega}(A_\phi)$. In the case of a purely
toroidal magnetic field, on the other hand, the perfect-conductor
condition is trivially satisfied and no condition is set on the
rotation law, so that differentially rotating models can be built
as in the unmagnetized case.

It is thus possible to consider differentially rotating configurations
following the procedure for the unmagnetized case. The magnetic
field $\bmath{B}'$ as measured by the fluid comoving observer
${\cal{O}}_{1}$ now becomes
\begin{equation}
\begin{array}{ll}
\label{EQ:EB5}
B'_{\alpha} \!\!\!\! & = \displaystyle
     - \fras{1}{2} \eta_{\alpha\beta\gamma\delta}
    F^{\gamma\delta} u^{\beta} \\[0.85ex]
    & = \displaystyle \frac{\mathit{\Gamma} \mathit{\Phi}\sin\theta}{\mathit{\Psi}^2}
    \left( \mathit{\Omega} \left[\spar{A_\theta}{r} -
    \spar{A_r}{\theta}\right], 0, 0, - \left[\spar{A_r}{\theta} -
    \spar{A_\theta}{r} \right] \right) \,.
\end{array}
\end{equation}
Note that $B'_{t} + \mathit{\Omega} B'_{\phi} = 0$ and that
$B'_{\phi} = \mathit{\Gamma} B_{\phi}$.

\subsection{Equation of state}
\label{ss:eos}

To solve the system of equations introduced in the previous sections,
we need a prescription relating the energy density $e$ and pressure
$p$ to the baryon number density $n$. The simplest of such relations
is offered by the polytropic EOS, in which case the following
identities hold:
\begin{equation} 
\label{EQU:POL1} 
   e(n) = m_{\rmn{b}} n +
  \kappa \frac{n^{\gamma}}{\gamma - 1} \,,
\end{equation}
\begin{equation}
\label{EQU:POL2} 
   p(n) = \kappa n^{\gamma} \,,
\end{equation}
where $\kappa$ is the polytropic constant and $\gamma$ the adiabatic
index. Combining equation~(\ref{EQ:EM06}), equations~(\ref{EQU:POL1})
and (\ref{EQU:POL2}), the log-enthalpy $H$ and $n$ can be expressed
as functions of each other, namely, as
\begin{equation}
\label{EQU:POL3} 
   H(n) = \ln\left(1 + \frac{\kappa}{m_{\rmn{b}}}
  \frac{\gamma}{\gamma - 1} n^{\gamma - 1}\right) \,,
\end{equation}

\begin{equation}
  \label{EQU:POL4} 
   n(H) = \left( \frac{\gamma - 1}{\gamma}
  \frac{m_{\rmn{b}}}{\kappa}(e^{H} - 1) \right) \,.
\end{equation}
Hereafter, we will assume $\gamma=2$ and refer to this EOS as to Pol2.
In addition to the Pol2 EOS, we have considered a sample of
seven single-constituent one-parameter EOSs treating the neutron star
matter as a perfect fluid and derived from very different models of
the ground state of cold (zero temperature) dense matter
in this work as they are provided by the \textsc{lorene} library.

More specifically, we have considered the APR EOS~\citep{Akmal1998},
the BBB2 EOS~\citep{Baldo1997}, the BN1H1 EOS \citep{Balberg1997}, the
BPAL12 EOS \citep{Bombaci1996}, the FPS EOS~\citep{Pandharipande1989},
the Sly4 EOS~\citep{Douchin2001}, and the GNH3
EOS~\citep{Glendenning1985}. All of these realistic EOSs are provided
in tabulated form, which requires the interpolation of listed
thermodynamical quantities $n$, $e$, and $p$ between contiguous
sampling points. Thermodynamical consistency of the interpolated
values is ensured by an interpolation procedure based on Hermite
polynomials introduced by~\citet{Swesty1996}, which is crucial for the
accuracy of the resulting numerical models~\citep{Salgado1994}.

The properties of the spherical non-rotating and unmagnetized reference
models with a gravitational mass of $M = 1.400 \, \mathrm{M}_{\sun}$ can be
found in Table~\ref{t:distcoef}. The softest EOS of this sample is the
BPAL12 EOS, which yields the smallest circumferential radius of
$R_\rmn{circ}=10.06 \, \rmn{km}$, whereas the stiffest one is the GNH3
EOS, which yields the largest circumferential radius of
$R_\rmn{circ}=14.20 \, \rmn{km}$.

\begin{table*}
\begin{tabular}{ccccccc}
\hline
                       & $\gamma$               & $2$        & $5/3$      & $3/2$     & $4/3$       & $9/7$ \\
\hline
Sood and Trehan (1972) & $(\epsilon/h)$ & $-0.21377$ & $-0.14284$ & $-0.09292$ & $-0.03374$ & $-0.01724$ \\
This work              & $(\epsilon_{\rmn{s}}/h)$ & $-0.21376$ & $-0.14218$ & $-0.09292$ & $-0.03374$ & $-0.01724$ \\
\hline\\
\end{tabular}
\caption{\label{t:polnewt} Comparison of results for the normalized
  oblateness $\epsilon_{\rmn{s}}/h$ computed in the Newtonian limit
  for various polytropic EOSs with results from~\citet{Sood1972}
  [values for $\epsilon/h$ from table~I of~\citet{Sood1972} where
  $\epsilon=(r_{\rmn{e}} - r_{\rmn{p}})/r_{0}$ and $r_{0}$ is the
  radius of the unperturbed model]. The perturbation parameter
  $h=\lambda^{2}_{0}/(4\upi^{2})$ is a measure of the strength of the
  magnetic field.}
\end{table*}

\subsection{Global quantities}
\label{ss:global}

A number of global quantities which characterize the numerical models
presented in this study can be computed and will be needed in the
course of this investigation. These are given as follows. The
gravitational mass:
\begin{equation}
\label{e:mgrav}
M \equiv \int N \mathit{\Psi}^{2} \mathit{\Phi} \left( E + S
  + \frac{2}{N} N^{\phi} J_{\phi} \right) r^{2} \sin\theta \, \rmn{d}r
  \, \rmn{d}\theta \, \rmn{d}\phi \,,
\end{equation}
the total angular momentum:
\begin{equation}
\label{e:jtot}
J \equiv \int \mathit{\Psi}^{2} \mathit{\Phi} \, J_{\phi} \, r^{2}
  \sin\theta \, \rmn{d}r \, \rmn{d}\theta \, \rmn{d}\phi \,,
\end{equation}
the total magnetic energy:
\begin{equation}
\label{e:em} 
\mathscr{M} \equiv \int \mathit{\Psi}^{2} \mathit{\Phi} \, \mathcal{E}
  r^{2}\sin\theta \, \rmn{d}r \, \rmn{d}\theta \, \rmn{d}\phi \,,
\end{equation}
the total kinetic energy:
\begin{equation} 
\label{e:tcin} 
T \equiv \fras{1}{2}\mathit{\Omega} J
\end{equation}
and the gravitational binding energy:
\begin{equation} 
\label{e:wgrav} 
W \equiv M - T - M_{\mathrm{p}} - \mathscr{M} \,,
\end{equation}
where the total proper mass $M_{\mathrm{p}}$ is defined as
\begin{equation} 
\label{e:mp} 
M_{\mathrm{p}} \equiv \int \mathit{\Psi}^{2} \mathit{\Phi} \,
  \mathit{\Gamma} e \, r^{2} \sin\theta \,
  \rmn{d}r \, \rmn{d}\theta \, \rmn{d}\phi \,,
\end{equation}
while the total baryon mass of the star is given by
\begin{equation} 
\label{e:mb} 
M_{\rmn{b}} \equiv \int \mathit{\Psi}^{2} \mathit{\Phi} \, \mathit{\Gamma} n \,
  r^{2} \sin\theta \, \rmn{d}r \, \rmn{d}\theta \, \rmn{d}\phi \,.
\end{equation}

Other important quantities that are more closely related to the
deformation of the star are the circumferential radius
$R_{\rmn{circ}}$, which is defined through the stellar equatorial
circumference as measured by the observer $\mathcal{O}_{0}$ and
is related to the equatorial coordinate radius $R$
according to
\begin{equation}
\label{e:rcirc}
R_{\rmn{circ}} \equiv \mathit{\Phi}(R,\upi/2) \,
  R \,,
\end{equation}
the surface deformation (or apparent oblateness)
\begin{equation}
\label{e:oblat} 
\epsilon_{\rmn{s}} \equiv r_{\rmn{e}}/r_{\rmn{p}} - 1 \,,
\end{equation}
where $r_{\rmn{e}}$, $r_{\rmn{p}}$ are the equatorial and polar
coordinate radii, respectively, and the quadrupole
distortion $\epsilon$ of the star~\citep{Bonazzola1996}:
\begin{equation}
\label{e:dist}
\epsilon \equiv -\frac{3}{2} \frac{\mathscr{I}_{zz}}{I} \,,
\end{equation}
where $\mathscr{I}_{zz}$ is the quadrupole moment measured in some
asymptotically Cartesian mass-centred coordinate
system~\citep{Thorne1980}, while the moment of inertia $I=I_{zz}$ is
defined as $I \equiv J/\mathit{\Omega}$. We stress that no moment of inertia
other than the latter can be defined in a meaningful way for
axisymmetric rotating bodies and that the rotation must be rigid and
about the axis of symmetry. Furthermore, we note that $-(3/2)
{\mathscr{I}_{zz}}$ differs from the standard quadrupole moment $Q$
and that in the case of QI coordinates, the latter can be read off
from the asymptotic expansion of $\ln N$ according to
\begin{equation} 
\label{e:quad} 
\ln N = - \frac{M}{r}
  + \frac{M^{2}}{12 r^{3}}
  + \frac{Q}{r^{3}} P_{2}(\cos\theta)
  + \cdots \,.
\end{equation}
Thorne's quadrupole moment $\mathscr{I}_{zz}$ is the relevant quantity
when the gravitational-wave emission from a distorted star has to be
determined, and the corresponding values are extracted from the
asymptotic expansion of certain components of the metric tensor
$\bmath{g}$ in QI coordinates following the procedure presented
in~\citet{Bonazzola1996}.

\section{Numerical Implementation and Tests}
\label{se:numerical}

\subsection{Numerical scheme}
\label{ss:numsch}

The analytic scheme presented in Section~\ref{se:theoretical} has been
implemented by extending an existing multidomain, surface-adaptive
spectral method for unmagnetized relativistic stars presented in
\citet*{Bonazzola1998},\citet{Gourgoulhon1999} and part of the
\textsc{lorene}\footnote{\url{http://www.lorene.obspm.fr}} C++ class
library for numerical relativity. We refer to~\citet{Bonazzola1998}
and \citet{Gourgoulhon1999} for a detailed description of
the numerical method and of the tests carried out. The numerical
solution is computed by iteration, starting from crude initial
conditions of a parabolic log-enthalpy profile, no rotation, flat
space and no magnetic field. Convergence is monitored by computing $||
H_{{n}} - H_{{n-1}} || / || H_{{n-1}} ||$ where $|| H_{{n-1}} ||$ is
the sum of the absolute values of the log-enthalpy $H$ over all
collocation points at iteration step $n-1$. A convergent solution is
assumed to be found when the normalized log-enthalpy residual
decreases below a prescribed threshold, which is usually chosen to be
of the order of $10^{-8}$ or smaller. The computational domain is
divided into a spherical nucleus containing the star and two shells
covering the exterior, where the last one maps on to a finite interval
the whole exterior space, from a certain radius $R$ up to spatial
infinity. The default number of collocation points used in our study
has been $n_{\theta}=17$ in the angular direction and $\bmath{n}_{R}=(33,33,17)$
in the radial direction, where the different numbers refer to the
nucleus and the two shells, respectively. However, depending on the
circumstances, the numerical resolution has been increased when
necessary, \ie up to $n_{\theta}=129$ and $\bmath{n}_{R}=(257,257,129)$
in the case of huge models with a matter distribution strongly deviating
from a spherical one (\cf Table~\ref{t:modbxxl} and Fig.~\ref{f:pol2xxl}).

\subsection{Comparison with Newtonian results}
\label{ss:numtst}

The results of the numerical scheme in the Newtonian limit have been
compared with those of an earlier investigation in the Newtonian
regime. \citet{Sood1972} have in fact computed linear perturbations of
polytropic stars induced by a toroidal magnetic field corresponding to
the Newtonian limit of equation~(\ref{EQ:EM10}), namely
\begin{equation} 
\label{EQ:NI1} 
\frac{B_{\phi}}{r\sin\theta} = \lambda_{0} \,\rho r\sin\theta \,,
\end{equation}
where $\rho$ denotes the Newtonian mass density. In
Table~\ref{t:polnewt}, we show normalized values of the resulting
oblateness for different adiabatic exponents produced with the present
code and the corresponding results from~\citet{Sood1972}. We note that
~\citet{Sood1972} measured the deformation as $\epsilon_{\rmn{s}}=
(r_{\rmn{e}} - r_{\rmn{p}})/r_{0}$ where $r_{0}$ is the radius of the
unperturbed model. Values computed according to this definition
coincide with $\epsilon_{\rmn{s}}=r_{\rmn{e}}/r_{\rmn{p}}-1$ within
the rounding error of tabulated values. Clearly, for the perturbation
parameter $h=\lambda^{2}_{0}/(4\upi^{2})$, the respective values of
$\epsilon_{\rmn{s}}/h$ agree to within $10^{-5}$, except for
$\gamma=5/3$, for which the difference is about $10^{-3}$. However,
for $\gamma=5/3$, certain functions intervening in the solution of the
perturbed Lane-Emden equation become singular at the surface of the
star~\citep{Das1977}, suggesting a larger numerical error for the
$\gamma=5/3$ case as computed by~\citet{Sood1972}.

\begin{figure*}
\begin{center}
\includegraphics[angle=0,width=7.0cm]{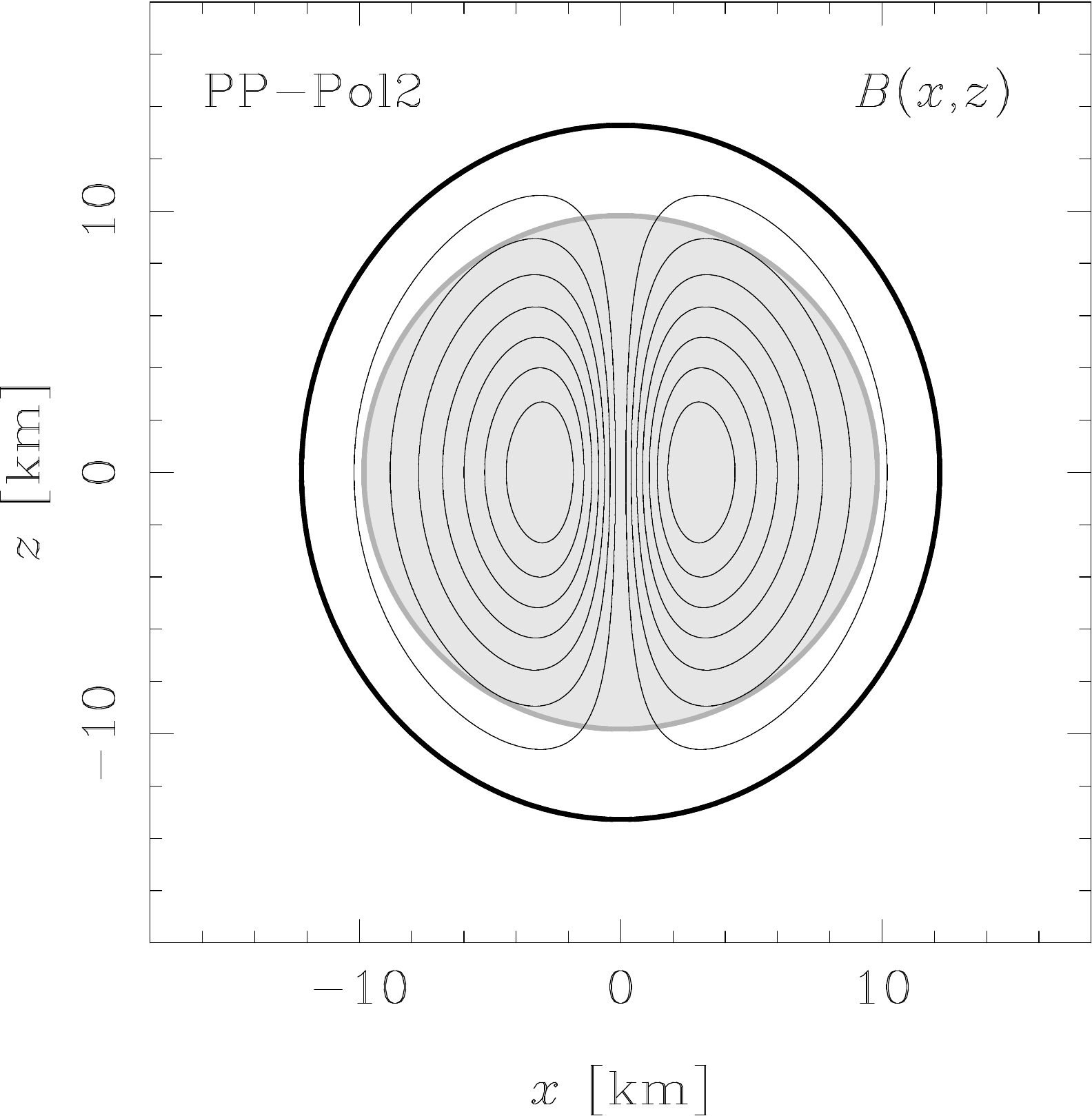} 
\hskip 1.5cm
\includegraphics[angle=0,width=7.0cm]{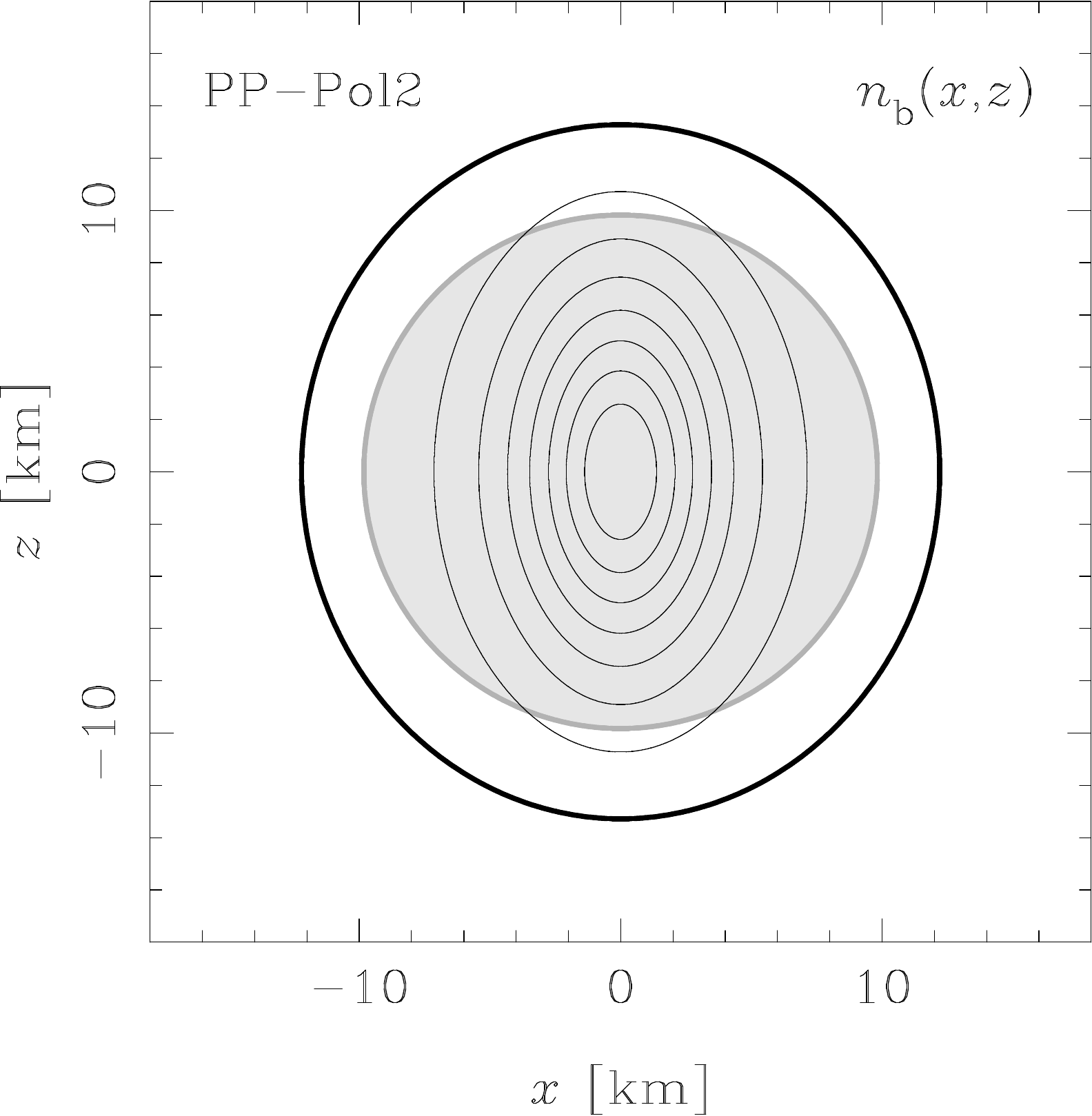}
\caption{Isocontours of magnetic field strength (left-hand panel) and baryon
  number density (right-hand panel) in the $(x,z)$ plane of model
  \texttt{PP-Pol2} of a non-rotating star with a gravitational mass of
  $M = 1.400 \, \mathrm{M}_{\sun}$ and a circumferential radius of
  $R_{\rmn{circ}}=12.00 \, \rmn{km}$ in the unmagnetized case built
  with a polytropic EOS with $\gamma = 2$ for a maximum average
  magnetic field strength of $\langle B^{2} \rangle^{1/2} =
  2.917 \times 10^{17}$ G. The grey disc indicates the dimensions of
  the unmagnetized reference model. Physical properties of model
  \texttt{PP-Pol2} are listed in Table~\ref{t:modbmax}.}
\label{f:pol2max}
\end{center}
\end{figure*}

\subsection{Virial identities}
\label{ss:numvir}

In order to monitor the global error of the numerical models computed
with the new code we have calculated the two general-relativistic
virial identities GRV2 \citep{Bonazzola1973,Bonazzola1994} and
GRV3~\citep{Gourgoulhon1994}. We recall that these identities have to
be fulfilled by any solution to the Einstein
equations~(\ref{EQ:EE1A})--(\ref{EQ:EE1D}) and are not enforced during the
iterative procedure that leads to the solution. The GRV2 identity, in
particular, has been shown to be directly related to the global error
of a numerical solution~\citep{Bonazzola1993}, while the GRV3 identity
is an extension of the virial theorem of Newtonian physics to general
relativity [see~\citet{Nozawa1998} for details of the practical
implementation]. For strongly magnetized non-rotating models based on
a polytropic EOS with $\gamma=2$ and the moderate number of collocation
points specified in Section~\ref{ss:numsch}, the violations to the identities
GRV2 and GRV3 have been found to be of the order of $10^{-6}$ or better.
For the huge models listed in Table~\ref{t:modbxxl}, corresponding values
are of the order of $10^{-4}$ which required the number of collocation
points to be multiplied by a factor of 8 compared to their default values.
In the case of rapid rotation, on the other hand, the
numerical grid is not perfectly adapted to the stellar surface, and
the spectral approximation suffers from Gibbs phenomena, as reported
in~\citet{Bonazzola1993}. However, the corresponding values of the
identities GRV2 and GRV3 are still of the order of $10^{-5}$.
For realistic EOSs, values of GRV2 and GRV3 are of the order of $10^{-4}$
or better.

\section{Non-rotating Magnetized Models}
\label{se:statmag}

Non-rotating models of neutron stars with a toroidal magnetic field
always generate \textit{static} space--times for which the time Killing
vector $\bmath{e}_{0}$ is hypersurface-orthogonal. Note that this
assumption does not always hold in the case of a poloidal magnetic
field because the additional presence of an electric field gives rise
to an azimuthal Poynting vector [\cf~\citet{Bonazzola1993}], which
introduces a non-zero angular momentum even when the star does not
rotate. As a consequence, the shift vector component $N^\phi$ does not
vanish.

Although we are interested in the construction of rapidly rotating
models (that we will present in Section~\ref{se:rotmag}), non-rotating
configurations allow us to investigate the effects of the magnetic
field without the influence of rotationally induced effects. As a
representative example, Fig.~\ref{f:pol2max} shows isocontours of the
magnetic field strength and baryon number density for the non-rotating
model \texttt{PP-Pol2} built with the Pol2 EOS and for the maximum
field strength (the physical properties of model \texttt{PP-Pol2} are
listed in Table~\ref{t:modbmax}). The unmagnetized reference model
with a circumferential radius of $R_\rmn{circ} = 12.00 \, \rmn{km}$ is
symbolized by a grey disc. In agreement with the simple structure
function defined in equation~(\ref{EQ:EM10}), the toroidal magnetic field
shown in the left-hand panel vanishes on the axis of symmetry, reaches a
maximum value of $B_\rmn{max} = 7.408 \times 10^{17} \, \rmn{G}$ in
the equatorial plane deep inside the star and decreases towards the
surface of the star, where it vanishes, so that the magnetic field is
fully contained inside the star. The forces exerted by the magnetic
field can be visualized through the magnetic potential $\tilde{M}$,
which shows a distribution similar to that of the magnetic field
strength and that, for this reason, we do not report here. It should
be noted that the magnetic potential $\tilde{M}$ is {\em repulsive},
since it is positive everywhere by construction. This is to be
contrasted with the magnetic potential associated with the purely
poloidal magnetic field adopted in~\citet{Bocquet1995}, that was
instead {\em attractive}. As a result, the corresponding forces are
the opposite, despite the isocontours appear to be very similar.
Note also that in the present case of a purely toroidal magnetic field
defined according to equation~(\ref{EQ:EM10}), the isocontours of
$\tilde{M}$ coincide with the flow lines of the
electric current $\bmath{j}$, which are nested loops in the meridional
plane. Finally, we note that $\tilde{M}$ vanishes at the surface of
the star, implying that the latter coincides with an isosurface of the
gravitational potential $\nu$, as required by equation~(\ref{EQ:EM13})
when $\ln\mathit{\Gamma} = 0$ and $\tilde{M} = 0$.

The forces that can be derived from $\tilde{M}$ vanish on the axis of
symmetry and reach their maximum in the equatorial plane. Between the
centre of the star and the maximum of $\tilde{M}$, they are directed
inwards, inducing an approximately cylindrical compression of the
central region of the star, which responds by a prolate deformation
visible in the right-hand panel of Fig.~\ref{f:pol2max}. In the outer
layers of the star, on the other hand, the magnetic forces are
directed radially outward, pushing them away from the centre, which
results in a growth of the dimensions of the star whose
circumferential radius has increased now to a value of $R_\rmn{circ} =
14.34 \, \rmn{km}$.

\begin{figure*}
\begin{center}
\includegraphics[angle=-0,width=7.0cm]{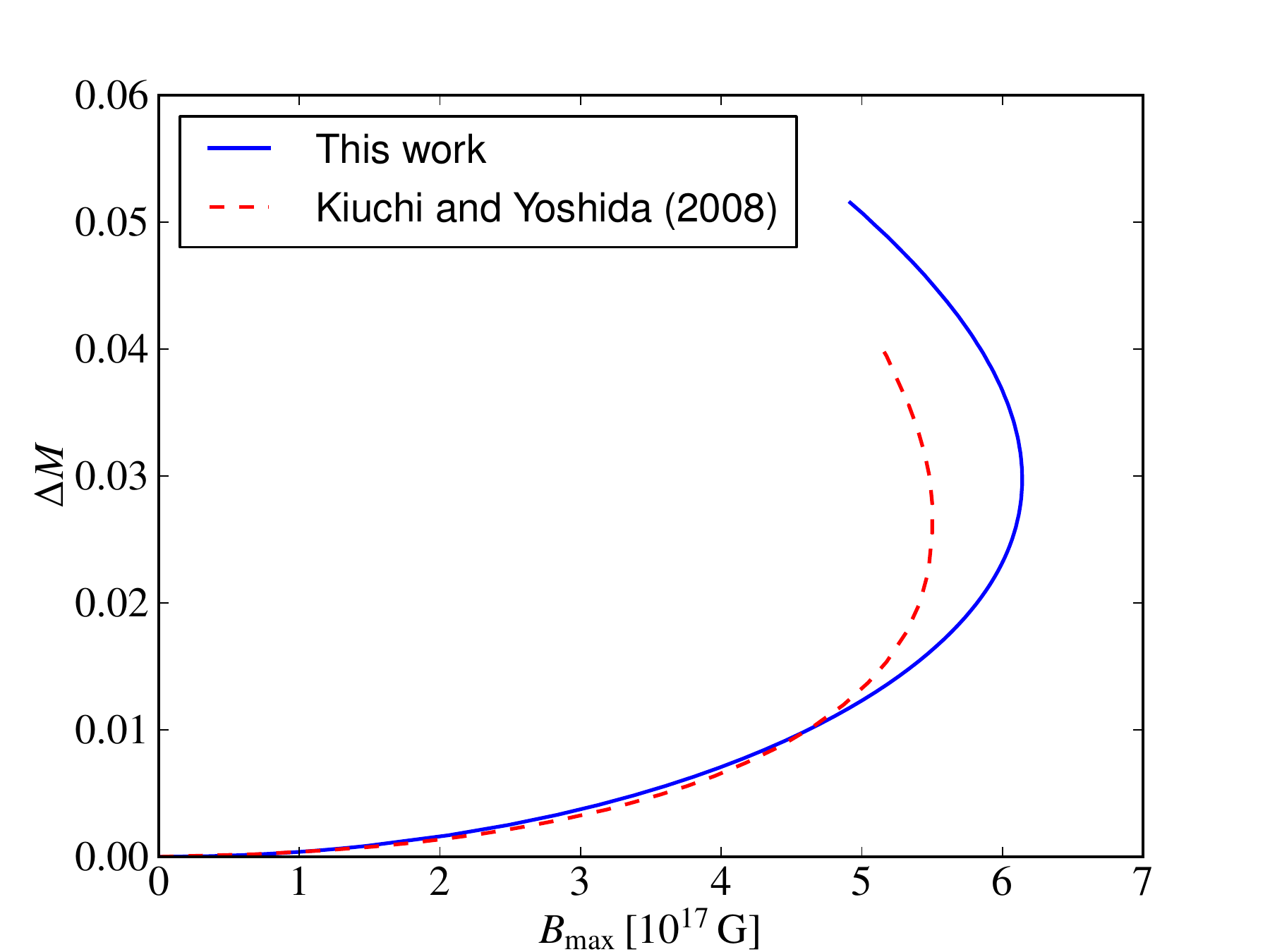}
\hskip 1.5cm
\includegraphics[angle=-0,width=7.0cm]{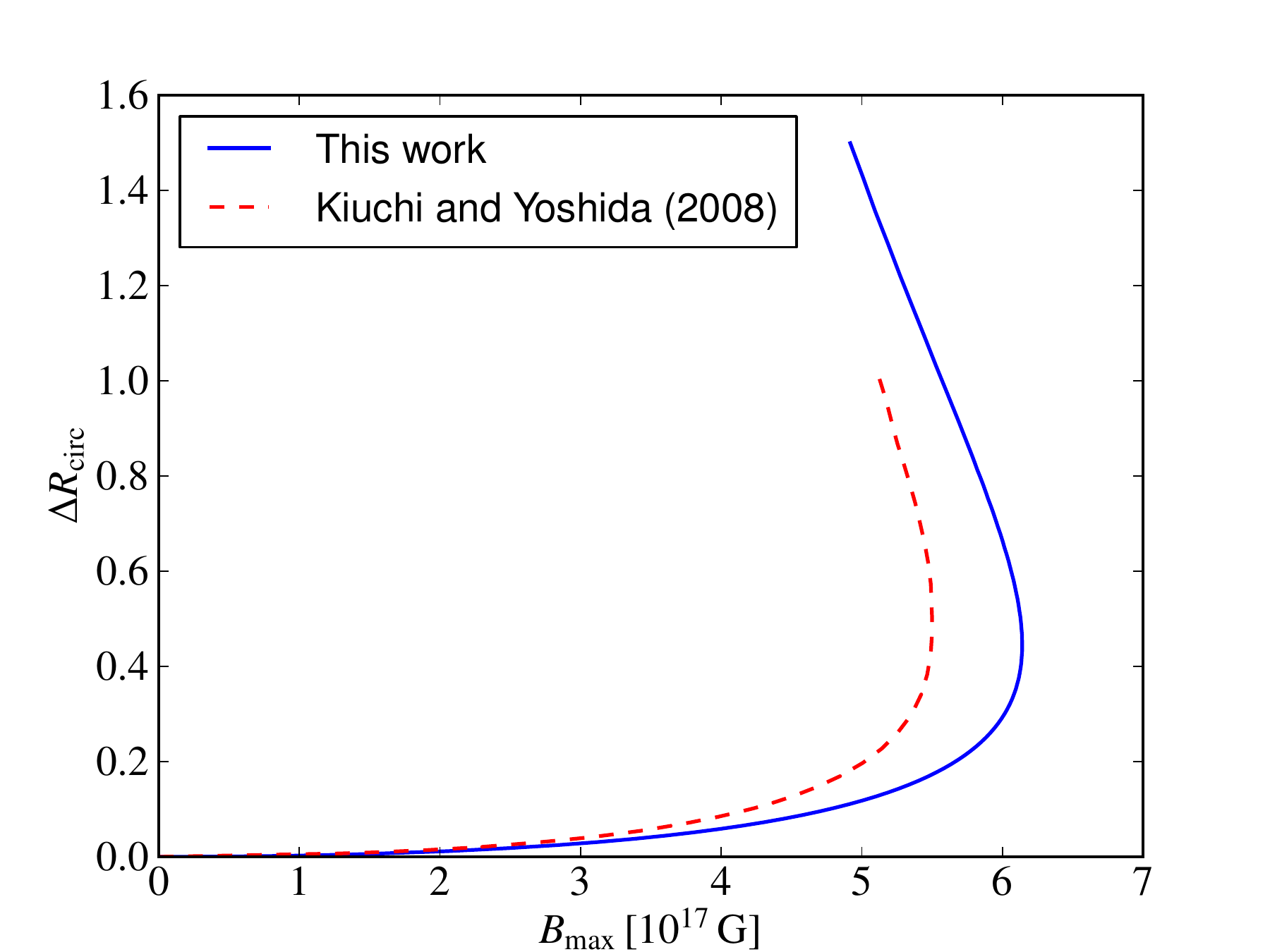} \\[2ex]
\includegraphics[angle=-0,width=7.0cm]{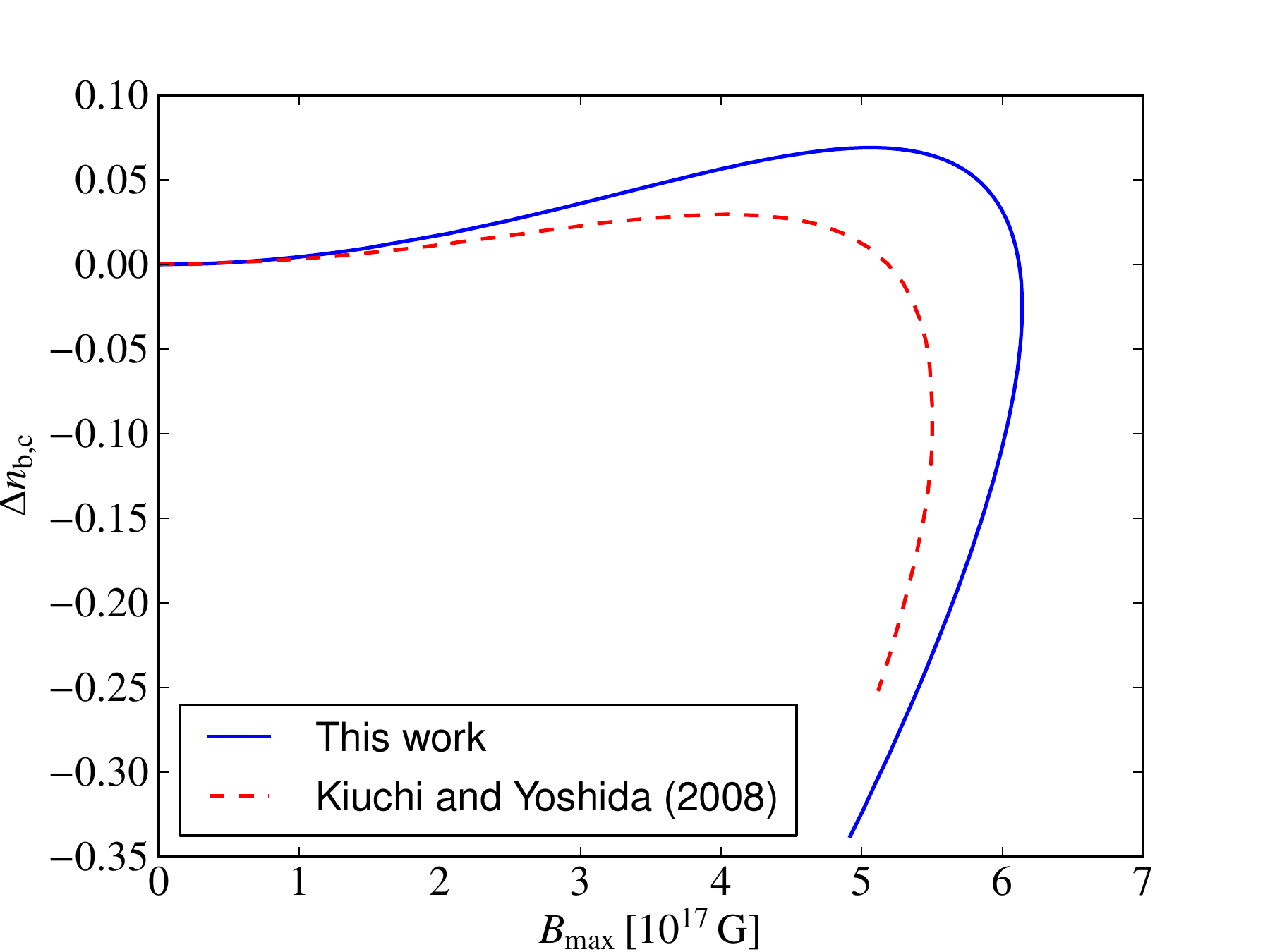}
\hskip 1.5cm
\includegraphics[angle=-0,width=7.0cm]{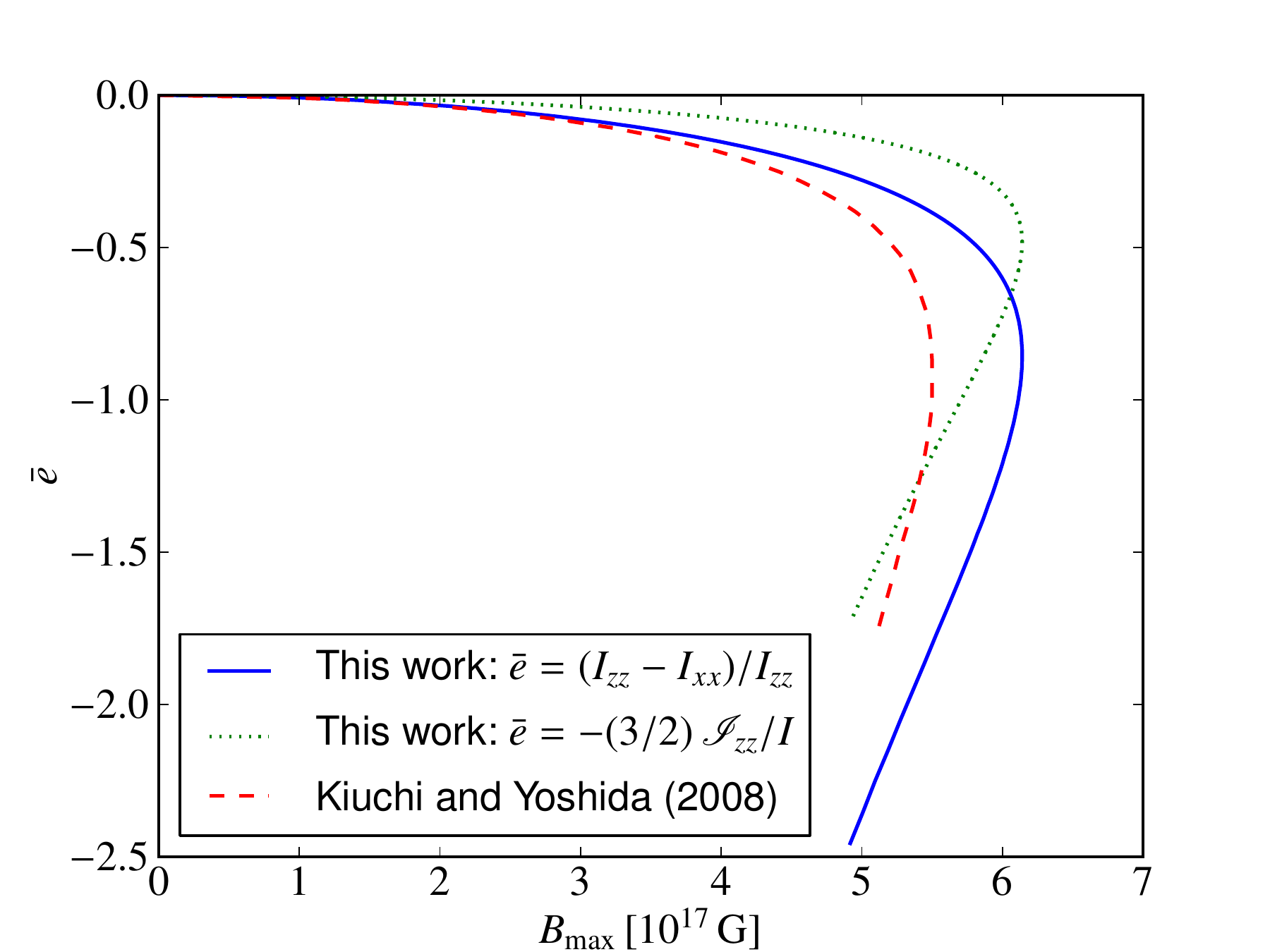}
\caption{Comparison of the gravitational mass $M$, circumferential
  radius $R_\rmn{circ}$, central baryon number density $n_\rmn{b,c}$
  and mean deformation $\bar{e}$ along with the maximum magnetic field
  strength $B_\rmn{max}$ attained inside the star as a function of the
  magnetization parameter $\lambda_{0}$ for a Pol2 EOS model with a baryon
  mass of $M_\rmn{b}=1.680 \, \mathrm{M}_{\sun}$ and a circumferential radius
  of $R_\rmn{circ}=14.30 \, \rmn{km}$ in the unmagnetized case with
  results of \citet{Kiuchi2008}. The relative variation of some
  quantity $\chi$ is defined as $\Delta \chi(B_\rmn{max}, M_{\rmn{b}})
  \equiv (\chi(B_\rmn{max}, M_{\rmn{b}})-\chi(0, M_{\rmn{b}}))/\chi(0,
  M_{\rmn{b}})$.}
\label{f:comp-kiuchi}
\end{center}
\end{figure*}

Both the apparent shape and the matter distribution are prolate,
corresponding to negative values of the surface deformation and of the
quadrupole distortion, namely $\epsilon_\rmn{s} =-0.0806$ and
$\epsilon=-0.1986$, respectively. The value of $\epsilon_\rmn{s}$ is
significantly smaller than that of $\epsilon$, and this due to the
fact that terms of higher order in the multipole expansion of the
gravitational potential $\nu$ fall off rapidly while, at the same
time, the surface of the star must coincide with an isosurface of
$\nu$. This interpretation is supported by the fact that this
difference is significantly smaller for small perturbations around
this model, as confirmed by comparing values of respective distortion
coefficients $b_{B}$ and $c_{B}$ reported in Section~\ref{se:distcoef}.

The strength of the magnetic field is controlled by the magnetization
parameter $\lambda_{0}$ (\cf equation~\ref{EQ:EM12}), and the physical
quantities associated with our models can be parametrized
accordingly, in particular the maximum magnetic field strength
$B_\rmn{max}$. In Fig.~\ref{f:comp-kiuchi}, we compare our results
with those reported in fig.~6 of \citet{Kiuchi2008},\footnote{The
  data from fig.~6 of \citet{Kiuchi2008} have been read off with
  the utility \textsc{g3data} see
  \url{https://github.com/pn2200/g3data}.} considering in
particular the variation of the gravitational mass $M$, of the
circumferential radius $R_\rmn{circ}$, of the central baryon number
density $n_\rmn{b,c}$ and of the mean deformation $\bar{e}$. The
comparison is made as a function of the maximum magnetic field
strength $B_\rmn{max}$ and for a non-rotating reference model built
with the Pol2 EOS and having a baryon mass of $M_\rmn{b}=1.680 \,
\mathrm{M}_{\sun}$ and a circumferential radius of $R_\rmn{circ} = 14.30 \,
\rmn{km}$ in the unmagnetized case. For each of these quantities
$\chi$, we define the difference as $\Delta \chi(B_\rmn{max},
M_{\rmn{b}}) \equiv [\chi(B_\rmn{max}, M_{\rmn{b}})-\chi(0,
  M_{\rmn{b}})]/\chi(0, M_{\rmn{b}})$. We note that
in~\citet{Kiuchi2008} the mean deformation is defined as $\bar{e}
\equiv (I_{zz}-I_{xx})/I_{zz}$, where $I_{xx}, I_{zz}$ are the moments
of inertia relative to the corresponding axes (\cf equation~3.12
of~\citealt{Kiuchi2008}); this measure is similar but different from
our measure of the quadrupole distortion $\epsilon$.\footnote{More
  specifically, the definition of $\bar{e}$ is valid only in a
  Newtonian framework, where $(I_{zz}-I_{xx})/I_{zz} = (I-I_{xx})/I =
  Q_{\rm Newt}/I$, and $Q_{\rm Newt}$ is the Newtonian quadrupole
  moment.}

The most prominent feature of this comparison is that $B_\rmn{max}$ is
not a monotonic function of $\lambda_{0}$, thus being responsible for the
presence of a turning point located at a magnetic field strength of
$B_\rmn{max} = 6.141 \, \times 10^{17} \, \rmn{G}$. While our results
(blue solid lines) agree qualitatively with those of \citet{Kiuchi2008}
(red dashed lines), we find significant quantitative differences for all
of the quantities considered, the largest one being an increase in
$n_\rmn{b,c}$ of 7 per cent against the smaller 3 per cent increase
found by \citet{Kiuchi2008}. Another important difference is in the
values of the maximum magnetic field strength, which is
$B_\rmn{max} = 6.141 \, \times 10^{17} \, \rmn{G}$ for us and
$B_\rmn{max} = 5.503 \, \times 10^{17} \, \rmn{G}$ for
~\citet{Kiuchi2008}. The differences between the two calculations are
smaller when comparing the mean deformations $\bar{e}$,
but only for moderate values of $B_\rmn{max}$. Furthermore, the
measure of $\bar{e}$ is considerably different from the quadrupole
deformation $\epsilon$ (green dotted line), suggesting that
the definition of $\bar{e}$ is not suitable for estimating the
gravitational-wave emission of a distorted star because it
overestimates corresponding values by about a factor of 2 for this
reference model. Similar differences in the values of $\bar{e}$ and
$\epsilon$ have been found also for the unmagnetized model with a
baryon mass of $M_\rmn{b}=1.780 \, \mathrm{M}_{\sun}$ rotating at $\mathit{\Omega} =
3.230 \times 10^{3} \, \rmn{s}^{-1}$ from table~IV of
\citet{Kiuchi2008}; in this case, the respective values of the mean
deformation of $\bar{e}=0.07764$ and $\bar{e}=0.07462$, differ by
about 4 per cent, the quadrupole distortion for the same model is
$\epsilon=0.03870$ which is again only half of the value of $\bar{e}$.

\begin{figure*}
\begin{center}
\includegraphics[angle=0,width=7.0cm]{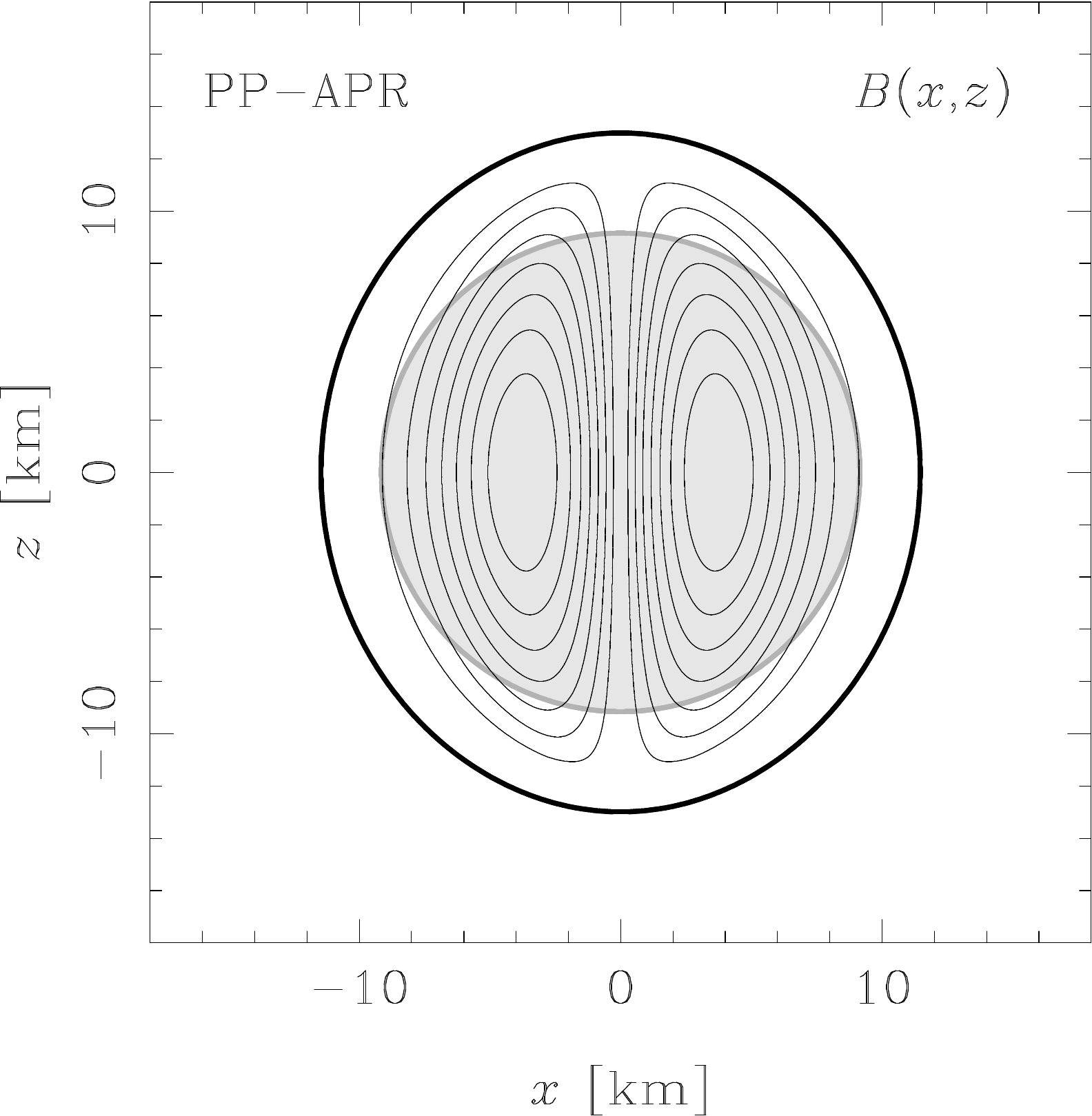}
\hskip 1.5cm
\includegraphics[angle=0,width=7.0cm]{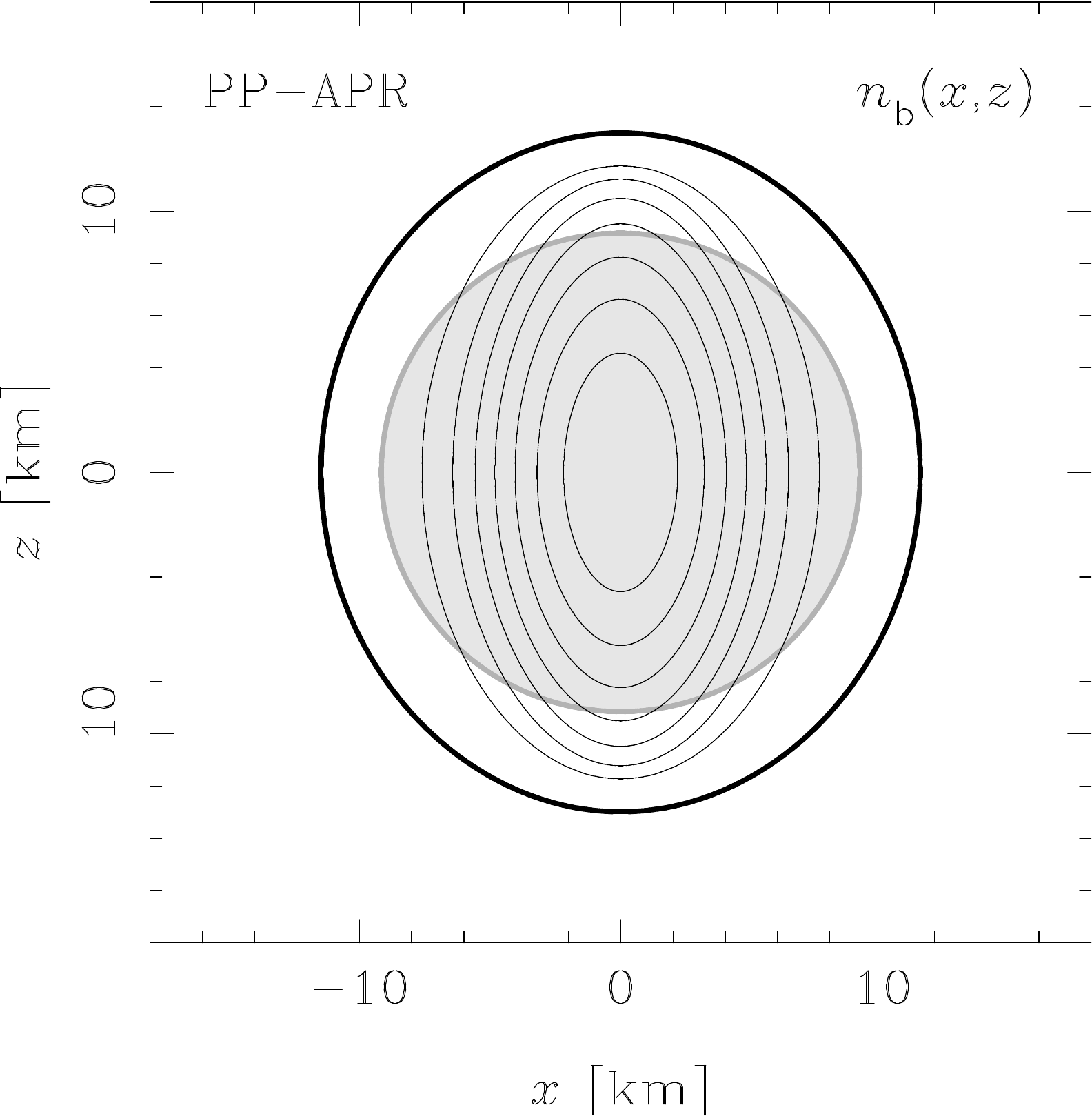}
\caption{Isocontours of magnetic field strength (left-hand panel) and
  baryon number density (right-hand panel) in the $(x,z)$ plane of
  model \texttt{PP-APR} of a non-rotating star with a gravitational
  mass of $M = 1.400 \, \mathrm{M}_{\sun}$ and a circumferential radius of
  $R_{\rmn{circ}}=11.34 \, \rmn{km}$ in the unmagnetized case built
  with the APR EOS for a maximum average magnetic field strength of
  $\langle B^{2}\rangle^{1/2} = 3.597 \times 10^{17}$ G.
  The grey disc indicates the dimensions of the unmagnetized
  reference model. The physical properties of model \texttt{PP-APR}
  are listed in Table~\ref{t:modbmax}.}
\label{f:aprmax}
\end{center}
\end{figure*}

\begin{table*}
\caption{\label{t:statbmax} Non-rotating neutron star models computed
  with different EOSs at the maximum field strength limit. The
  gravitational mass is $M = 1.400 \, \mathrm{M}_{\sun}$ for each EOS,
  respectively, in the unmagnetized and non-rotating case for which
  properties are listed in Table~\ref{t:distcoef}.
  $B_{\rmn{max}}$ is the maximum value of the magnetic field strength,
  $\langle B^{2} \rangle^{1/2}$ the root mean square average of the
  magnetic field strength determined over the volume of the star,
  $n_{\rmn{b,c}}$ the central baryon number density
  $( \, \times \, 0.1 \, \rmn{fm}^{-3})$, $M$ the gravitational mass,
  $R_{\rmn{circ}}$ the circumferential radius, $I$ the moment of inertia,
  $\mathscr{M}/|W|$ the ratio of the total magnetic energy
  $\mathscr{M}$ to the potential energy $W$, $\epsilon_{\rmn{s}}$ the
  surface deformation, $\epsilon$ the quadrupole distortion, and
  GRV2/GRV3 the estimates of the global error of respective models
  based on the relativistic virial identities introduced in
  Section~\ref{ss:numvir}.}
\begin{tabular}{@{}lcccccccccccll}
\hline
 EOS & $B_{\rmn{max}}$ & $\langle B^{2} \rangle^{1/2}$ &
 $n_{\rmn{b, c}}$ & $M$ &
 $R_{\rmn{circ}}$ & $I$ & $\mathscr{M}/|W|$ &
 $\epsilon_{\rmn{s}}$ & $\epsilon$ & $|\,$GRV2$\,|$ & $|\,$GRV3$\,|$ \\
 & $( \, \times \, 10^{17} \, \rmn{G})$ & $( \, \times \, 10^{17} \, \rmn{G})$ & $( \, \times \, 0.1 \, \rmn{fm}^{-3})$ & $(\mathrm{M}_{\sun})$ & $(\rmn{km})$ &
 $( \, \times \, 10^{38} \, \rmn{kg} \, \rmn{m}^{2})$ & $ $ & \\
\hline
 Pol2   & $7.408$ & $2.917$ & $8.409$ & $1.427$ & $14.34$ & $1.278$ & $0.3186$ & $-0.0806$ & $-0.1986$ & $7 \times 10^{-11}$ & $3 \times 10^{-9}$ \\
 APR    & $8.046$ & $3.597$ & $6.036$ & $1.438$ & $13.58$ & $1.136$ & $0.2922$ & $-0.1176$ & $-0.3045$ & $5 \times 10^{-7}$ & $4 \times 10^{-7}$ \\
 BBB2   & $8.475$ & $3.781$ & $7.132$ & $1.439$ & $13.31$ & $1.093$ & $0.3133$ & $-0.1138$ & $-0.2899$ & $3 \times 10^{-6}$ & $4 \times 10^{-6}$ \\
 BN1H1  & $6.414$ & $2.830$ & $7.174$ & $1.432$ & $15.14$ & $1.319$ & $0.2810$ & $-0.1086$ & $-0.3065$ & $2 \times 10^{-5}$ & $2 \times 10^{-5}$ \\
 BPAL12 & $12.10$ & $4.790$ & $13.91$ & $1.438$ & $11.83$ & $0.897$ & $0.4589$ & $-0.0782$ & $-0.1729$ & $3 \times 10^{-6}$ & $4 \times 10^{-6}$ \\
 FPS    & $9.090$ & $4.019$ & $7.908$ & $1.440$ & $12.99$ & $1.046$ & $0.3308$ & $-0.1122$ & $-0.2809$ & $1 \times 10^{-5}$ & $2 \times 10^{-5}$ \\
 GNH3   & $5.090$ & $2.161$ & $4.182$ & $1.427$ & $17.06$ & $1.605$ & $0.1984$ & $-0.1083$ & $-0.3313$ & $5 \times 10^{-6}$ & $1 \times 10^{-5}$ \\
 SLy4   & $7.523$ & $3.341$ & $5.972$ & $1.436$ & $14.04$ & $1.191$ & $0.2848$ & $-0.1156$ & $-0.3076$ & $6 \times 10^{-6}$ & $3 \times 10^{-6}$ \\
\hline
\end{tabular}
\end{table*}

In addition to the polytropic Pol2 EOS, we have computed non-rotating
models for the sample of realistic EOSs discussed in
Section~\ref{ss:eos} and for reference models having a gravitational
mass of $M = 1.400 \, \mathrm{M}_{\sun}$ in the unmagnetized case
(basic physical properties of the unmagnetized reference models are
collected in Table~\ref{t:distcoef}). Because we
cannot show equilibrium models for all of these models, we have
decided to use as alternative reference EOS the APR one and to
complement model \texttt{PP-Pol2} at the maximum field strength limit
with the corresponding model \texttt{PP-APR}, which nicely illustrates
the impact of a prototypical realistic EOS on the resulting
equilibrium model.

Figure~\ref{f:aprmax} shows isocontours of magnetic field strength and
baryon number density for model \texttt{PP-APR} at the maximum
field strength limit, whose physical properties are collected in
Table~\ref{t:modbmax} (again, the dimensions of the unmagnetized
reference model with a circumferential radius of $R_\rmn{circ}=11.34
\, \rmn{km}$ are indicated by a grey disc). The maximum magnetic field
strength attains a value of $B_\rmn{max}=8.046 \times 10^{17} \,
\rmn{G}$ in the equatorial plane. Because the APR EOS is stiffer than
the Pol2 EOS, the matter distribution appears less condensed and the
peaks of the magnetic field have moved slightly outwards. Moreover,
both the surface deformation $\epsilon_{\mathrm{s}}=-0.1176$ and the
quadrupole distortion $\epsilon=-0.3045$ are larger than the
corresponding values of model \texttt{PP-Pol2} by $\sim 50$ per cent.
In contrast with model \texttt{PP-Pol2}, a substantial fraction of
the stellar interior below the surface appears to be field-free.
This is easy to understand: according to equation~(\ref{EQ:EM10}),
in fact, the amplitude of the toroidal magnetic field is proportional
to $(e+p)$, so that the presence of a low-density crust as in the APR
EOS suppresses the presence of a toroidal magnetic field in the outer
layers of the star.\footnote{As customary, we consider the crust not
  as a solid but as a perfect fluid in analogy with the fluid
  treatment of the core.} The contour of the unmagnetized reference
model not only emphasizes the significant prolate deformation of the
maximum field strength model, but also the important increase in the
dimensions of the star.

Overall, the non-rotating models built with realistic EOSs show a
behaviour which is qualitatively similar to that of the polytropic
Pol2 EOS as shown in Table~\ref{t:statbmax} and, for increasing
magnetization, all realistic EOSs exhibit a maximum value of the
magnetic field strength $\langle B^{2} \rangle^{1/2}$, beyond which it
then decreases. The smallest value is obtained for the GNH3 EOS, with
$\langle B \rangle \equiv \langle B^{2} \rangle^{1/2}=2.161 \times
10^{17}$ G; the largest one is instead obtained for the BPAL12 EOS,
with $\langle B^{2} \rangle^{1/2}=4.790 \times 10^{17}\, \rmn{G}$.
Relevant data for all maximum field strength models are collected in
Table~\ref{t:statbmax} and reveal that the values of the maximum
magnetic field strength decrease with increasing circumferential radii
$R_\rmn{circ}$, and which itself is a telltale of the stiffness of the
EOS. As a result, the soft BPAL12 EOS with a circumferential radius of
only $R_\rmn{circ} = 11.83 \, \rmn{km}$ yields the model with the
highest maximum magnetic field strength, while the stiff GNH3 EOS the
model with the lowest one.

We note that in Fig.~\ref{f:comp-kiuchi}, we have employed the peak
magnetic field strength $B_{\rmn{max}}$ and not the average
magnetic field strength $\langle B^{2}\rangle^{1/2}$, which reaches
its turning point already at a lower magnetization level. For this
reason, all models based on a realistic EOS are located in a region
where $\Delta n_{\rmn{b,c}} > 0$, as can be seen by a comparison with
the related values of the unmagnetized models compiled in
Table~\ref{t:distcoef}. The maximum values of the peak magnetic field
strength $B_{\rmn{max}}$ are sensibly larger than those of the average
magnetic field strength $\langle B^{2}\rangle^{1/2}$, attaining their
maximum value for the model built with the FPS EOS with $B_{\rmn{max}}
= 1.210 \times 10^{18} \, \rmn{G}$.

\begin{figure*}
\begin{center}
\includegraphics[angle=-0,width=7.0cm]{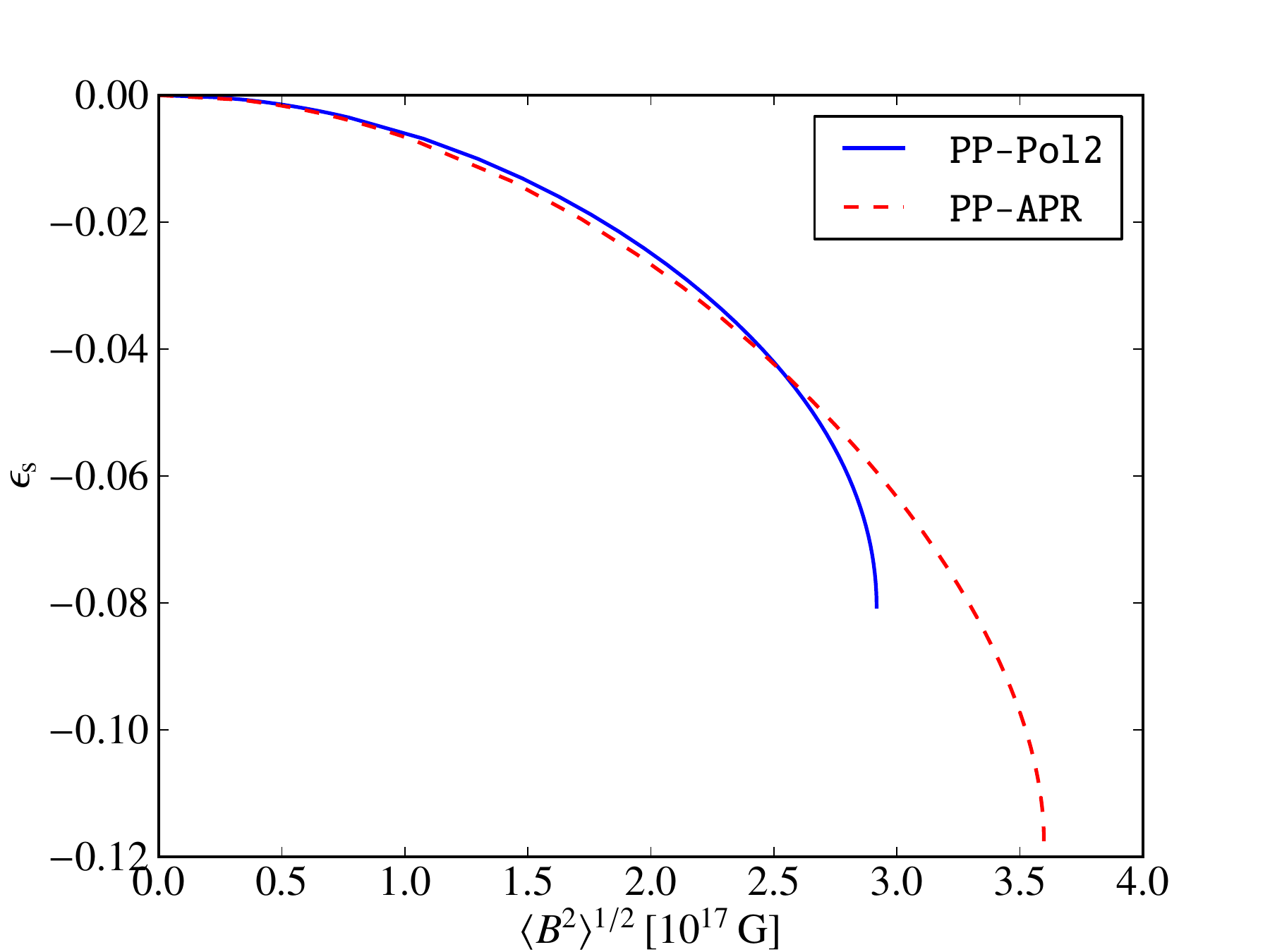}
\hskip 1.5cm
\includegraphics[angle=-0,width=7.0cm]{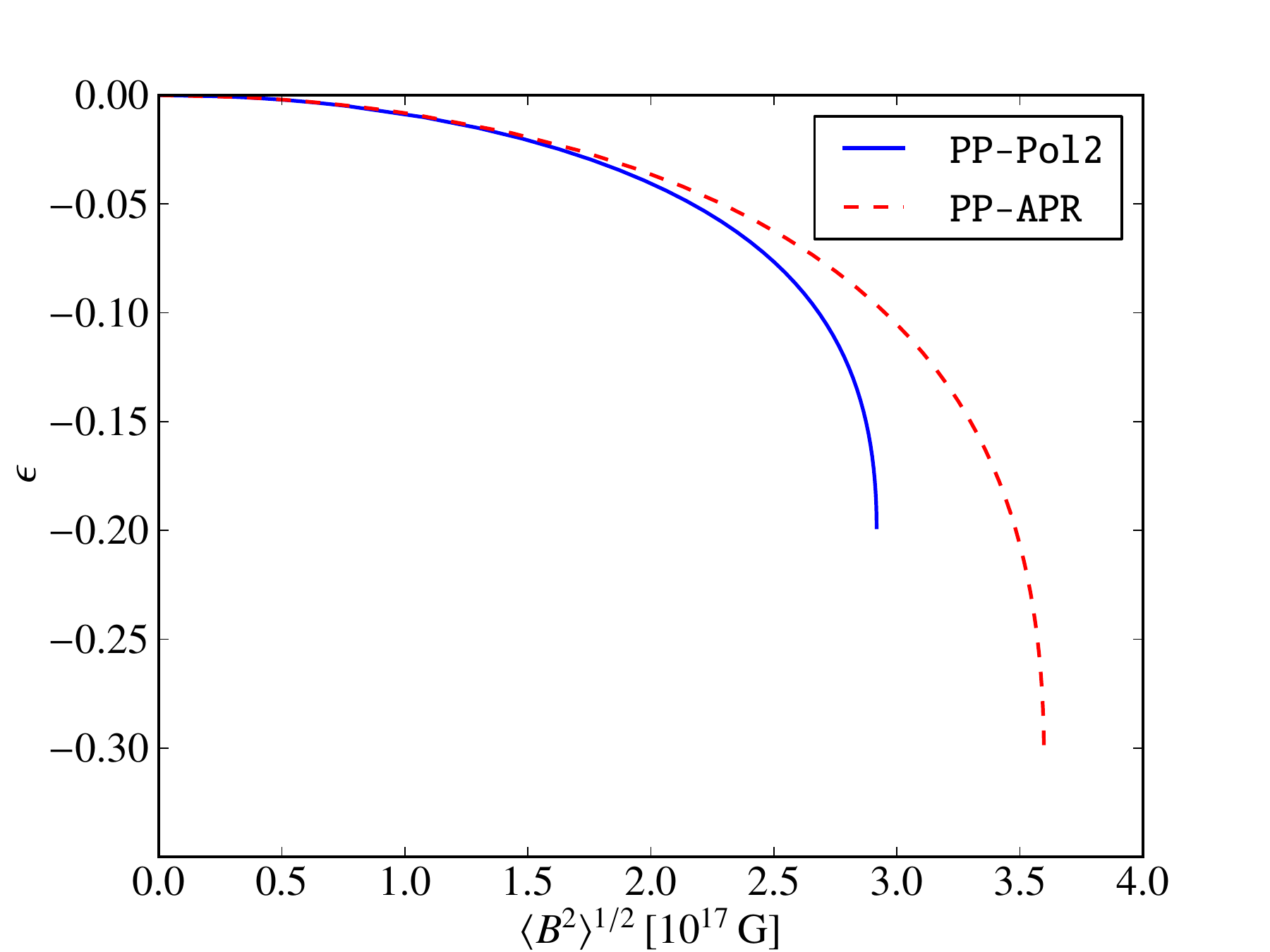}
\caption{Surface deformation $\epsilon_{\rmn{s}}$ (left-hand panel) and
  quadrupole distortion $\epsilon$ (right-hand panel) for non-rotating
  models built with the Pol2 EOS and the APR EOS as a function of the
  root mean square magnetic field strength $\langle B^{2}
  \rangle^{1/2}$ from the unmagnetized limit up to the maximum field
  strength models \texttt{PP-Pol2} and \texttt{PP-APR},
  respectively. The physical properties of models \texttt{PP-Pol2} and
  \texttt{PP-APR} are listed in Table~\ref{t:modbmax}.}
\label{f:eps-nrot}
\end{center}
\end{figure*}

\begin{figure*}
\begin{center}
\includegraphics[angle=-90,width=7.0cm]{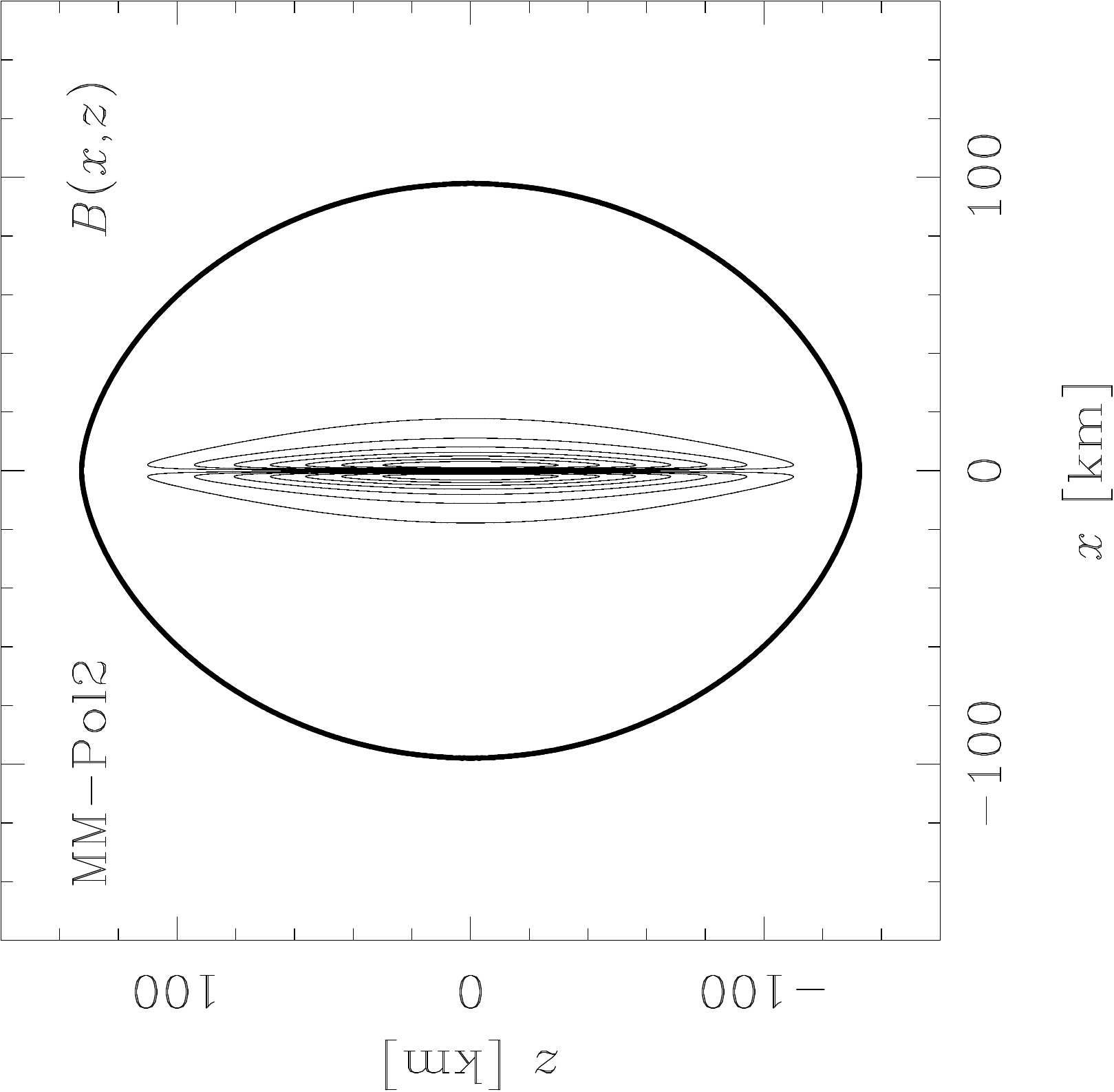}
\hskip 1.5cm
\includegraphics[angle=-90,width=7.0cm]{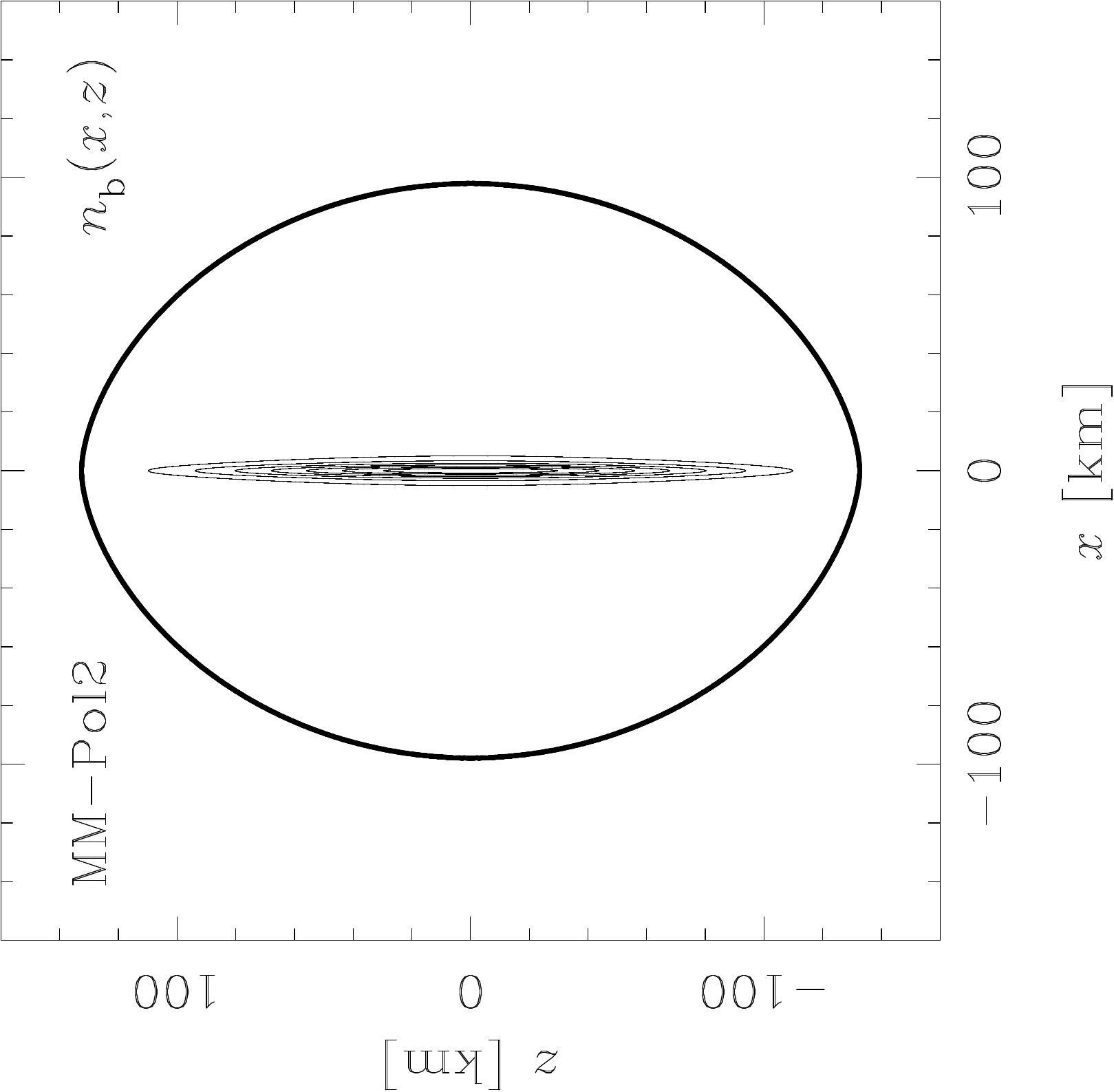}
\caption{Isocontours of magnetic field strength (left-hand panel) and baryon
  number density (right-hand panel) in the $(x,z)$ plane of model
  \texttt{MM-Pol2} of a non-rotating star with a gravitational mass of
  $M = 1.400 \, \mathrm{M}_{\sun}$ and a circumferential radius of
  $R_{\rmn{circ}}=12.00 \, \rmn{km}$ in the unmagnetized case built
  with the Pol2 EOS for an average magnetic field strength of
  $\langle B^{2}\rangle^{1/2} = 0.1400 \times 10^{17}$
  G. The physical properties of model \texttt{MM-Pol2} are listed in
  Table~\ref{t:modbxxl}.}
\label{f:pol2xxl}
\end{center}
\end{figure*}

\begin{figure*}
\begin{center}
\includegraphics[angle=-90,width=7.0cm]{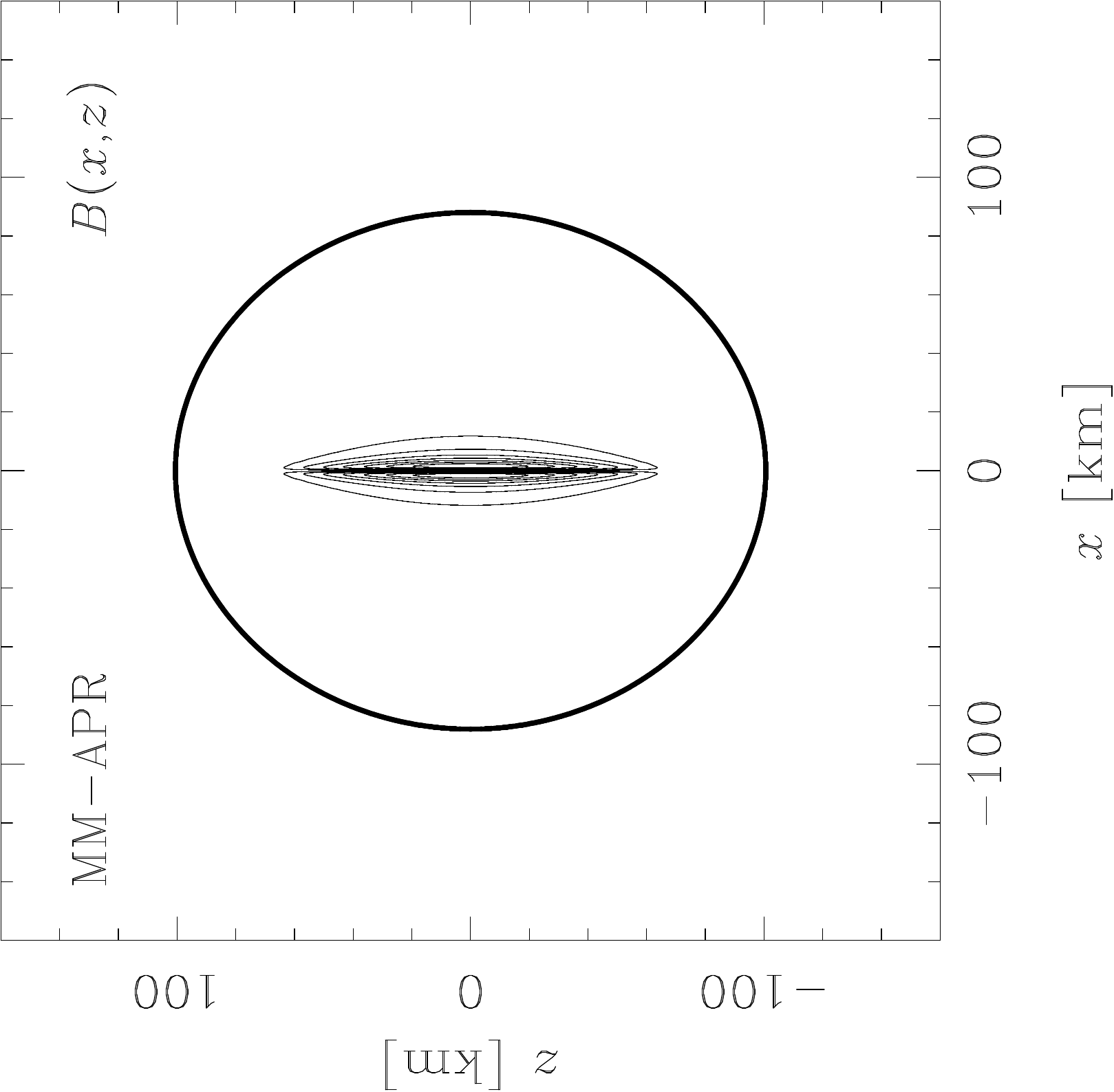}
\hskip 1.5cm
\includegraphics[angle=-90,width=7.0cm]{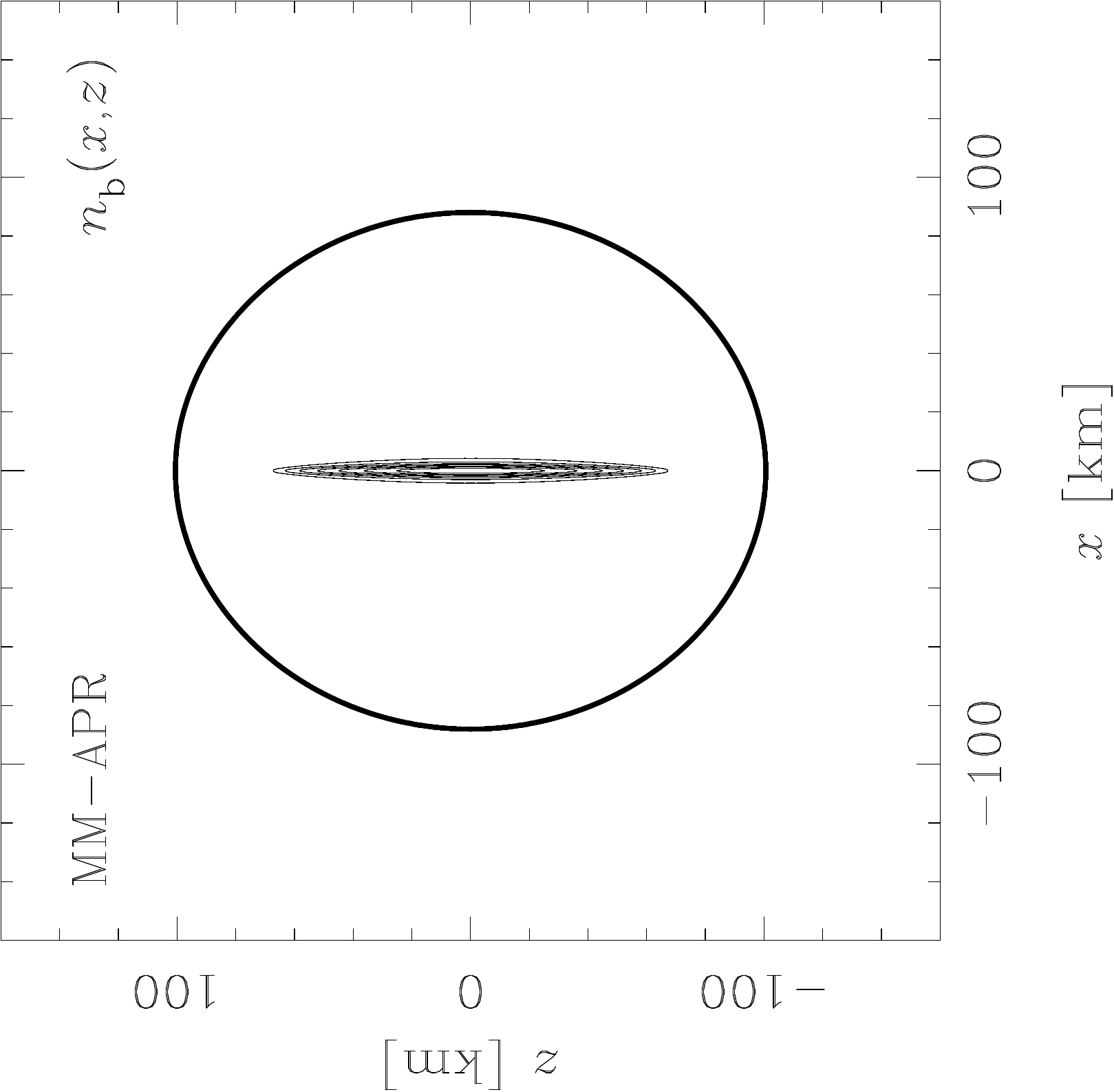}
\caption{Isocontours of magnetic field strength (left-hand panel) and baryon
  number density (right-hand panel) in the $(x,z)$ plane of model \texttt{MM-APR}
  of a non-rotating star with a gravitational mass of $M =
  1.400 \, \mathrm{M}_{\sun}$ and a circumferential radius of
  $R_{\rmn{circ}}=11.34 \, \rmn{km}$ in the unmagnetized case built
  with the APR EOS for an average magnetic field strength of $\langle
  B^{2}\rangle^{1/2} = 0.2313 \times 10^{17}$ G. The physical
  properties of model \texttt{MM-APR} are listed in
  Table~\ref{t:modbxxl}.}
\label{f:aprxxl}
\end{center}
\end{figure*}

The slight increase of the gravitational mass is the result of
different contributions, which can be illustrated by the example of
the \texttt{PP-Pol2} model. More specifically, the increase by
$0.02682 \, \mathrm{M}_{\sun}$ in the gravitational mass of the maximum
field strength model with respect to the unmagnetized model. This
increase is the sum of a positive contribution of $0.03036 \,
\mathrm{M}_{\sun}$ due to the total magnetic energy, of a negative contribution
of $-0.02160 \, \mathrm{M}_{\sun}$ due to the internal energy lost because of
the increased volume, and of a positive contribution of $0.01806 \,
\mathrm{M}_{\sun}$ by which the magnetized model is less gravitationally bound
than its unmagnetized counterpart. Furthermore, all maximum
field strength models exhibit smaller moments of inertia than in the
unmagnetized case because of the lateral compression of the
stellar core. Finally, we note that after reaching a minimum value,
the moments of inertia increase continuously with the magnetization,
as the growth of the dimensions of the star becomes significant.

The values of the surface deformation $\epsilon_{\rmn{s}}$ and of the
quadrupole distortion $\epsilon$ obtained for our sample of maximum
field strength models also show a noticeable dependence on the
stiffness of the EOS. In particular, the value of $\epsilon$ decreases
(its absolute value increasing) with the increase of the
circumferential radius $R_\rmn{circ}$. The Pol2 EOS model appears
somehow at variance with this trend, but we believe this is caused by
our choice of a comparatively large circumferential radius of
$R_\rmn{circ} = 12.00 \, \rmn{km}$ in the unmagnetized case, and which
does not reflect the rather soft character of this EOS. Lower and
upper bounds for the quadrupole distortion are set by the two extremal
EOSs of our sample: the model built with the soft EOS BPAL12 with a
radius of less than $12.00\, \rmn{km}$ exhibits the smallest
quadrupole distortion despite of the largest ratio of magnetic energy
to binding energy, whereas the large GNH3 model with a radius of
$17.06\, \rmn{km}$ shows the largest quadrupole distortion for the
lowest ratio of magnetic energy to binding energy of the whole sample.

Much of what discussed so far is summarized in Fig.~\ref{f:eps-nrot},
where we present the dependence of the surface deformation
$\epsilon_{\rmn{s}}$ and of the quadrupole distortion $\epsilon$ on
the average magnetic field strength $\langle B^{2} \rangle^{1/2}$, for
the representative Pol2 EOS and the APR EOSs. Note that up to $\langle
B^{2} \rangle^{1/2} \simeq 2.0 \times 10^{17}\,{\rm G}$, the behaviour
of both distortions is essentially the same for the two EOSs and an
almost perfectly linear function\footnote{This is not straightforward
  to deduce from Fig.~\ref{f:eps-nrot}, which reports
  $\epsilon_{\rmn{s}}$ and $\epsilon$ as a function of $\langle B
  \rangle$. However, it is very apparent when plotting the distortions
  as a function $\langle B^{2} \rangle$ (not reported here for
  compactness).} of $\langle B^{2} \rangle$. This is not particularly
surprising and indeed earlier Newtonian studies~\citep{Wentzel1960,
  Ostriker1969} have suggested to parametrize the quadrupolar
distortion $\epsilon_{\rm Newt}$ induced in a self-gravitating
incompressible fluid by a toroidal magnetic field as a simple function
of the ratio of the magnetic energy to the binding energy, and hence
as a function which is quadratic in the magnetic field strength.
Fig.~\ref{f:eps-nrot} suggests therefore that this behaviour is
preserved also in general relativity and up to very large magnetic
fields, thus offering the possibility of expressing the
magnetically induced deformation in terms of a simple algebraic expression
with coefficients which will correct the previously known Newtonian
ones. A more detailed discussion of this point is presented in
Section~\ref{se:distcoef} and in Appendix~\ref{ap:newtonian}.

We conclude this section by commenting on a novel and particularly
interesting result of our analysis of non-rotating configurations,
namely, that no physical or numerical limit was found for the
magnetization level. More specifically, after suitably increasing the
number of collocation points according to the desired level of
accuracy, we were able to find solutions with very large values of the
surface and quadrupolar deformations. As an example, we could obtain
convergent solutions of the non-rotating reference model of
\citet{Kiuchi2008} with a baryon mass of $M_\rmn{b}=1.680 \,
\mathrm{M}_{\sun}$ and a circumferential radius of $R_\rmn{circ}=14.30 \,
\rmn{km}$ in the unmagnetized case, now strongly magnetized and
attaining a circumferential radius of $R_\rmn{circ} = 101.5 \,
\rmn{km}$ and a related value of $\Delta R_\rmn{circ} = 6.098$. While
the average magnetic field strength $\langle B^{2} \rangle^{1/2} =
0.1461 \times 10^{17}$ G is about an order of magnitude smaller than
the maximum value of $\langle B^{2} \rangle^{1/2} = 2.241 \times
10^{17}$ G reached at a circumferential radius of $R_\rmn{circ} =
17.09 \, \rmn{km}$, the ratio of magnetic energy to binding energy is
much higher with values of $\mathscr{M}/|W|=0.4868$ and
$\mathscr{M}/|W|=0.1308$, respectively. The surface deformation of
this model reaches a value of $\epsilon_{\mathrm{s}}=-0.2403$ and the
quadrupole distortion an impressive value of $\epsilon=-6.946$.
While clearly unrealistic for ordinary neutron stars, young
proto-neutron stars may, in principle, attain these magnetizations
and possibly these sizes~\citep{Villain2004}; clearly, this
represents a possibility that deserves a more careful investigation.

\begin{table*}
\caption{\label{t:modbxxl}The same as in Table~\ref{t:statbmax} but
  for models with large magnetizations of $\mathscr{M}/|W|
  \approx 0.5$.}
\begin{tabular}{@{}lcccccccccccll}
\hline
 Model & $B_{\rmn{max}}$ & $\langle B^{2} \rangle^{1/2}$ &
 $n_{\rmn{b, c}}$ & $M$ &
 $R_{\rmn{circ}}$ & $I$ & $\mathscr{M}/|W|$ &
 $\epsilon_{\rmn{s}}$ & $\epsilon$ & $|\,$GRV2$\,|$ & $|\,$GRV3$\,|$ \\
 & $(\, \times \, 10^{17} \, \rmn{G})$ & $( \, \times \, 10^{17} \, \rmn{G})$ & $( \, \times \, 0.1 \, \rmn{fm}^{-3})$ & $(\mathrm{M}_{\sun})$ & $(\rmn{km})$ &
 $( \, \times \, 10^{38} \, \rmn{kg} \, \rmn{m}^{2})$ & $ $ & \\
\hline
 \texttt{MM-Pol2}   & $2.514$ & $0.1400$ & $1.832$ & $1.508$ & $100.0$ & $9.024$ & $0.5006$ & $-0.2609$ & $-7.803$ & $2 \times 10^{-5}$ & $1 \times 10^{-4}$ \\
 \texttt{MM-APR}    & $6.166$ & $0.2313$ & $4.219$ & $1.544$ & $90.26$ & $4.909$ & $0.5095$ & $-0.1248$ & $-4.810$ & $8 \times 10^{-6}$ & $1 \times 10^{-4}$\\
\hline
\end{tabular}
\end{table*}
\begin{table*}
\caption{\label{t:modbmax} Neutron star models at the maximum
  field strength limit. The gravitational mass is $M = 1.400 \,
  \mathrm{M}_{\sun}$ for each EOS, respectively, in the unmagnetized and
  non-rotating case for which properties are listed in
  Table~\ref{t:distcoef}. $B_{\rmn{max}}$ is the maximum value of the
  magnetic field strength, $\langle B^{2} \rangle^{1/2}$ the root mean
  square value of the magnetic field strength determined over the
  volume of the star, $n_{\rmn{b,c}}$ the central baryon number density
  $( \, \times \, 0.1 \, \rmn{fm}^{-3})$, $M$ the gravitational mass, $R_{\rmn{circ}}$ the
  circumferential radius, $I$ the moment of inertia, $\mathit{\Omega}$ the angular
  velocity,  $T/|W|$ the absolute value of the
  ratio of total kinetic energy $T$ and potential energy
  $W$, $\mathscr{M}/|W|$ the absolute value of the
  ratio of total magnetic energy $\mathscr{M}$ and potential energy
  $W$,  
  $\epsilon_{\rmn{s}}$ the surface deformation, $\epsilon$ the
  quadrupole distortion, and GRV2/GRV3 the estimates of the
  global error of respective models based on the relativistic virial
  identities introduced in Section~\ref{ss:numvir}.}
\begin{tabular}{lcccccccccrrcll}
\hline
 Model & $\!\!B_{\rmn{max}}\!\!$ & $\!\!\langle B^{2} \rangle^{1/2}\!\!$ &
 $n_{\rmn{b, c}}\!\!$ & $\!\!M\!\!$ &
 $R_{\rmn{circ}}\!\!$ & $\!\!I\!\!$ & $\!\!\mathit{\Omega}\!\!$ & $\!\!T/|W|\!\!$ & $\!\!\mathscr{M}/|W|\!\!$ &
 $\epsilon_{\rmn{s}}\!\!$ & $\!\!\epsilon\!\!$ & $\!\!|\,$GRV2$\,|\!\!$ & $\!\!|\,$GRV3$\,|$ \\
 & $\!\!\!\!( \, \times \, 10^{17} \, \rmn{G})\!\!\!$ & $\!\!\!\!( \, \times \, 10^{17} \, \rmn{G})\!\!\!$ & $\!\!\!\!( \, \times \, 0.1 \, \rmn{fm}^{-3})\!\!\!\!$ & $\!\!(\mathrm{M}_{\sun})\!\!$ & $\!\!(\rmn{km})\!\!$ & $\!\!\!\!( \, \times \, 10^{38} \, \rmn{kg} \, \rmn{m}^{2})\!\!\!$ & $\!\!\!\!( \, \times 10^{3} \, \rmn{s}^{-1})\!\!\!\!$ & \\
\hline
 \texttt{PP-Pol2}   & $\!\!7.408\!\!$ & $\!\!2.917\!\!$ & $\!\!8.409\!\!$ & $\!\!1.427\!\!$ & $\!\!14.34\!\!$ & $\!\!1.278\!\!$ & $\!\!0.000\!\!$ & $\!\!0.00000\!\!$ & $\!\!0.1326\!\!$ & $\!\!-0.0806\!\!$ & $\!\!-0.1986\!\!$ & $\!\!4 \times 10^{-11}\!\!$ & $\!\!3 \times 10^{-9}$ \\
 \texttt{PO-Pol2}   & $\!\!5.898\!\!$ & $\!\!2.324\!\!$ & $\!\!7.077\!\!$ & $\!\!1.427\!\!$ & $\!\!16.35\!\!$ & $\!\!1.549\!\!$ & $\!\!3.969\!\!$ & $\!\!0.03200\!\!$ & $\!\!0.1056\!\!$ & $\!\!0.1521\!\!$ & $\!\!-0.0684\!\!$ & $\!\!1 \times 10^{-10}\!\!$ & $\!\!2 \times 10^{-10}$ \\
 \texttt{OO-Pol2}   & $\!\!3.264\!\!$ & $\!\!1.393\!\!$ & $\!\!5.470\!\!$ & $\!\!1.424\!\!$ & $\!\!19.05\!\!$ & $\!\!1.971\!\!$ & $\!\!5.050\!\!$ & $\!\!0.07136\!\!$ & $\!\!0.0422\!\!$ & $\!\!0.5861\!\!$ & $\!\!0.0856\!\!$ & $\!\!7 \times 10^{-7}\!\!$ & $\!\! 4 \times 10^{-6}$ \\
 \texttt{PP-APR}    & $\!\!8.046\!\!$ & $\!\!3.597\!\!$ & $\!\!6.036\!\!$ & $\!\!1.438\!\!$ & $\!\!13.58\!\!$ & $\!\!1.136\!\!$ & $\!\!0.000\!\!$ & $\!\!0.00000\!\!$ & $\!\!0.1754\!\!$ & $\!\!-0.1176\!\!$ & $\!\!-0.3045\!\!$ & $\!\!5 \times 10^{-7}\!\!$ & $\!\!4 \times 10^{-7}$ \\
 \texttt{PO-APR}    & $\!\!7.361\!\!$ & $\!\!3.255\!\!$ & $\!\!5.861\!\!$ & $\!\!1.437\!\!$ & $\!\!14.64\!\!$ & $\!\!1.247\!\!$ & $\!\!4.219\!\!$ & $\!\!0.02671\!\!$ & $\!\!0.1494\!\!$ & $\!\!0.0703\!\!$ & $\!\!-0.1622\!\!$ & $\!\!2 \times 10^{-6}\!\!$ & $\!\!1 \times 10^{-6}$ \\
 \texttt{OO-APR}    & $\!\!5.548\!\!$ & $\!\!2.551\!\!$ & $\!\!5.493\!\!$ & $\!\!1.432\!\!$ & $\!\!16.31\!\!$ & $\!\!1.431\!\!$ & $\!\!6.004\!\!$ & $\!\!0.06276\!\!$ & $\!\!0.0878\!\!$ & $\!\!0.4591\!\!$ & $\!\!0.0072\!\!$ & $\!\!2 \times 10^{-6}\!\!$ & $\!\!3 \times 10^{-6}$ \\
\hline
\end{tabular}
\end{table*}
\begin{table*}
\caption{\label{t:distcoef} Static reference models without magnetic
  field and with a gravitational mass of $M = 1.400 \,
  \mathrm{M}_{\sun}$. $n_{\rmn{b,c}}$ is the central baryon number density
  $( \, \times \, 0.1 \, \rmn{fm}^{-3})$, $M_{\rmn{b}}$ the baryon mass,
  $R_{\rmn{circ}}$ the circumferential radius, $I$ the moment of
  inertia, $b_{B}$ the magnetic distortion coefficient defined in
  equation~(\ref{e:distcoef}), $b_{\mathit{\Omega}}$ the rotational distortion
  coefficient defined in equation~(\ref{e:distcoef}), $c_{B}$ the magnetic
  distortion coefficient defined in equation~(\ref{e:distcoef}),
  $c_{\mathit{\Omega}}$ the rotational distortion coefficient defined in
  equation~(\ref{e:distcoef}), and GRV2/GRV3 the estimates of the global
  error of respective models based on the relativistic virial
  identities introduced in Section~\ref{ss:numvir}. }
\begin{tabular}{@{}lcccccccccll}
\hline
 EOS & $n_{\rmn{b,c}}$ & $M_{\rmn{b}}$ & $R_{\rmn{circ}}$ & $I$ & $b_{B}$ & $b_{\mathit{\Omega}}$ & $c_{B}$ & $c_{\mathit{\Omega}}$ & $|\,$GRV2$\,|$ & $|\,$GRV3$\,|$ \\
 & $( \, \times \, 0.1 \, \rmn{fm}^{-3})$ & $(\mathrm{M}_{\sun})$ & $(\rmn{km})$ & $( \, \times \, 10^{38} \, \rmn{kg} \, \rmn{m}^{2})$ & & & & \\
\hline
 Pol2   & $7.942$ & $1.523$ & $12.00$ & $1.280$ & $5.860 \times 10^{-7}$ & $6.137 \times 10^{-9}$ & $8.338 \times 10^{-7}$ & $2.456 \times 10^{-9}$ & $4 \times 10^{-13}$ & $3 \times 10^{-12}$ \\
 APR    & $5.538$ & $1.553$ & $11.34$ & $1.306$ & $6.545 \times 10^{-7}$ & $5.579 \times 10^{-9}$ & $7.922 \times 10^{-7}$ & $2.481 \times 10^{-9}$ & $6 \times 10^{-5}$ & $8 \times 10^{-5}$ \\
 BBB2   & $6.368$ & $1.555$ & $11.13$ & $1.256$ & $5.809 \times 10^{-7}$ & $5.229 \times 10^{-9}$ & $6.989 \times 10^{-7}$ & $2.282 \times 10^{-9}$ & $6 \times 10^{-5}$ & $8 \times 10^{-5}$ \\
 BN1H1  & $5.018$ & $1.529$ & $12.90$ & $1.590$ & $1.137 \times 10^{-6}$ & $8.215 \times 10^{-9}$ & $1.547 \times 10^{-6}$ & $3.845 \times 10^{-9}$ & $5 \times 10^{-6}$ & $3 \times 10^{-6}$ \\
 BPAL12 & $11.88$ & $1.549$ & $10.06$ & $0.971$ & $2.619 \times 10^{-7}$ & $3.528 \times 10^{-9}$ & $3.178 \times 10^{-7}$ & $1.267 \times 10^{-9}$ & $2 \times 10^{-5}$ & $3 \times 10^{-5}$ \\
 FPS    & $6.949$ & $1.559$ & $10.85$ & $1.200$ & $4.912 \times 10^{-7}$ & $4.821 \times 10^{-9}$ & $5.798 \times 10^{-7}$ & $2.076 \times 10^{-9}$ & $5 \times 10^{-5}$ & $7 \times 10^{-5}$ \\
 GNH3   & $3.654$ & $1.512$ & $14.20$ & $1.814$ & $1.577 \times 10^{-6}$ & $1.078 \times 10^{-8}$ & $2.362 \times 10^{-6}$ & $5.132 \times 10^{-9}$ & $1 \times 10^{-4}$ & $1 \times 10^{-4}$ \\
 SLy4   & $5.376$ & $1.546$ & $11.72$ & $1.367$ & $7.422 \times 10^{-7}$ & $6.141 \times 10^{-9}$ & $9.310 \times 10^{-7}$ & $2.756 \times 10^{-9}$ & $9 \times 10^{-5}$ & $1 \times 10^{-4}$ \\
\hline
\end{tabular}
\end{table*}

In order to illustrate the extreme effects of the toroidal magnetic
field at high levels of magnetization, we have computed model
\texttt{MM-Pol2} of the Pol2 EOS reference model with a gravitational
mass of $M = 1.400 \, \mathrm{M}_{\sun}$ and a circumferential radius of
$R_\rmn{circ}=12.00 \, \rmn{km}$ in the non-rotating and unmagnetized
case, now inflated to a circumferential radius of $R_\rmn{circ}=100.0
\, \rmn{km}$. Such an `extra-large' model is shown in
Fig.~\ref{f:pol2xxl}, which, as the previous ones, reports the
isocontours of magnetic field strength (left-hand panel) and baryon number
density (right-hand panel) for an average magnetic field strength of $\langle
B^{2}\rangle^{1/2} = 0.1400 \times 10^{17}$ G. The physical properties
of model \texttt{MM-Pol2} are listed in Table~\ref{t:modbxxl}. In this
case, therefore, the surface and quadrupolar deformations reach the
extreme values of $\epsilon_\rmn{s}=-0.2609$ and $\epsilon=-7.803$,
respectively.

Although new and somewhat surprising, these results are not totally
unexpected, and we note that already for a purely poloidal magnetic
field, \citet{Cardall2001} doubted the existence of a mass-shedding
limit in the non-rotating case. This leads to the conclusion that the
non-convergence limit encountered by \citet{Kiuchi2008} at
$R_\rmn{circ} = 28.85 \, \rmn{km}$ may be due to their numerical
scheme. Furthermore, to confirm that the extreme distortions reported
for model \texttt{MM-Pol2} are not a peculiarity of the Pol2 EOS, we
have found similar equilibria also for model \texttt{MM-APR}, built
with the APR EOS with a gravitational mass of $M=1.400 \, \mathrm{M}_{\sun}$ and
a circumferential radius of $R_\rmn{circ}=11.34 \, \rmn{km}$ in the
unmagnetized case. This is shown in Fig.~\ref{f:aprxxl} and exhibits
the same spindle-shaped matter distribution of the stellar core for a
circumferential radius of $R_\rmn{circ}=90.26 \, \rmn{km}$ and an
average magnetic field strength of $\langle B^{2}\rangle^{1/2} =
0.2313 \times 10^{17}$ G. The physical properties of model
\texttt{MM-APR} are listed in Table~\ref{t:modbxxl}, but we note here
that the surface and quadrupolar deformations reach the extreme values
of $\epsilon_\rmn{s}=-0.1248$ and of $\epsilon=-4.810$, respectively.
These values are smaller than those of model \texttt{MM-Pol2} (despite
the ratio of magnetic energy to binding energy is similar and
$\mathscr{M}/|W| \approx 0.5$) but shows quite clearly that extremely
large deformations are a feature of non-rotating models, independently
of the EOS.

\section{Rotating Magnetized Models}
\label{se:rotmag}

Having investigated non-rotating models in the previous section, we
next turn to models that include rotation, which is well known to
induce deformations of the stars that are the opposite of those
discussed so far, namely, of introducing an oblateness generically
both in the surface deformation and in the quadrupole distortion. In
particular, we are concerned with determining the limits of the space
of solutions in terms of both the magnetization and the rotation
rate. To probe in detail such a space of solutions we consider two
rotating models corresponding to maximum field strength configurations
of the Pol2 EOS. The first of these reference models, which we refer
to as \texttt{PO-Pol2}, has a mean magnetic field strength of $\langle
B^{2}\rangle^{1/2}=2.324 \times 10^{17} \, \rmn{G}$ and rotates at an
angular velocity of $\mathit{\Omega} = 3.969 \times 10^{3} \, \rmn{s}^{-1}$,
thus with a moderate ratio of kinetic energy to binding energy of
$T/|W| = 0.03200$. The second model, that we refer to as
\texttt{OO-Pol2}, has a mean magnetic field strength of $\langle
B^{2}\rangle^{1/2}=1.393 \times 10^{17} \, \rmn{G}$ and rotates
rapidly at an angular velocity of $\mathit{\Omega} = 5.050 \times 10^{3} \,
\rmn{s}^{-1}$, with a ratio $T/|W| = 0.07136$ (see
Table~\ref{t:modbmax} for a complete list of the physical properties).

\begin{figure*}
\begin{center}
\includegraphics[angle=0,width=7.0cm]{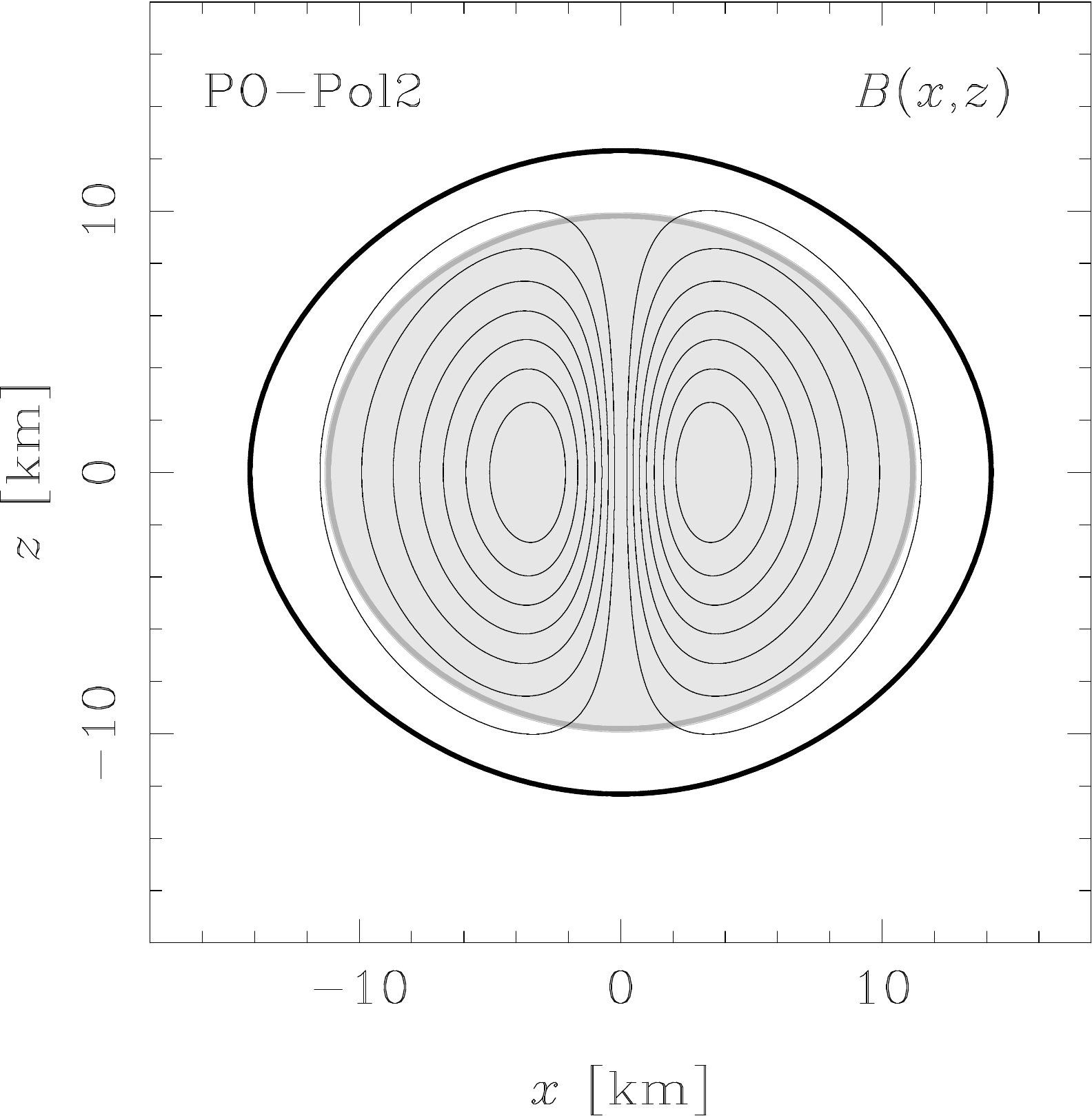}
\hskip 1.5cm
\includegraphics[angle=0,width=7.0cm]{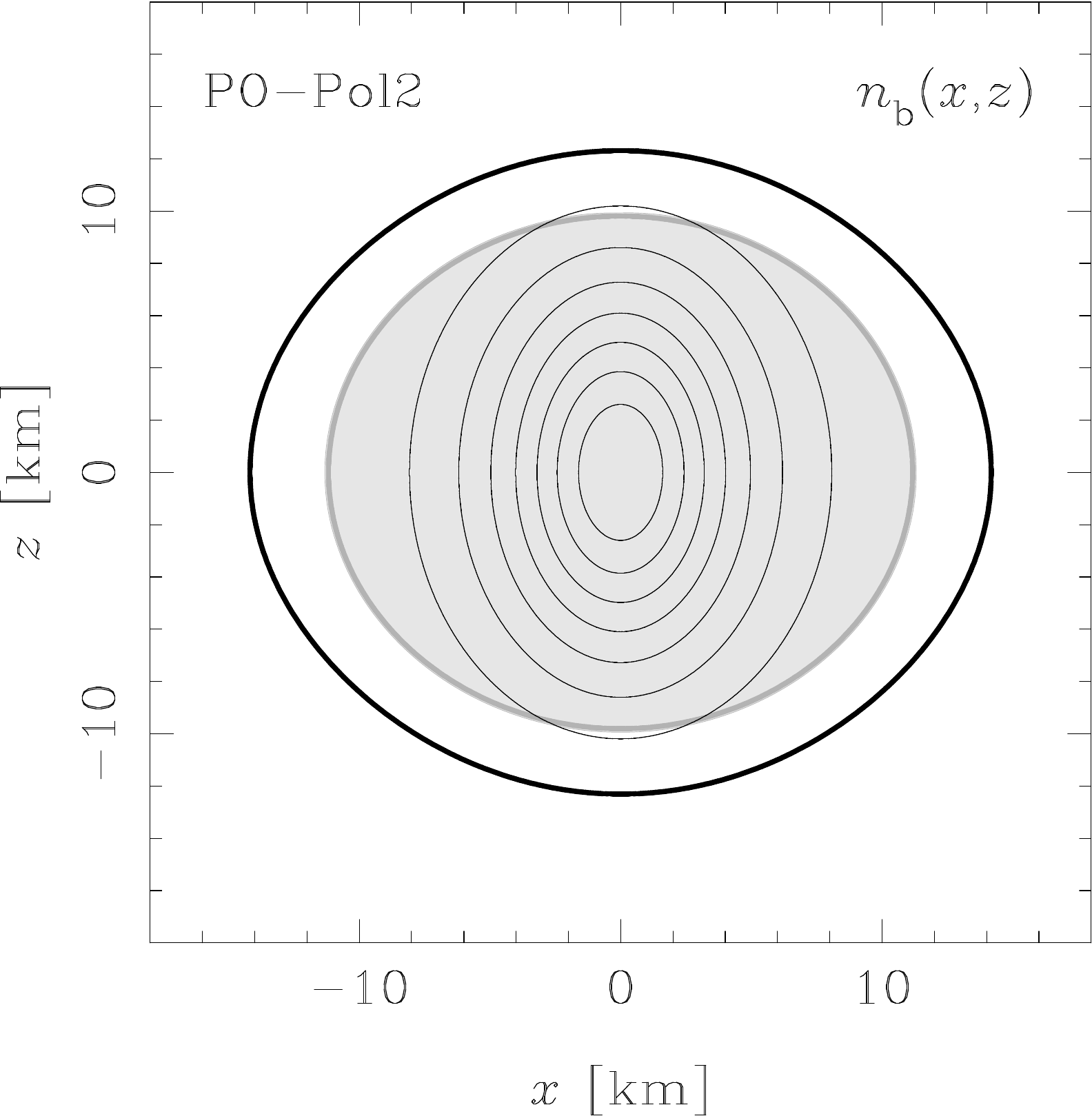}
\vskip 0.75cm
\includegraphics[angle=0,width=7.0cm]{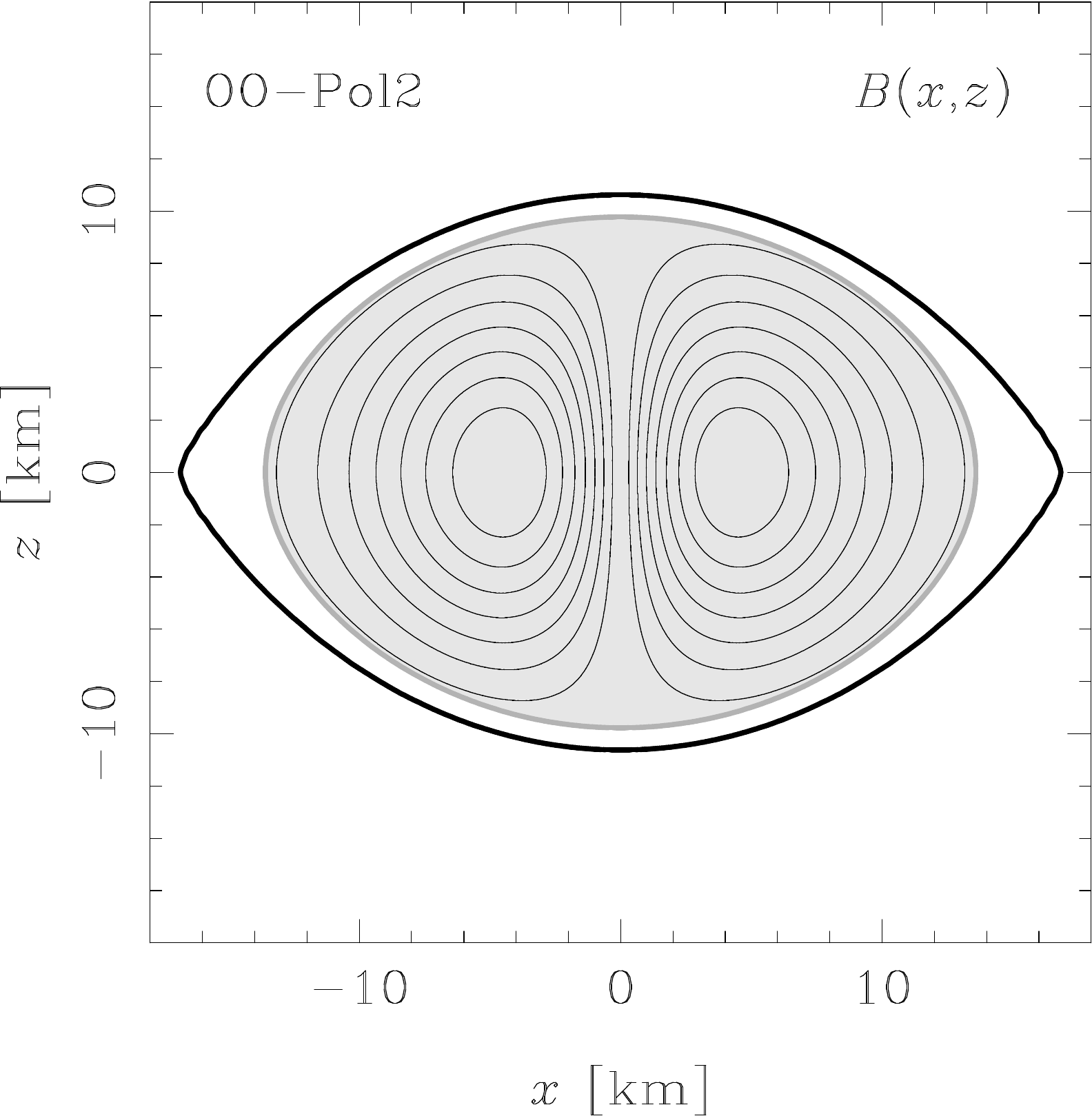}
\hskip 1.5cm
\includegraphics[angle=0,width=7.0cm]{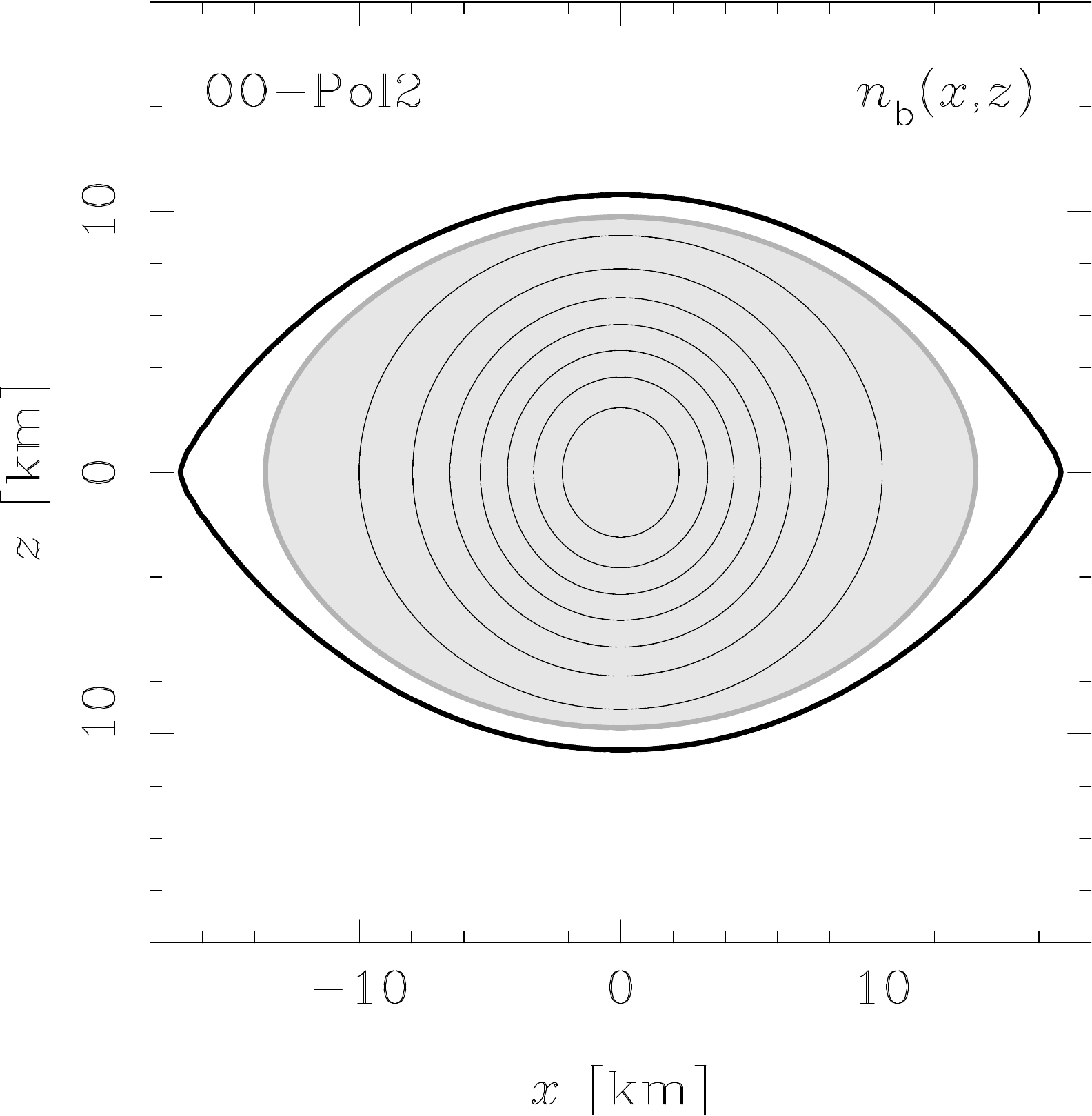}
\caption{Isocontours of magnetic field strength (left-hand panel) and baryon
  number density (right-hand panel) in the $(x,z)$ plane of the \texttt{PO-Pol2}
  and the \texttt{OO-Pol2} models of a star built with a polytropic
  EOS with $\gamma = 2$ with a gravitational mass of $M =
  1.400 \, \mathrm{M}_{\sun}$ and a circumferential radius of
  $R_{\rmn{circ}}=12.00 \, \rmn{km}$ in the unmagnetized and
  non-rotating case which is now rotating at $\mathit{\Omega} = 3.969 \times
  10^{3} \, \rmn{s}^{-1}$ (top) and $\mathit{\Omega} = 5.050 \times 10^{3} \,
  \rmn{s}^{-1}$ (bottom), respectively. The grey disc indicates the
  dimensions of the unmagnetized reference model. The physical
  properties of models \texttt{PO-Pol2} and \texttt{OO-Pol2} are
  listed in Table~\ref{t:modbmax}.}
\label{f:pol2rot}
\end{center}
\end{figure*}

The magnetic field strength and the baryon number density of both
models are shown in Fig.~\ref{f:pol2rot}, with the top row referring
to \texttt{PO-Pol2} and the bottom one to \texttt{OO-Pol2}. The
denomination of these two models becomes apparent when considering the
corresponding surface and quadrupole deformations. While in fact the
qualitative properties of \texttt{PO-Pol2} seem to resemble those of
the non-rotating model \texttt{PP-Pol2}, the surface of this model is
markedly flattened and indeed with a positive oblateness of
$\epsilon_\rmn{s}=0.1521$. The matter distribution inside the star,
however, is still prolately deformed, as confirmed by a negative value
of the quadrupole distortion, namely $\epsilon=-0.0684$. On the other
hand, for model \texttt{OO-Pol2} both the surface deformation and the
quadrupole distortion are positive, \ie $\epsilon_\rmn{s}=0.5861$ and
$\epsilon=0.0856$, even though the isocontours of the baryon number
density are still prolate towards the centre of the
star. Interestingly, for model \texttt{PO-Pol2}, for which the
magnetic field is still dominating, the ratio of the magnetic energy
to the binding energy $\mathscr{M}/|W|$ exceeds that of the kinetic
energy to the binding energy $T/|W|$, whereas for model
\texttt{OO-Pol2}, the opposite is true. Hence, while this condition
does not hold in all cases, we can take the inequality $\mathscr{M}/T
\gtrsim 1$ as a first approximate criterion for the production of a
negative quadrupole distortion. A more quantitative discussion on
this will be presented in Section~\ref{se:distcoef}.

In our sampling of the space of parameters we have computed a total
of more than 900 models of rotating and magnetized equilibrium
configurations of the Pol2 EOS, which are \textit{uniquely} labelled
by the values of the angular velocity $\mathit{\Omega}$ and of the
magnetization parameter $\lambda_{0}$. The latter extends up to a
maximum obtained for a non-rotating model with a radius of
$R_{\rmn{circ}}=19.45 \, \rmn{km}$ and a ratio of magnetic energy to
binding energy of $\mathscr{M}/|W| = 0.2448$.\footnote{Note that the
  `extra-large' models shown in Figs~\ref{f:pol2xxl} and
  \ref{f:aprxxl} are located beyond the truncation limit and thus do
  not belong to what we consider as the parameter space.} Overall,
the space of physical solutions is delimited by four boundaries:
(1) the non-rotating limit with $\mathit{\Omega} = 0$; (2) the
unmagnetized limit with $B = 0$; (3) the (self-imposed) `truncation
limit' with respect to the magnetization parameter $\lambda_{0}$;
(4) the mass-shedding limit, beyond which no rotating solution does
exist.

We have already noted that for non-rotating models the mean
magnetic field strength $\langle B^{2}\rangle^{1/2}$ is not a
monotonic function of the magnetization parameter $\lambda_{0}$ and that
after attaining a maximum value of $\langle B^{2}\rangle^{1/2}=2.917
\times 10^{17} \, \rmn{G}$, it decreases continuously as the
magnetization is increased. This behaviour has been found also for
rotating models, thus implying a non-uniqueness for models in the
space $(\mathit{\Omega}^2, \langle B^{2} \rangle)$. Such a degeneracy could be
avoided by replacing the average magnetic field strength with a
quantity that grows monotonically with the magnetization, \eg the
circumferential radius $R_{\rmn{circ}}$ or the ratio of magnetic
energy to binding energy $\mathscr{M}/|W|$. However, because of their
fundamental astrophysical importance, we have chosen to report the
results for the surface deformation $\epsilon_\rmn{s}$ and the
quadrupole distortion $\epsilon$ in
terms of the mean magnetic field strength $\langle B^{2}\rangle^{1/2}$
as the ordering quantity and of the angular velocity $\mathit{\Omega}$.

\begin{figure*}
\begin{center}
\includegraphics[angle=-0,width=7.0cm]{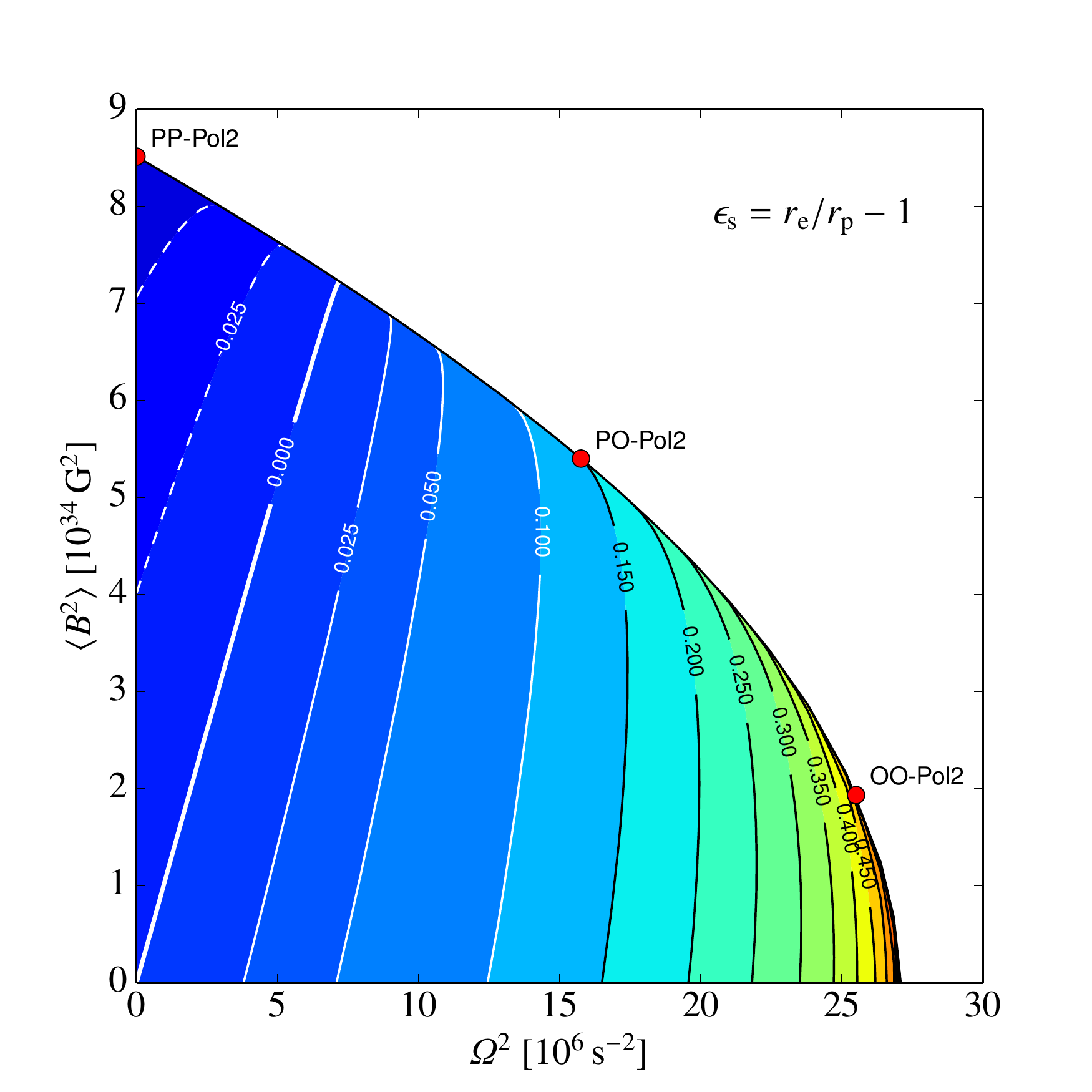}
\hskip 1.5cm
\includegraphics[angle=-0,width=7.0cm]{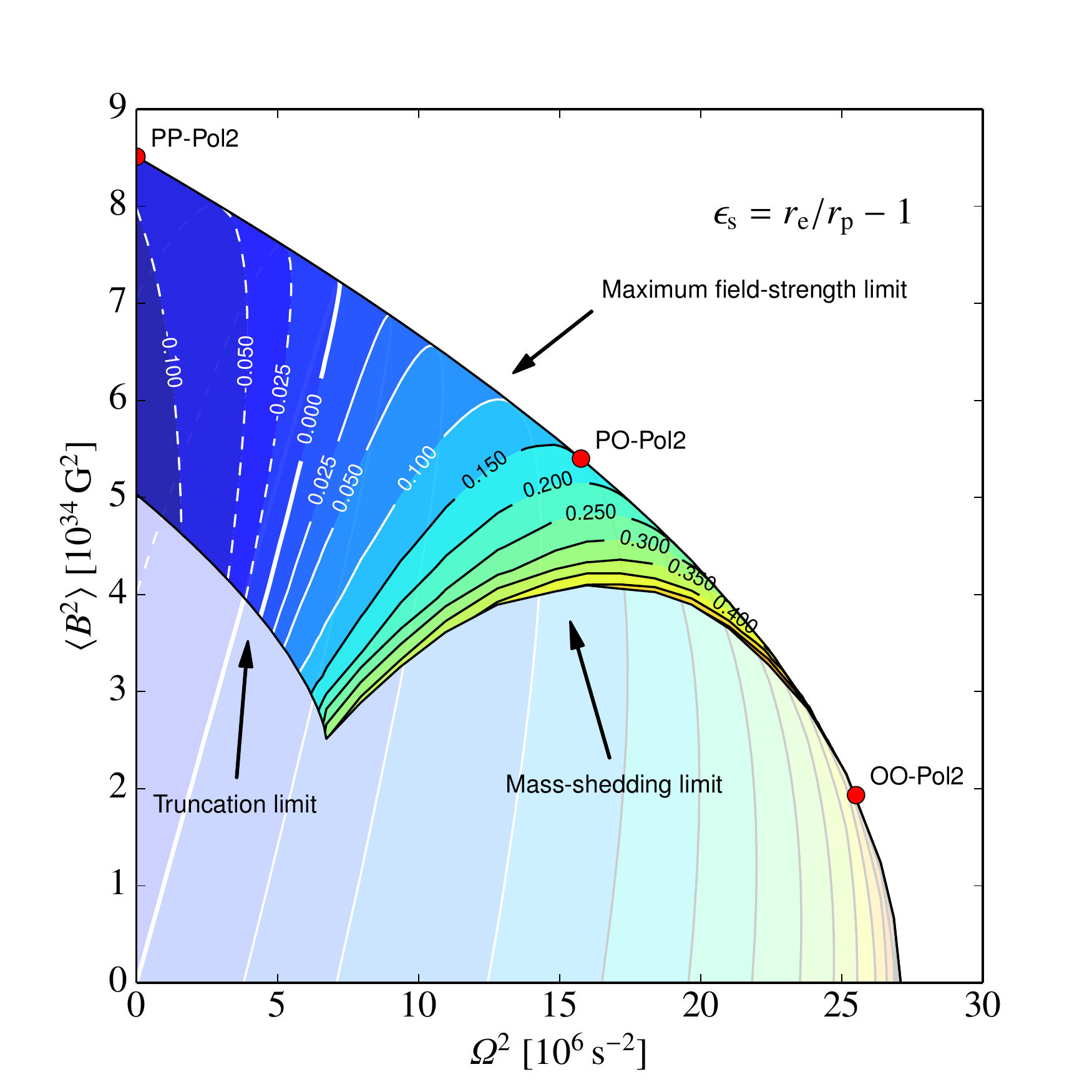}
\caption{Isocontours of the surface deformation $\epsilon_{\rmn{s}}$
  for the Pol2 EOS
  reference model with a gravitational mass of $M = 1.400 \, \mathrm{M}_{\sun}$
  and a circumferential radius of $R_\rmn{circ} = 12.00 \, \rmn{km}$
  in the unmagnetized and non-rotating case. The curved line which
  connects the non-rotating model with a maximum average magnetic field
  strength of $\langle B^{2}\rangle ^{1/2} = 2.917 \times 10^{17}$ G
  with the unmagnetized model rotating at the mass-shedding limit of
  $\mathit{\Omega} = 5.205 \times 10^{3} \, \rmn{s}^{-1}$ separates the lower
  part of the solution space (left-hand panel) where the mean
  magnetic field strength increases as a function of the field
  strength parameter $\lambda_{0}$ from the upper part of the solution
  space (right-hand panel) where the mean magnetic field strength decreases
  as a function of the latter. In the right-hand panel, we have also shown
  the lower sheet lying underneath.}
\label{f:ell}
\end{center}
\end{figure*}

The presence of this degeneracy implies that when evaluated in the
space of parameters $(\mathit{\Omega}^2, \langle B^{2} \rangle)$, the
distortions $\epsilon_{\rm s}$ and $\epsilon$ will select a
two-dimensional surface which can be split up along the turning points
of maximum magnetic field strength into a lower sheet, where $\langle
B^{2}\rangle^{1/2}$ increases as a function of the magnetization
parameter $\lambda_{0}$, and an upper sheet, where the opposite
happens. The left-hand panel of Fig.~\ref{f:ell}, in particular, shows the
surface deformation $\epsilon_{\rmn{s}}$ for the lower part of the
solution space, while the right-hand panel is its continuation beyond the
turning point of $\langle B^{2}\rangle^{1/2}$ and thus represents the
upper sheet of the surface. Since non-rotating magnetized models always
have a prolate shape, \ie $\epsilon_{\rmn{s}} < 0$, while rotating
unmagnetized models always have an oblate one, \ie $\epsilon_{\rmn{s}}
> 0$, it follows that rotating models between these limiting cases are
divided into \textit{prolate} ones and \textit{oblate} ones by a
neutral line $\epsilon_{\rmn{s}} = 0$. According to its definition in
equation~(\ref{e:oblat}), the neutral line $\epsilon_{\rmn{s}} = 0$ does
not require the stellar interior to be spherically symmetric and,
indeed, when moving towards large values of the circumferential radius
$R_{\rmn{circ}}$, models with comparable values of the equatorial and
polar coordinate radii $r_{\rmn{e}}, r_{\rmn{p}}$ (\ie with
$\epsilon_{\rmn{s}} \to 0$), show a progressively more pronounced
`diamond shape' caused by an increasingly spindle-shaped matter
distribution inside the star.

\begin{figure*}
\begin{center}
\includegraphics[angle=-0,width=7.0cm]{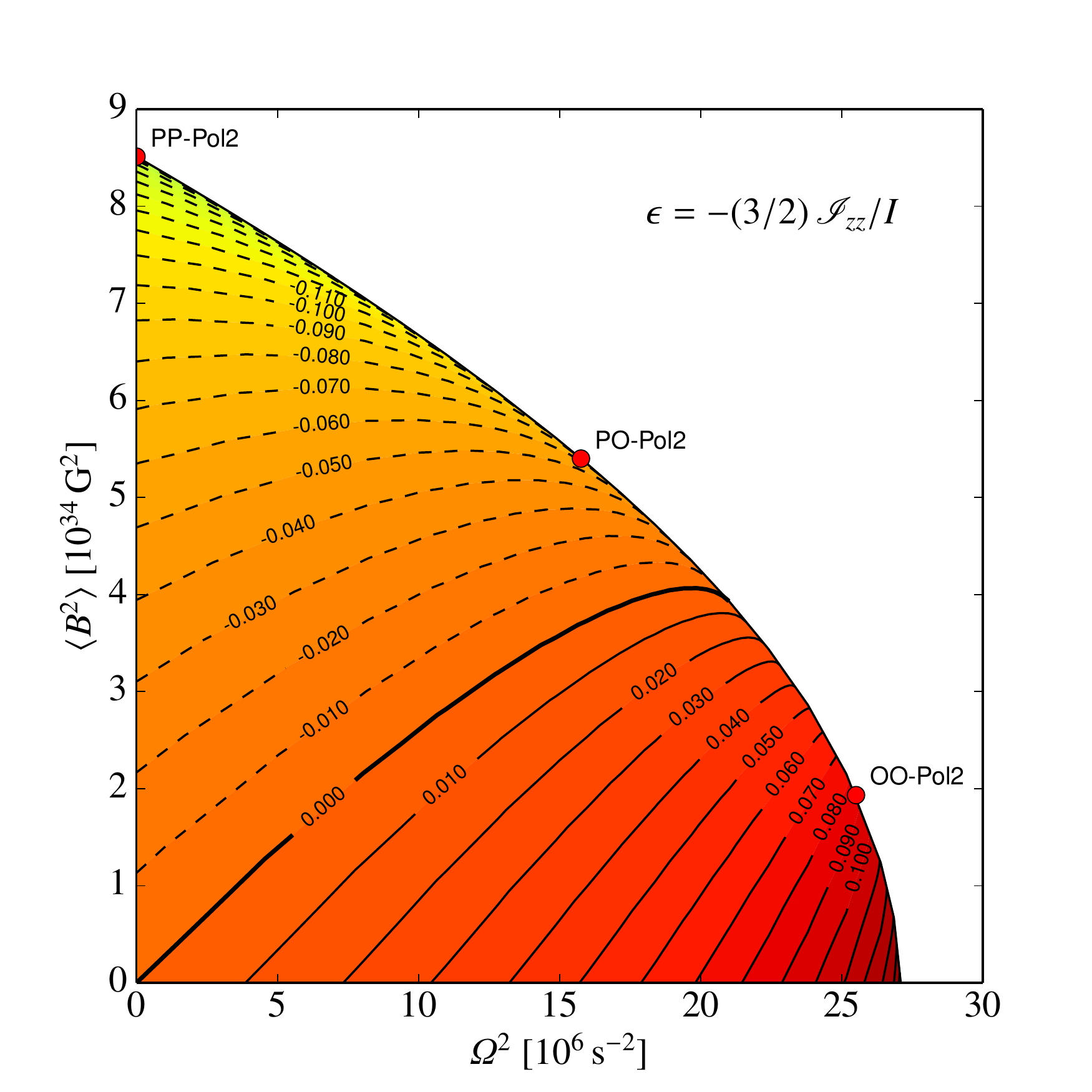}
\hskip 1.5cm
\includegraphics[angle=-0,width=7.0cm]{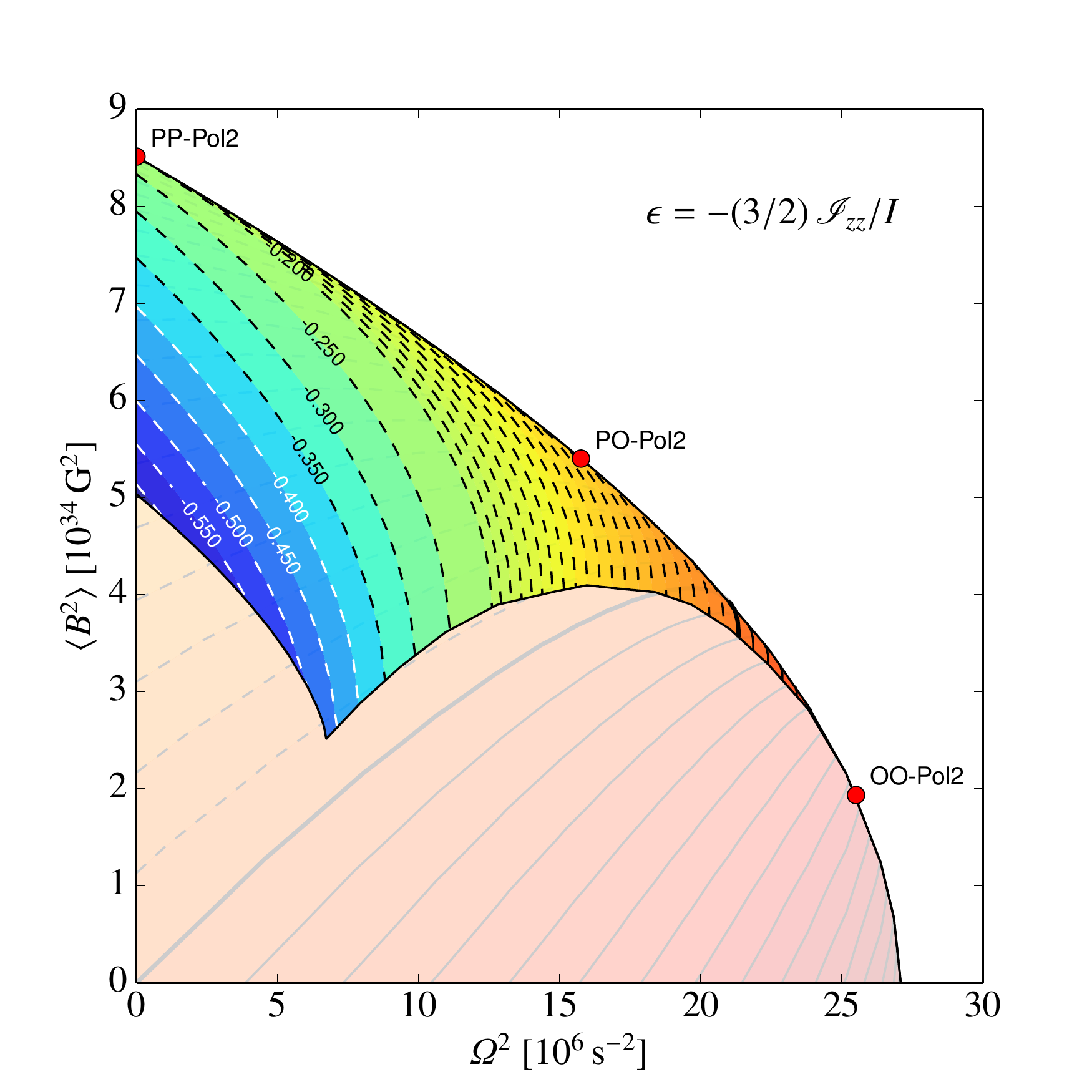}
\caption{Isocontours of the quadrupole distortion $\epsilon$ for the
  Pol2 EOS reference
  model with a gravitational mass of $M = 1.400 \, \mathrm{M}_{\sun}$ and a
  circumferential radius of $R_\rmn{circ} = 12.00 \, \rmn{km}$ in the
  unmagnetized and non-rotating case. The curved line which connects
  the non-rotating model with a maximum average magnetic field strength
  of $\langle B^{2} \rangle^{1/2} = 2.917 \times 10^{17}$ G with the
  unmagnetized model rotating at the mass-shedding limit of $\mathit{\Omega} =
  5.205 \times 10^{3} \, \rmn{s}^{-1}$ separates the lower part of the
  solution space (left-hand panel) where the mean magnetic field strength
  increases as a function of the field strength parameter $\lambda_{0}$
  from the upper part of the solution space (right-hand panel) where the
  mean magnetic field strength decreases as a function of latter.
  In the right-hand panel we have also shown the lower sheet lying underneath.}
\label{f:eps}
\end{center}
\end{figure*}

Unlike the magnetic potential, the centrifugal one is not confined to
the star, and its influence increases moving away from the axis of
rotation, reaching its minimum at the equator of the star. Note that
$\tilde{M} = 0$ at the surface of the star, so that the latter is
now an isosurface of the function $\nu - \ln\mathit{\Gamma}$ (\cf
  equation~\ref{EQ:EM13}). From Fig.~\ref{f:ell}, we further infer that
for any fixed angular velocity $\mathit{\Omega} \geq \mathit{\Omega}_{0}$,
where $\mathit{\Omega}_{0} = 2.594 \times 10^{3} \, \rmn{s}^{-1}$ is
the maximum angular velocity at the magnetization truncation limit,
the mass-shedding limit is reached for some oblate shape and
$\epsilon_{\rmn{s}} > 0$. Since the toroidal magnetic field is a
source of additional pressure, augmenting the magnetization not only
increases the deformation of the surface and of the matter
distribution, but it also causes an expansion, particularly of the
circumferential radius $R_\rmn{circ}$. At a sufficient level of
magnetization, it will therefore be possible to reach the mass-shedding
angular velocity, \ie the rotation frequency such that the condition of
geodesic motion at the equator is satisfied, and the star will develop
the characteristic cusp at the equator. Because this angular frequency
is smaller than the corresponding one for an unmagnetized model having
the same rest-mass, the toroidal magnetic field indirectly sets a reduced
limit of the spin frequency of these objects. In Fig.~\ref{f:ell}, this
mass-shedding limit corresponds to the lower right-hand boundary of the
upper part of the solution space (see right-hand panel) and along this
limit, the Newtonian condition of geodesic motion, $\mathit{\Omega}^{2}
R^{3}_\rmn{circ} = \rmn{const}$, is fulfilled. Note also that with the
exception of the unmagnetized one, all mass-shedding models appear to
belong to the upper sheet of the solution space. For models rotating
at angular velocities up to that of model \texttt{OO-Pol2}, \ie
$\mathit{\Omega} = 5.050 \times 10^{3} \, \rmn{s}^{-1}$, we have
verified this proposition directly by determining the turning point of
$\langle B^{2} \rangle^{1/2}$; for even more rapidly rotating models
for which the numerical determination of the maximum field strength
limit has not been conclusive, it is supported by extrapolating the
boundary between lower and upper part of the solution space
beyond model \texttt{OO-Pol2} towards the unmagnetized mass-shedding limit.

Using a small set of additional rotating models computed beyond the
magnetization truncation limit, we have started exploring the
behaviour of the equilibrium models in these rather extreme
conditions. Overall, we have found that the angular velocity of
mass-shedding models decreases progressively, while the boundary
associated with the mass-shedding limit and the neutral line
$\epsilon_{\rmn{s}} = 0$ converge to the point
$(\mathit{\Omega}^{2},\langle B^{2}\rangle) = (0,0)$. These results suggest
therefore that the solution at $(\mathit{\Omega}^{2},\langle B^{2}\rangle) =
(0,0)$ in the upper sheet of the space of solutions corresponds to the
limit of a non-rotating model of infinite radius and vanishing mean
magnetic field strength. This fascinating suggestion clearly requires
a more extensive analysis to be confirmed; we postpone this to a
subsequent work.

Fig.~\ref{f:ell} also reveals that unlike in the unmagnetized rotating
case with the same angular velocity $\mathit{\Omega}$, all magnetized
mass-shedding configurations behave qualitatively identically when the
magnetization is altered, namely, lowering their magnetization moves
them away from the mass-shedding limit. Accordingly, rotation becomes
sub-critical, and the characteristic cusp at the equator disappears,
which is accompanied by a decrease of the surface
deformation. Slowly rotating models pass through different stages as
the magnetization is lowered from the mass-shedding limit to the given
angular velocity $\mathit{\Omega}$ down to the limit of vanishing
magnetization. More specifically, we find that: (i) below the
mass-shedding limit, $\epsilon_{\rmn{s}}$ remains positive, but
decreases continuously until the neutral line $\epsilon_{\rmn{s}} = 0$
is crossed for the first time; (ii) models then become increasingly
prolate because the toroidal magnetic field becomes the
principal source of deformation until they reach some negative minimum
value of $\epsilon_{\rmn{s}}$; (iii) as the magnetization is further
decreased, $\epsilon_{\rmn{s}}$ increases progressively, and models
eventually become oblate again when they cross the neutral line
$\epsilon_{\rmn{s}} = 0$ for a second time, and rotation prevails
over a decreasing toroidal magnetic field. For models rotating at
moderate angular velocities of $3 \times 10^{3} \, \rmn{s}^{-1} \la
\mathit{\Omega} \la 5 \times 10^{3} \, \rmn{s}^{-1}$, the behaviour is similar
to that of slowly rotating ones, except that $\epsilon_{\rmn{s}}$
remains positive, so that the models always exhibit an oblate shape
regardless of the level of magnetization. Finally, rapidly rotating
models with $\mathit{\Omega} \ga 5 \times 10^{3} \, \rmn{s}^{-1}$ remain close
to the magnetized mass-shedding limit regardless of the magnetization
level.

Generally speaking, for any fixed angular velocity $\mathit{\Omega}$,
the line $\mathit{\Omega}=\rmn{const}$ is tangential to exactly one
level curve of $\epsilon_{\rmn{s}}$. The touching point determines
the magnetization level for which the surface deformation attains
its minimum. For decreasing angular velocity, these minima move to higher
magnetizations and smaller values of $\epsilon_{\rmn{s}}$. The angular
velocity $\mathit{\Omega}'$ for which the maximum magnetic field strength model
exhibits $\epsilon_{\rmn{s}}=0.007$ separates rotating models into two
groups: (1) models rotating at $\mathit{\Omega} > \mathit{\Omega}'$ are always oblate,
and the model of minimum (positive) surface deformation is located in
the lower part of the solution space; (2) models rotating at
$\mathit{\Omega} < \mathit{\Omega}'$ include a model of minimum surface deformation
$\epsilon_{\rmn{s}} \leq 0.007$ which is located in the upper part of
the solution space.

Fig.~\ref{f:eps} provides an equivalent representation of the data
shown in Fig.~\ref{f:ell} but this time in terms of the quadrupole
distortion $\epsilon$. Also in this case, the space of the numerical
solutions is split into an upper sheet and a lower one, where the
distinction is the same one as made for the surface distortion
$\epsilon_{\rmn{s}}$. For all models, $\epsilon$ increases
monotonically as a function of the magnetization, and the neutral line
$\epsilon = 0$ extends from the non-rotating and unmagnetized
reference model, up to a strongly magnetized mass-shedding model with
$\langle B^{2}\rangle ^{1/2} \approx 2 \times 10^{17} \, \rmn{G}$
rotating at an angular velocity of $\mathit{\Omega} \approx 4.5 \times 10^{3}\,
\rmn{s}^{-1}$, and with $T/|W| \approx 0.06$. Thus, even rapidly
rotating models can exhibit a prolate matter distribution provided the
magnetization is sufficiently strong. Although we have truncated the
solution space towards high magnetizations, the minimum
quadrupole distortion of $\epsilon=-0.6127$, which corresponds to the
strongest prolate deformation of our sample and has been obtained for
the non-rotating model at the magnetization truncation limit, exceeds
considerably the quadrupole distortion $\epsilon=0.1537$ of the unmagnetized
mass-shedding model rotating at $\mathit{\Omega}=5.205 \times 10^{3}\,
\rmn{s}^{-1}$ with $T/|W|=0.09380$. This significant difference can be
explained by the large ratio $\mathscr{M}/|W|=0.2448$ of the
magnetized model. Since no mass-shedding limit has been encountered in
the non-rotating magnetized case, $\mathscr{M}/|W|$ can possibly attain
arbitrarily high values and accordingly, the quadrupole distortion
$\epsilon$ can, in principle, grow without bounds, too. We leave the
assessment of this conjecture to a subsequent work.

\begin{figure*}
\begin{center}
\includegraphics[angle=0,width=7.0cm]{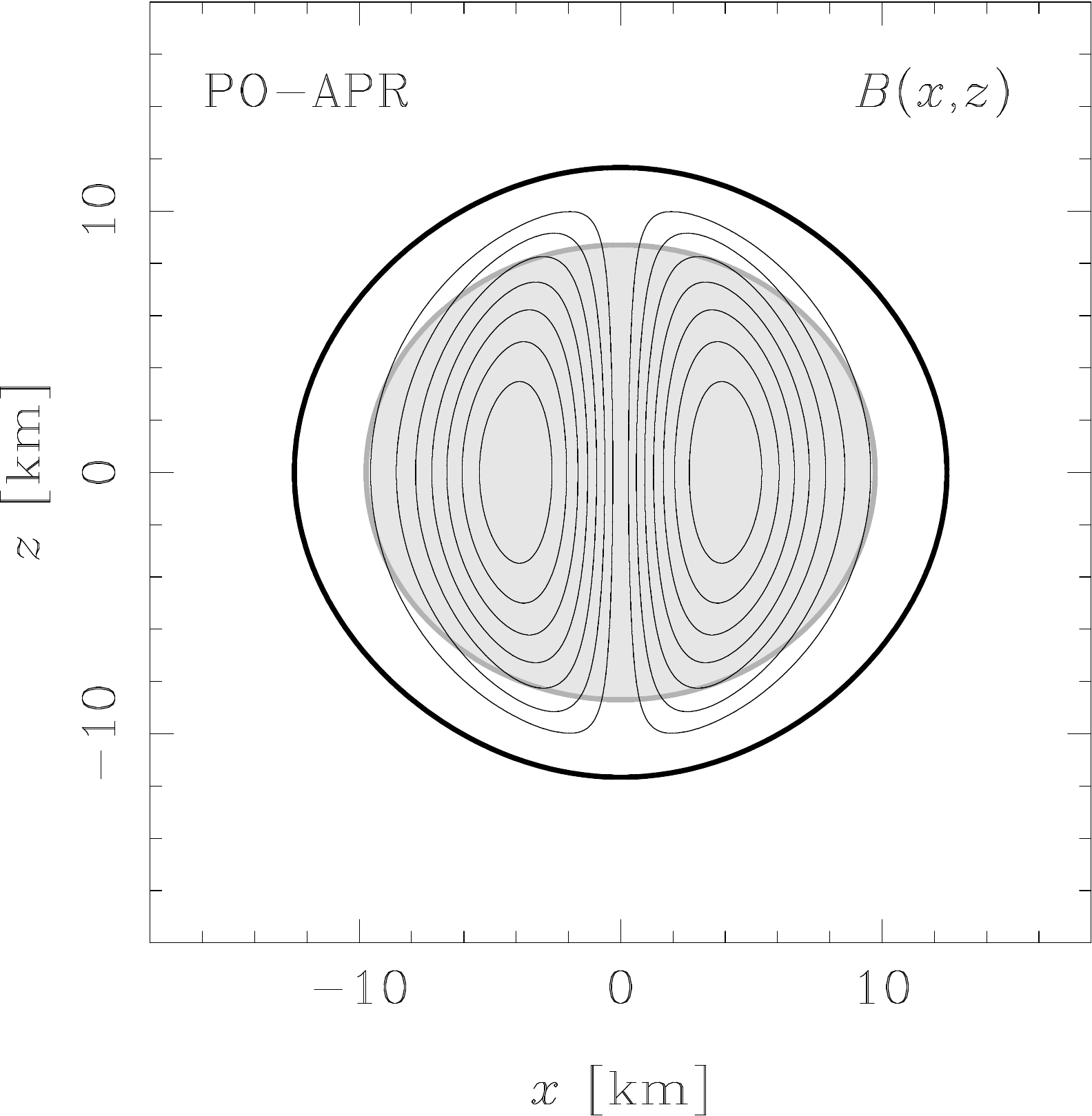}
\hskip 1.5cm
\includegraphics[angle=0,width=7.0cm]{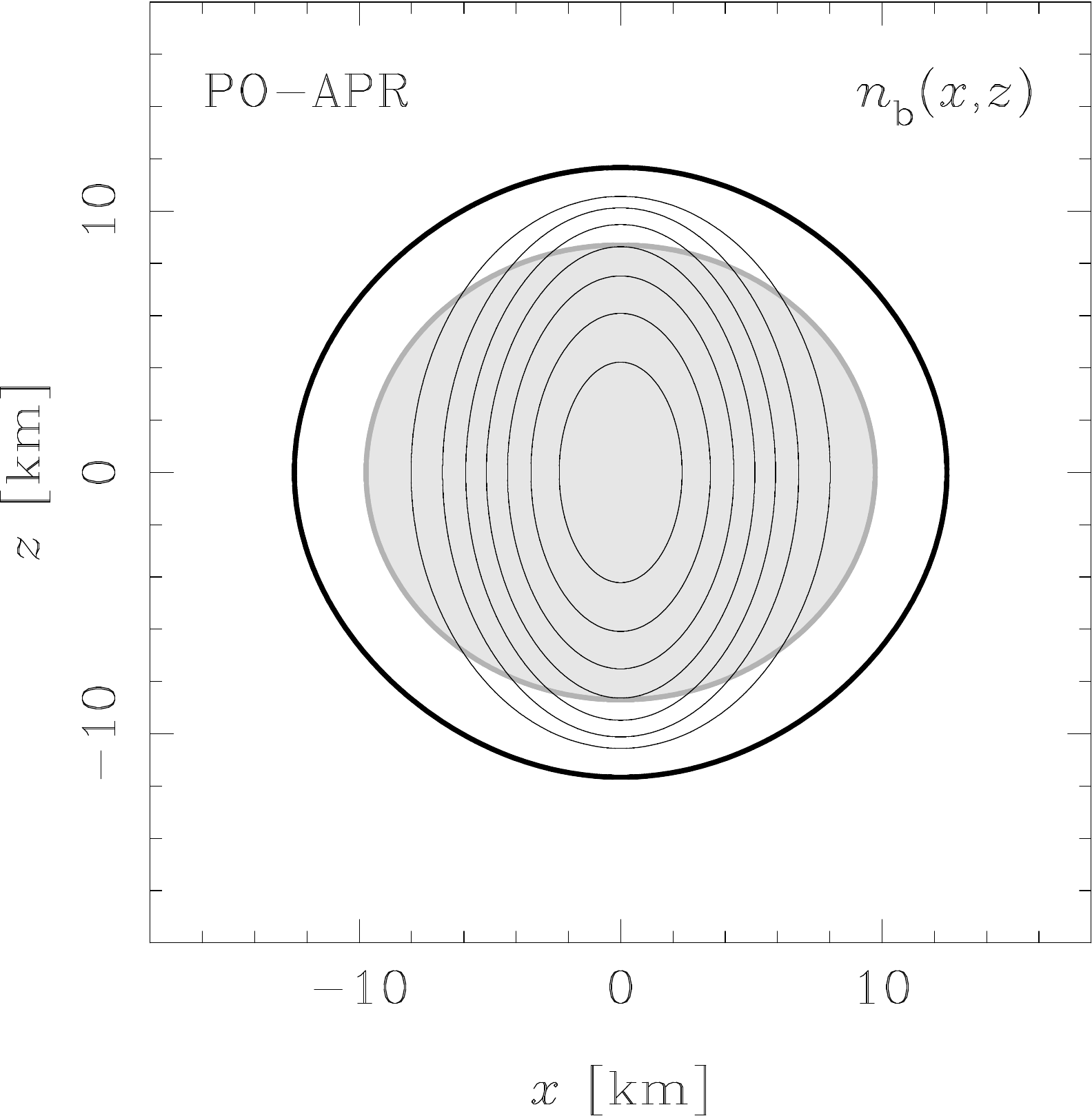} \\
\vskip 0.75cm
\includegraphics[angle=0,width=7.0cm]{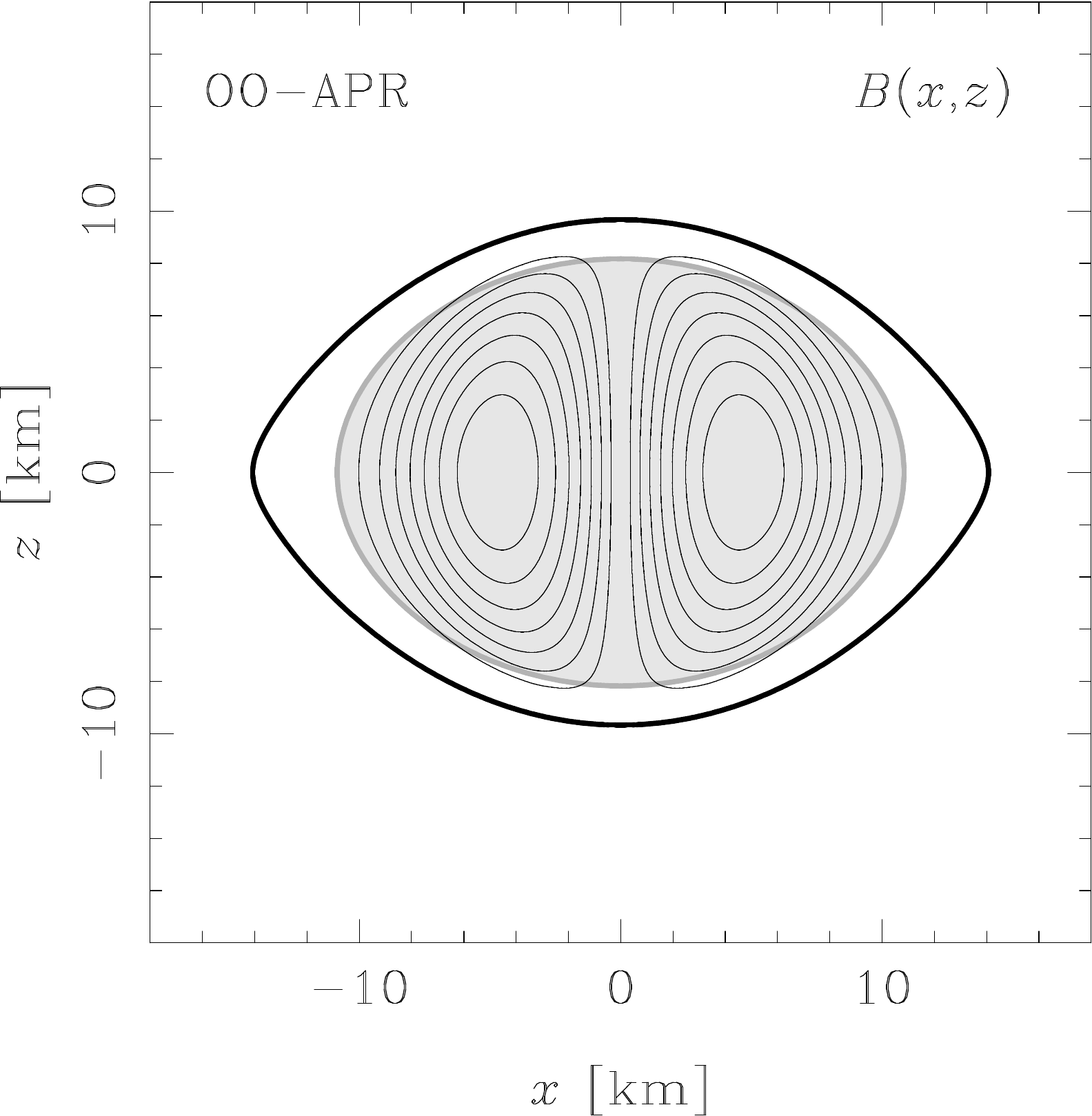}
\hskip 1.5cm
\includegraphics[angle=0,width=7.0cm]{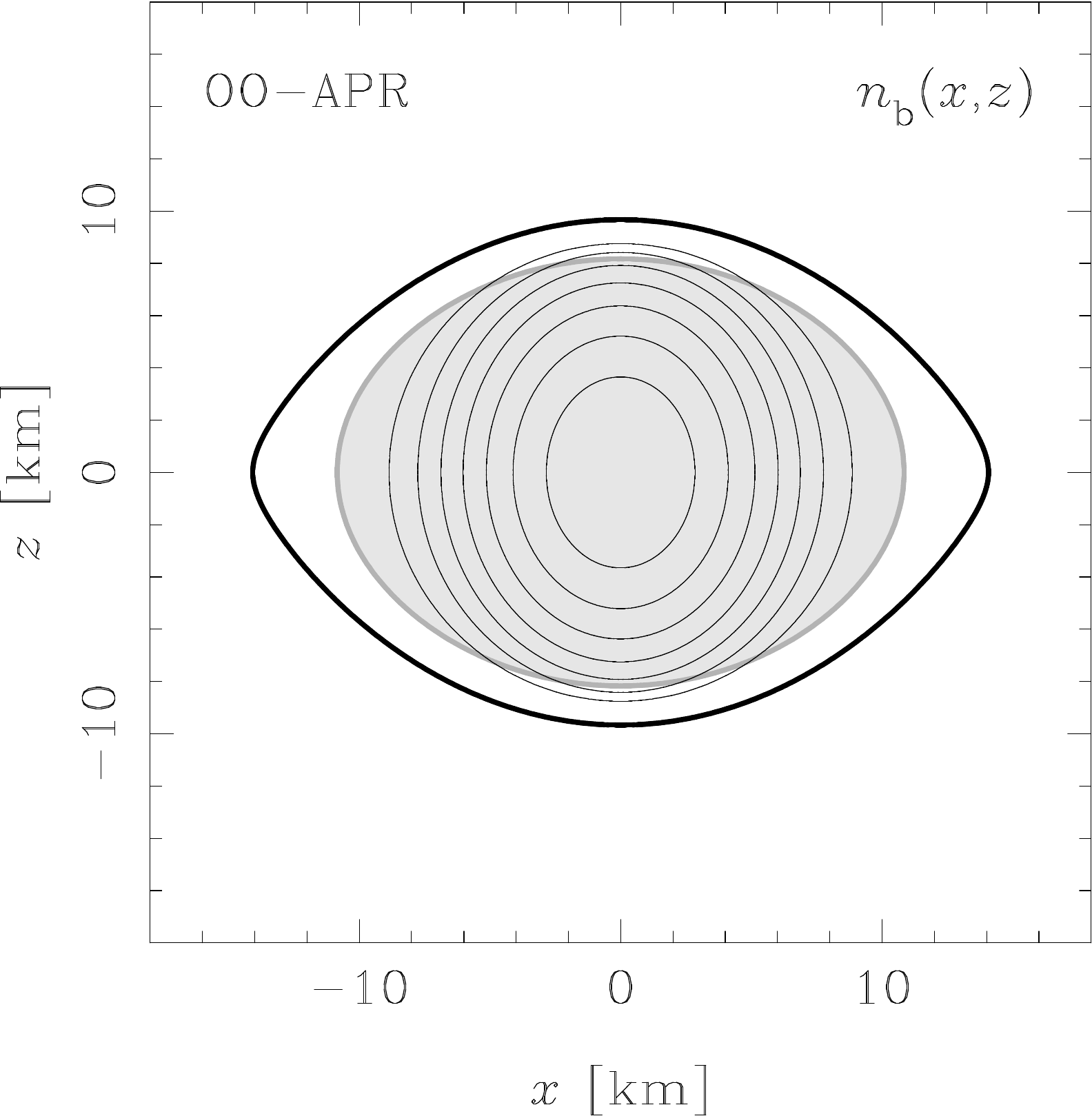}
\caption{Isocontours of magnetic field strength (left-hand panel) and baryon
  number density (right-hand panel) in the $(x,z)$ plane of the \texttt{PO-APR}
  and \texttt{OO-APR} models of a star built with the APR EOS with a
  gravitational mass of $M = 1.400 \, \mathrm{M}_{\sun}$ in the unmagnetized
  and non-rotating case which is now rotating at $\mathit{\Omega} = 4.219 \times
  10^{3} \, \rmn{s}^{-1}$ (top) and $\mathit{\Omega} = 6.004 \times 10^{3} \,
  \rmn{s}^{-1}$ (bottom), respectively. The grey disc indicates the
  dimensions of the unmagnetized reference model. Physical properties
  of models \texttt{PO-APR} and \texttt{OO-APR} are listed in
  Table~\ref{t:modbmax}.}
\label{f:aprrot}
\end{center}
\end{figure*}

We also note that whereas all mass-shedding models exhibit an oblate
surface deformation for which $\epsilon_\rmn{s} > 0$, only rapidly
rotating mass-shedding models with $\mathit{\Omega} \ga 4.5 \times 10^{3} \,
\rmn{s}^{-1}$ show a positive quadrupole distortion with $\epsilon >
0$. On the other hand, all mass-shedding models rotating at lower
angular velocities actually possess a prolate matter distribution and
$\epsilon < 0$ which, moreover, seems to grow without bounds for
increasing magnetization like in the non-rotating case. Altogether,
Figs~\ref{f:ell} and \ref{f:eps} reveal that the neutral lines
$\epsilon_\rmn{s} = 0$ and $\epsilon = 0$ differ
significantly, suggesting the division of the space of solutions of
magnetized and rotating models into three classes: (1) models for
which apparent shape and distortion of the matter distribution are
both prolate, thus with $\epsilon_\rmn{s} < 0$ and $\epsilon < 0$,
which we label \texttt{PP} for prolate--prolate; (2) models for which
the apparent shape is oblate and the distortion of the matter
distribution is prolate, thus with $\epsilon_\rmn{s} > 0$ and
$\epsilon < 0$, which we label \texttt{PO} for prolate--oblate; (3)
models for which apparent shape and distortion of the matter
distribution are both oblate, thus with $\epsilon_\rmn{s} > 0$ and
$\epsilon > 0$, which we label \texttt{OO} for oblate--oblate. As a
result, in contrast with the results of~\citet{Kiuchi2008}, rotating
models with a strong toroidal magnetic field do not necessarily
exhibit a negative quadrupole distortion $\epsilon$ (\cf discussion
  in Section~\ref{se:overview} and Fig.~\ref{f:cartoon}).

\begin{figure*}
\begin{center}
\includegraphics[angle=-0,width=7.0cm]{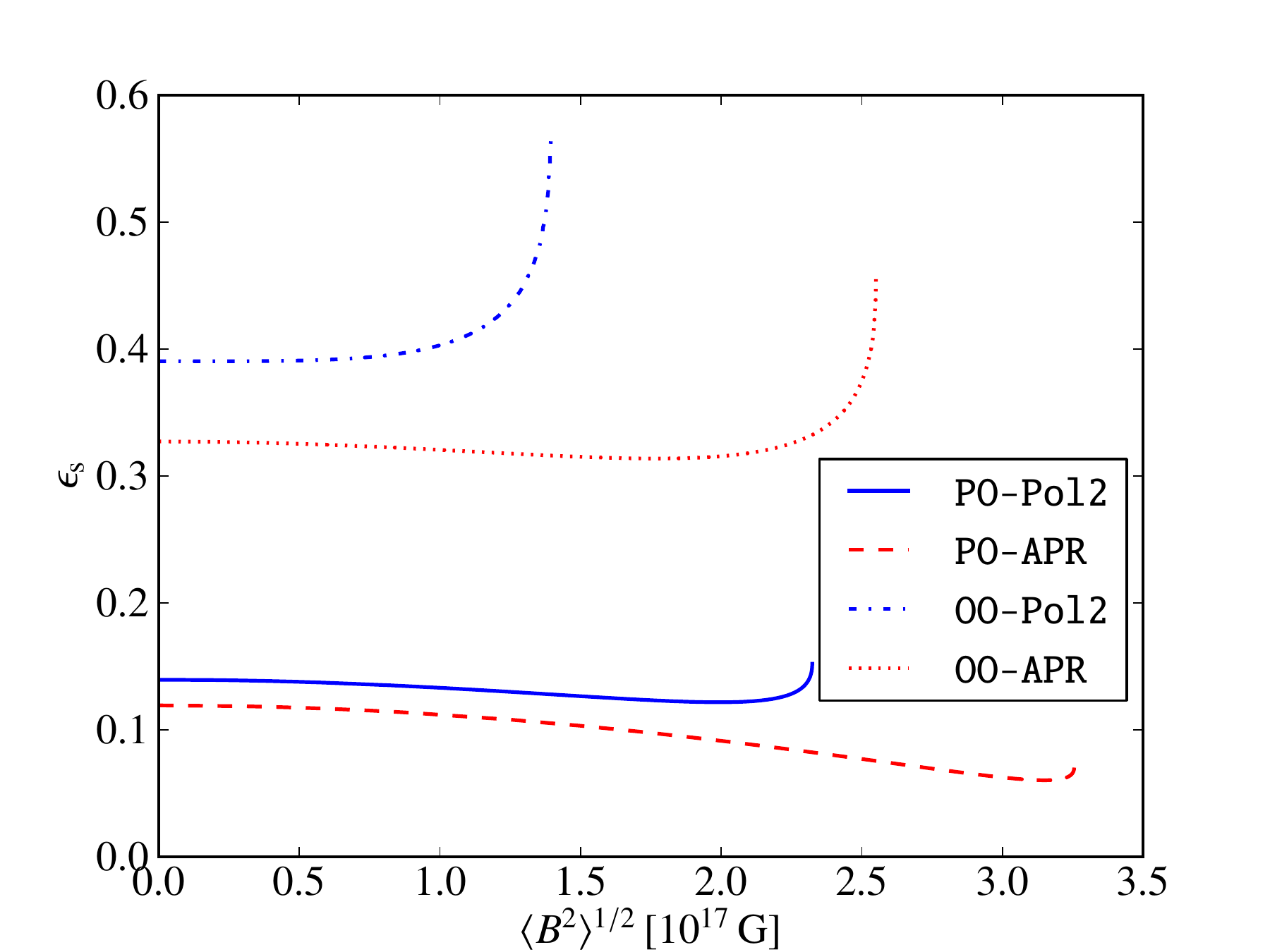}
\hskip 1.5cm
\includegraphics[angle=-0,width=7.0cm]{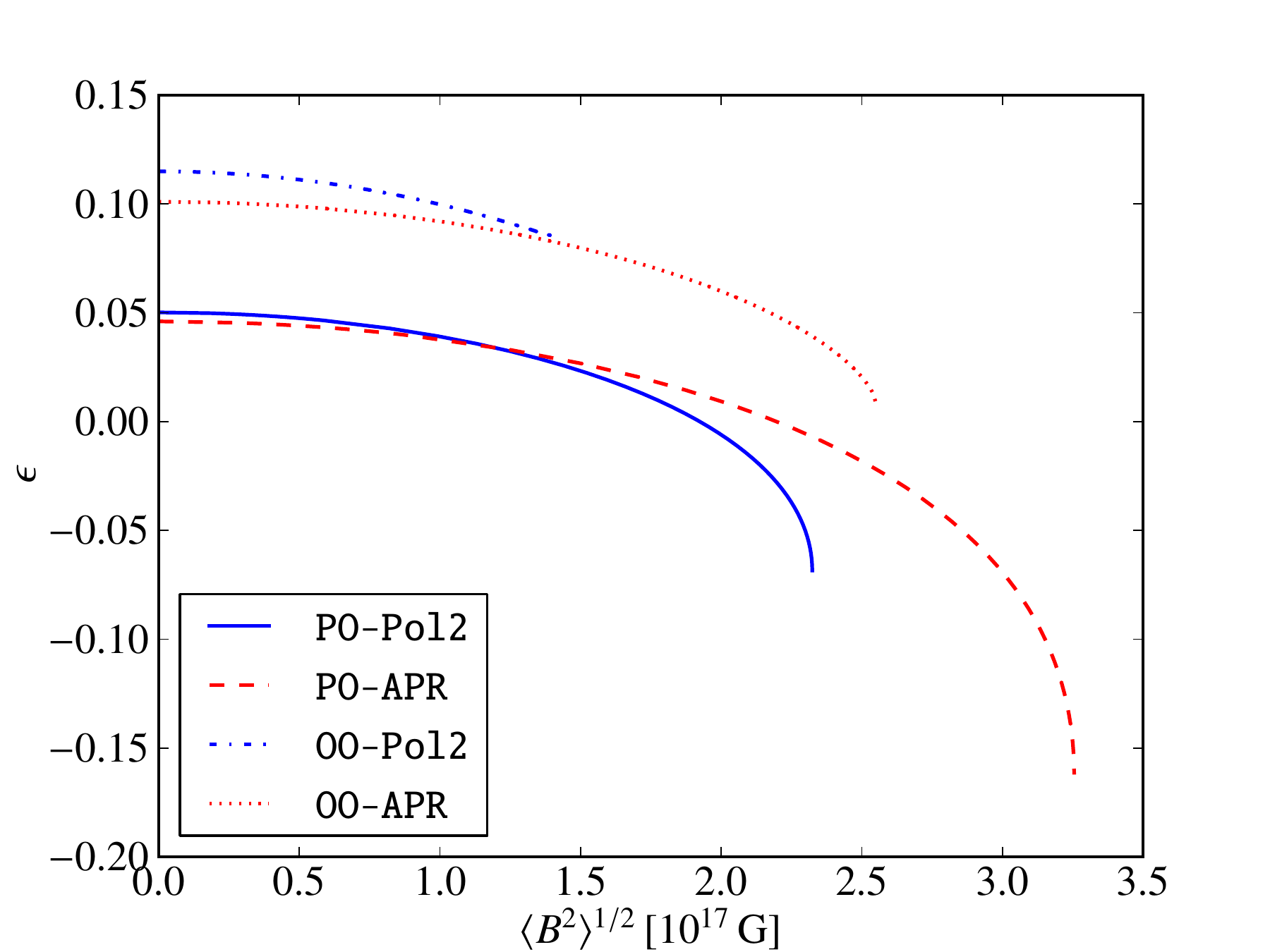}
\caption{Surface deformation $\epsilon_{\rmn{s}}$ (left-hand panel) and
  quadrupole distortion $\epsilon$ (right-hand panel) for rotating models built
  with the Pol2 EOS and the APR EOS as a function of the mean square
  magnetic field strength $\langle B^{2} \rangle$ from the
  unmagnetized limit up to the maximum field strength models
  \texttt{PO-Pol2} and \texttt{PO-APR} as well as \texttt{OO-Pol2} and
  \texttt{OO-APR}. The angular velocities are chosen such that
  $T/|W|=0.03$ and $T/|W|=0.07$, respectively, in the unmagnetized
  case.}
\label{f:eps-rot}
\end{center}
\end{figure*}

Representative models for each of the three classes located at the
maximum field strength limit have been presented in
Section~\ref{se:statmag}, namely model \texttt{PP-Pol2}, the non-rotating
configuration with a prolate apparent shape and a prolate matter
distribution, and earlier in this section, namely model
\texttt{PO-Pol2}, with a prolate apparent shape and an oblate matter
distribution, and model \texttt{OO-Pol2} with an oblate shape and an
oblate matter distribution. Within Figs~\ref{f:ell} and \ref{f:eps},
their positions in the lower and upper parts of the solution space
have been marked by red dots along the line of maximum magnetic field
strength related to $\langle B^{2} \rangle^{1/2}$.
Note that no model exists for which the apparent shape is
prolate and the distortion of the matter distribution oblate, \ie
$\epsilon_\rmn{s} < 0$ and $\epsilon > 0$; therefore, because of the
different nature of the forces caused by a toroidal magnetic field and
rotation, a class \texttt{OP} does not exist. While magnetic and
centrifugal forces distort the matter distribution at a comparable
level with respect to the involved amounts of magnetic and kinetic
energy, the magnetic potential is confined and does not act on the
surface of the star directly through equation~(\ref{EQ:EM13}). Therefore,
already in the magnetized and non-rotating case, the surface
deformation is always smaller than the quadrupole
distortion. Moreover, since centrifugal forces act more efficiently at
large distances from the rotation axis, their influence on the surface
of the star can be significant, and indeed the value of
$\epsilon_\rmn{s}=0.7242$ obtained at the unmagnetized mass-shedding
limit is considerably larger (in absolute terms) than the value
obtained for the non-rotating model at the upper magnetization limit of
$\epsilon_\rmn{s}=-0.1369$; the opposite is true when considering the
quadrupole distortion with respective values of $\epsilon=-0.6127$
and $\epsilon=0.1537$.

In order to assess the general validity of our results obtained in the
rotating case for the analytic Pol2 EOS, we have compared the two
rotating models \texttt{PO-Pol2} and \texttt{OO-Pol2} with the
corresponding models \texttt{PO-APR} and \texttt{OO-APR}, built with
the APR EOS for identical values of the $T/|W|$ ratio at the
unmagnetized limit, \ie $T/|W|=0.03$ and $0.07$. The physical
properties of all models are listed in Table~\ref{t:modbmax}, where it
is clear that the effects of rotation are less pronounced for the two
APR models because of the stiffer nature of the APR EOS. The APR
models admit a higher maximum value of $T/|W|=0.1052$ at the
unmagnetized mass-shedding limit, which should be compared with the
corresponding $T/|W|=0.09380$ for the softer Pol2 EOS. Moreover, as
discussed in Section~\ref{se:statmag}, the APR EOS reference model
supports a higher maximum field strength when compared to the Pol2 EOS
one, which causes magnetic effects to be stronger for the two models
\texttt{PO-APR} and \texttt{OO-APR}.

The toroidal magnetic field strength and the baryon number density for
the \texttt{PO-APR} and \texttt{OO-APR} models are reported in
Fig.~\ref{f:aprrot}, which shows that the matter distributions are less
condensed (the EOS is stiffer) and the peaks of the magnetic field
strength have moved slightly outward. Their prolate
deformation appears more pronounced in agreement with the higher ratio
$\mathscr{M}/|W|$, and the outer crust is easily discernible because
the magnetic field is essentially absent from this low density
region. The ratio $T/|W|$ for models \texttt{PO-APR} and
\texttt{OO-APR} is smaller than that of their unmagnetized counterparts,
in agreement with their smaller moments of inertia. In contrast, models
\texttt{PO-Pol2} and \texttt{OO-Pol2} are already located in the region
of increasing moments of inertia and show ratios $T/|W|=0.03200$ and
$0.07119$, which are larger than those of the unmagnetized models
rotating at the same angular velocities $\mathit{\Omega}$. 

Finally, Fig.~\ref{f:eps-rot} compares the dependence of
$\epsilon_\rmn{s}$ and $\epsilon$ on the magnetization levels for the
four reference rotating models (\cf Fig.~\ref{f:eps-nrot}, where the
comparison was made for the non-rotating models \texttt{PP-Pol2} and
\texttt{PP-APR}). Clearly, there is a good qualitative agreement
between the two EOSs with differences that are due mostly to the
higher maximum magnetic field strength and the higher maximum value of
$T/|W|$ supported by models \texttt{PO-APR} and \texttt{OO-APR}. Note
that model \texttt{OO-Pol2} is so close to the mass-shedding limit
that it is the only one for which models rotating at the same angular
velocity $\mathit{\Omega}$ show increasing $\epsilon_\rmn{s}$ from the
unmagnetized limit on. While all sequences in the left-hand panel of
Fig.~\ref{f:eps-rot} maintain an oblate shape, those associated with
models \texttt{PO-Pol2} and \texttt{PO-APR} show a transition from an
oblate matter distribution to a prolate one (right-hand panel of
Fig.~\ref{f:eps-rot}).

\section{Distortion coefficients}
\label{se:distcoef}

Despite the complex behaviour shown by the equilibrium models when
both the magnetic field strength and the rotation are varied, it is
possible to express such a behaviour through a very simple algebraic
expression. This was pointed out already by~\citet{Wentzel1960}
and~\citet{Ostriker1969}, who have considered this issue in earlier
Newtonian studies and have suggested to parametrize the quadrupole
distortion $\epsilon_{\rm Newt}$ induced by a toroidal magnetic field
and by rotation in a self-gravitating incompressible fluid,
respectively, as
\begin{equation} 
\label{e:repslin1} 
\epsilon_{\rm Newt} = \epsilon_{B} + \epsilon_{\mathit{\Omega}} =
    - a_{B} \, \frac{\mathscr{M}}{|W|} + a_{\mathit{\Omega}} \, \frac{T}{|W|} \,,
\end{equation}
where $a_{B} = a_{\mathit{\Omega}} = 3.750$. This approximation was adopted
also by~\citet{Cutler2002} in order to derive an estimate for the quadrupole
distortion of neutron stars, within a Newtonian framework. Since
neutron stars are highly relativistic objects, we next consider
whether an expression similar to equation~(\ref{e:repslin1}) can be derived
in a relativistic regime and the quantitative differences that then
emerge with respect to a Newtonian treatment.

\begin{figure*}
\begin{center}
\includegraphics[angle=-0,width=7.0cm]{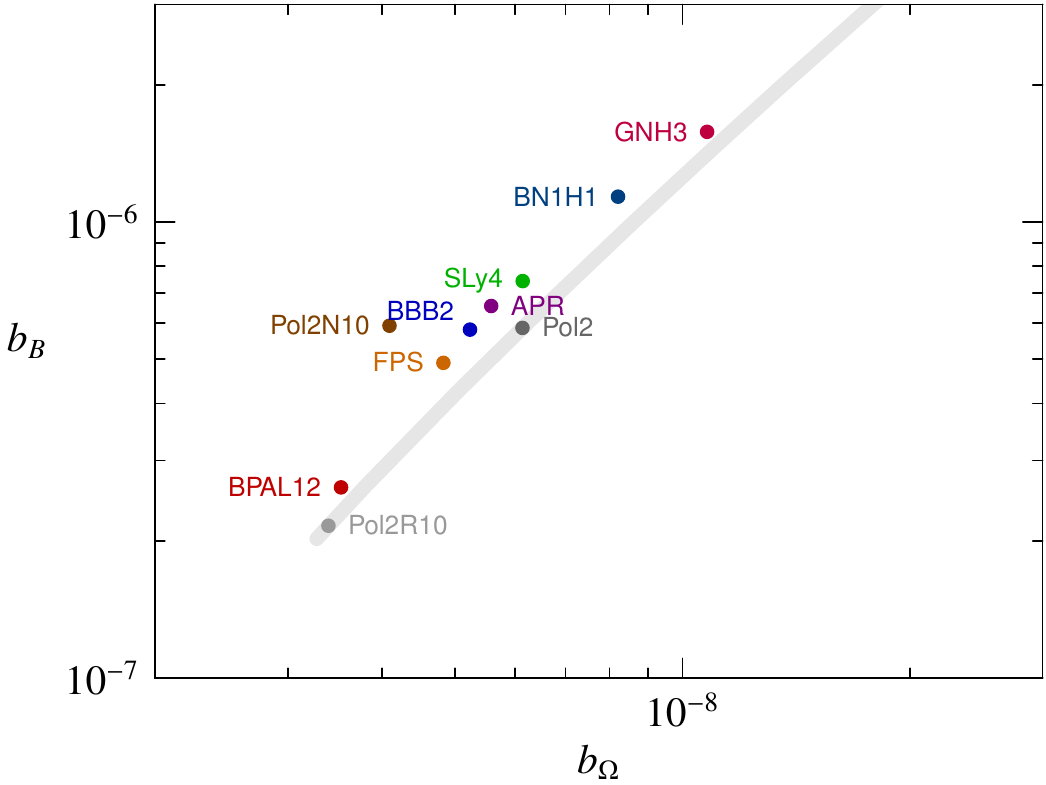}
\hskip 1.5cm
\includegraphics[angle=-0,width=7.0cm]{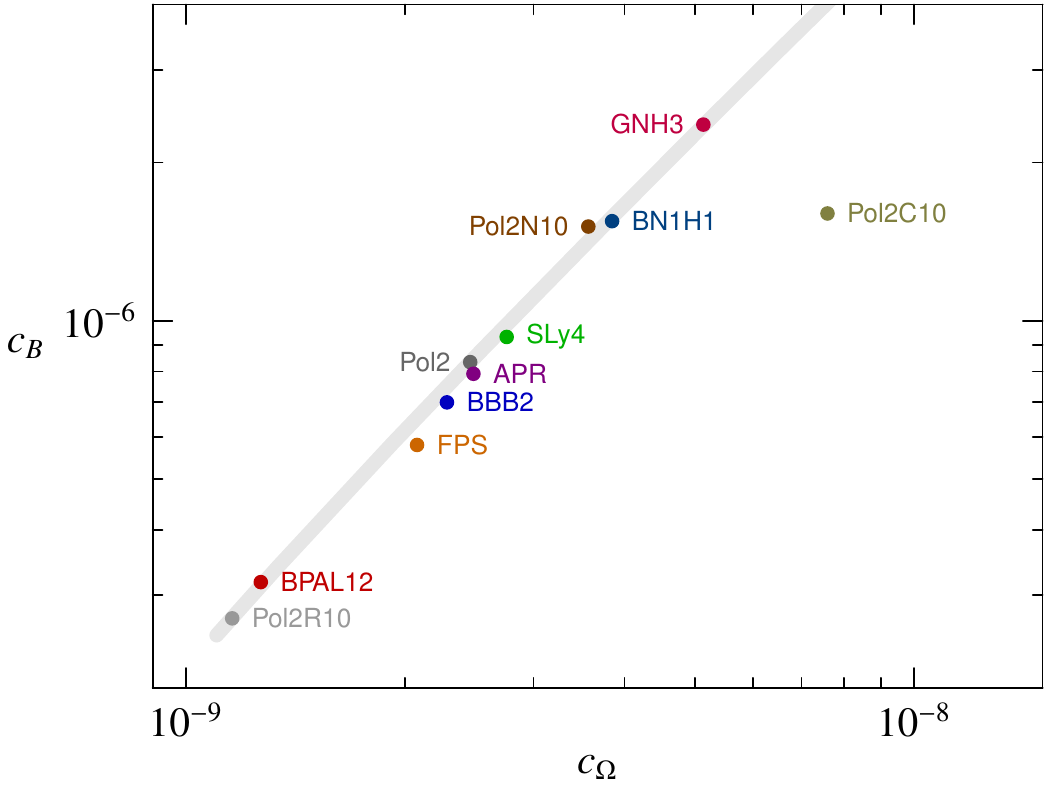}
\caption{Distortion coefficients $b_{B}$ and $b_{\mathit{\Omega}}$ for
  $\epsilon_{\rmn{s}}$ (left-hand panel) as well as $c_{B}$ and
  $c_{\mathit{\Omega}}$ for $\epsilon$ (right-hand panel) derived in the linear
  regime by perturbing a non-rotating and unmagnetized star with a
  gravitational mass of $M = 1.400 \, \mathrm{M}_{\sun}$.}
\label{f:c-distlin}
\end{center}
\end{figure*}

We start by recalling that already in Section~\ref{se:rotmag} we have
remarked how the surface deformation $\epsilon_\rmn{s}$ and the
quadrupole distortion $\epsilon$ for highly magnetized and rapidly
rotating models exhibit an almost linear dependence on $\mathit{\Omega}^2$ and
on $\langle B^{2} \rangle$. As a result, we can express these
quantities in terms of `deformation coefficients'
\begin{equation} 
\label{e:distcoef} 
\epsilon_\rmn{s} = - b_{B} \,
\langle B^{2}_{15} \rangle + b_{\mathit{\Omega}} \, \mathit{\Omega}^2 \,, 
\qquad \quad 
\epsilon = - c_{B} \, \langle B^{2}_{15}
      \rangle + c_{\mathit{\Omega}} \, \mathit{\Omega}^2 \,,
\end{equation}
where $B_{15}$ expresses the magnetic field strength in units of
$10^{15}\,\rm{G}$ and $\mathit{\Omega}$ is expressed in ${\rm s}^{-1}$. The
distortion coefficients $b_{B}, b_\mathit{\Omega}$ for the surface deformation
$\epsilon_\rmn{s}$ and the coefficients $c_{B}, c_{\mathit{\Omega}}$ for the
quadrupole distortion $\epsilon$ can be easily computed through the
directional derivatives along the coordinate axes of $\mathit{\Omega}^2$ and
$\langle B^{2} \rangle$ and are collected in Table~\ref{t:distcoef}
besides the basic properties of the unperturbed models for all of the
EOSs considered. The importance of these distortion coefficients is
that they provide all the information needed to compute the resulting
values of $\epsilon_\rmn{s}$ and $\epsilon$ by simply inserting
appropriate values for the angular velocity $\mathit{\Omega}$ and for the mean
magnetic field strength $\langle B_{15}^{2} \rangle^{1/2}$ in
equation~(\ref{e:distcoef}). In addition, the approximate expressions of
equation~(\ref{e:distcoef}) can cover a very large portion of the
physically realistic space of parameters and, for example, in the case
of the Pol2 EOS the phenomenological relations yield relative errors
$\lesssim 8$ per cent up to values of $\langle B^{2} \rangle^{1/2}=1 \times
10^{17} \, \rmn{G}$ and angular velocities of $\mathit{\Omega} = 2 \times
10^{3} \, \rmn{s}^{-1}$, which largely cover all known magnetars.

We note that the quadratic dependence expressed by
equation~(\ref{e:distcoef}) applies only to old and cold neutron stars
\textit{without} a superconducting proton phase, or to hot
proto-neutron stars.  However, for type II superconducting neutron
stars, it has been shown that below the first critical magnetic field
strength $H_\mathrm{c1} \approx 10^{15} \, \rmn{G}$, the magnetic
field is confined to flux tubes of strength $B_\mathrm{c1} \approx
10^{15} \, \rmn{G}$, which enlarge the anisotropic part of the average
electromagnetic stress tensor~\citep{Jones1975,Easson1977} by a factor
of $H_\mathrm{c1} / B$.  As a result, for magnetized models with
$\langle B \rangle < 10^{15} \, \rmn{G}$, a \textit{linear} dependence
of the matter distortion on the average magnetic field strength is
expected, and at $\langle B \rangle = 10^{15} \, \rmn{G}$, the
enhancement factor is $H_\mathrm{c1} / B = 1$, and this allows us to
match the normal-MHD and the superconducting case at this value of
$\langle B \rangle$.  As a result, equation~(\ref{e:distcoef}) has to be
modified taking the alternate form
\begin{equation} 
 \label{e:distcoef-ss} 
 \epsilon_\rmn{s} = - b_{B} \,
     \langle B_{15} \rangle + b_{\mathit{\Omega}} \, \mathit{\Omega}^2 \,,
  \qquad \quad 
     \epsilon = - c_{B} \,
     \langle B_{15} \rangle + c_{\mathit{\Omega}} \, \mathit{\Omega}^2\,,
 \end{equation}
In principle, the coefficients appearing in equation~(\ref{e:distcoef-ss})
should be calculated from the self-consistent equilibrium models of
magnetized superconducting relativistic stars, something which is
still rather difficult to do. In practice, however, it is possible to
take as coefficients the same as those in equation~(\ref{e:distcoef}),
bearing in mind, however, that the corresponding estimates are useful
only as a first approximation and as an improvement over the
corresponding coefficients by~\citet{Cutler2002}.

Beyond the second critical magnetic field strength $H_\mathrm{c2}$,
where $H_\mathrm{c2} > H_\mathrm{c1}$, superconductivity is
suppressed, and equation~(\ref{e:distcoef}) applies without restriction
yielding the usual quadratic scaling.

Fig.~\ref{f:c-distlin} offers a graphical representation of the
distortion coefficients for the surface deformation (left-hand panel) and
for the quadrupole distortion (right-hand panel). The coefficients $b_{B}$
and $c_{B}$ as well as $b_{\mathit{\Omega}}$ and $c_{\mathit{\Omega}}$
are computed for different EOSs and using the reference non-rotating
models with a gravitational mass of $M = 1.400 \, \mathrm{M}_{\sun}$.
Also shown as a grey-shaded band are the distortion coefficients of
non-rotating models built with a Pol2 EOS having the same gravitational
mass, but where the polytropic constant is varied to obtain different
circumferential radii $R_{\rm circ}$ (these have therefore $R_\rmn{circ}
\geq 9.891 \, \rmn{km}$).
The influence of the different EOSs is in fact most easily followed
through the associated circumferential radius $R_\rmn{circ}$, which
had already turned out to be a crucial quantity for the maximum field
strength models presented in Section~\ref{se:statmag}. As we explain
in Appendix~\ref{ap:newtonian}, using basic scaling considerations,
the magnetic distortion coefficients exhibit a dependence of the type
$b_{B}, c_{B} \propto R^{4}_\rmn{circ}$, whereas the rotational
distortion coefficients exhibit a dependence of the type $b_{\mathit{\Omega}},
c_{\mathit{\Omega}} \propto R^{3}_\rmn{circ}$.

Also shown for comparison in Fig.~\ref{f:c-distlin} are the Newtonian
reference model from~\citet{Cutler2002}, labelled \texttt{Pol2C10},
and its numerical equivalent as computed with our code in the
Newtonian limit, labelled as \texttt{Pol2N10}.\footnote{Only the
quadrupole distortion coefficients $c_{B}$ and $c_{\mathit{\Omega}}$ for
model \texttt{Pol2C10} were considered by ~\citet{Cutler2002}.}
Note that both Newtonian
models have similar values of $c_{B}$ and that the \texttt{Pol2N10}
data point is rather close to the reference band of relativistic Pol2
EOS models. However, the rotational distortion coefficient $c_{\mathit{\Omega}}$
reported in~\citet{Cutler2002} exceeds the correct Newtonian one of
model \texttt{Pol2N10} by almost a factor of 2. Furthermore, both
data points largely overestimate the correct relativistic result
obtained for model \texttt{Pol2R10} which is about a factor of 4
smaller and located at the lower end of the grey-shaded band of
relativistic Pol2 EOS models. This simple
example highlights therefore how a Newtonian treatment is inadequate
for the determination of realistic distortion coefficients for real
neutron stars.

\section{Conclusions}
\label{se:conclusions}

We have computed models of rotating relativistic stars with a toroidal
magnetic field under the assumption that the matter is a
single-constituent perfect fluid described by a one-parameter EOS
and behaves as a perfect conductor subject to the laws of
ideal MHD. We have investigated the combined effects
of a toroidal magnetic field and rotation on the apparent shape and
on the internal matter distribution, focusing in particular on the
quadrupole distortion, as this is the relevant quantity for the
gravitational-wave emission. Models of maximum field strength have
been computed for a sample of eight different nuclear matter equations
of state, together with the surface deformation and the quadrupole
distortion.

We have found that non-rotating models appear to admit arbitrary levels
of magnetization accompanied by a seemingly unlimited growth of size
and quadrupole distortion. In particular, we have been able to compute
a highly magnetized model for the Pol2 EOS with a baryon mass of
$M_{\rmn{b}} = 1.680 \, \mathrm{M}_{\sun}$, whose circumferential radius of
$R_\rmn{circ}=14.30 \, \rmn{km}$ in the unmagnetized case inflates to
$R_\rmn{circ}=101.5 \, \rmn{km}$ for an average magnetic field of
$\langle B \rangle = 0.1461 \times 10^{17} \, \rmn{G}$. These results
should be contrasted with those of~\citet{Kiuchi2008}, who reported a
loss of convergence for this model at a moderate level of
magnetization and corresponding to a value of merely
$R_\rmn{circ}=28.85 \, \rmn{km}$.

When considering rotating models we have instead found that the
increase in equatorial size introduced by the toroidal magnetic field
reduces the frequency at which mass shedding would otherwise appear in
unmagnetized models. Overall, the full space of solutions can be split
up into three distinct classes for which the surface distortion and
the quadrupole distortion are either prolate and prolate, oblate and
prolate or oblate and oblate, respectively.

We have also determined the relativistic distortion coefficients
whose absolute value depends mainly on the radius of the star for all
the EOSs considered. Using such coefficients it is possible to compute
the surface deformation and the quadrupole distortion by means of a
simple algebraic expression which is effective for all magnetizations
and rotation rates of known magnetars. Finally, a comparison with the
corresponding Newtonian distortion coefficients has shown that the
latter overestimate the quadrupole distortion induced by the toroidal
magnetic field by about a factor of 6 and the one induced by
rotation by about a factor of 3. Hence, they are inadequate for
strongly relativistic objects like neutron stars.

The results presented here relative to equilibrium configurations
provide the first basic steps to explore the stability properties of
magnetized stars, whose analysis in full general relativity has
recently seen a spur of activity~(\citealt{Ciolfi2011};
  \citealt*{Kiuchi2011}; \citealt{Lasky2011}; \citealt{Ciolfi2012};
  \citealt*{Lasky2012}; \citealt*{Zink2012}). We will investigate
the stability properties of our models in a forthcoming paper.

\section*{Acknowledgments}

We thank Eric Hirschmann for useful discussions.
This work was supported in part by the DFG grant SFB/Transregio 7.
JF gratefully acknowledges financial support from the Daimler und
Benz Stiftung.

\bibliographystyle{mn2e}
\bibliography{bibtex/references}

\begin{thebibliography}{}

\bibitem[\protect\citeauthoryear{{Akmal}, {Pandharipande} \&
  {Ravenhall}}{{Akmal} et~al.}{1998}]{Akmal1998}
{Akmal} A.,  {Pandharipande} V.~R.,    {Ravenhall} D.~G.,  1998, \prc, 58, 1804

\bibitem[\protect\citeauthoryear{{Balberg} \& {Gal}}{{Balberg} \&
  {Gal}}{1997}]{Balberg1997}
{Balberg} S.,  {Gal} A.,  1997, Nuclear Physics A, 625, 435

\bibitem[\protect\citeauthoryear{{Baldo}, {Bombaci} \& {Burgio}}{{Baldo}
  et~al.}{1997}]{Baldo1997}
{Baldo} M.,  {Bombaci} I.,    {Burgio} G.~F.,  1997, \aap, 328, 274

\bibitem[\protect\citeauthoryear{{Bardeen} \& {Piran}}{{Bardeen} \&
  {Piran}}{1983}]{Bardeen1983}
{Bardeen} J.~M.,  {Piran} T.,  1983, \physrep, 96, 205

\bibitem[\protect\citeauthoryear{{Bocquet}, {Bonazzola}, {Gourgoulhon} \&
  {Novak}}{{Bocquet} et~al.}{1995}]{Bocquet1995}
{Bocquet} M.,  {Bonazzola} S.,  {Gourgoulhon} E.,    {Novak} J.,  1995, \aap,
  301, 757

\bibitem[\protect\citeauthoryear{{Bombaci}}{{Bombaci}}{1996}]{Bombaci1996}
{Bombaci} I.,  1996, in {Bombaci} I.,  {Bonaccorso} A.,  {Fabrocini} A.,
  {Kievsky} A., {Rosati} S., {Viviani} M.,
  eds, Perspectives on Theoretical Nuclear Physics.
  Edizioni ETS, Pisa, p. 223

\bibitem[\protect\citeauthoryear{{Bonanno}, {Rezzolla} \& {Urpin}}{{Bonanno}
  et~al.}{2003}]{Bonanno:2003uw}
{Bonanno} A.,  {Rezzolla} L.,    {Urpin} V.,  2003, \aap, 410, L33

\bibitem[\protect\citeauthoryear{{Bonazzola}}{{Bonazzola}}{1973}]{Bonazzola197%
3}
{Bonazzola} S.,  1973, \apj, 182, 335

\bibitem[\protect\citeauthoryear{{Bonazzola} \& {Gourgoulhon}}{{Bonazzola} \&
  {Gourgoulhon}}{1994}]{Bonazzola1994}
{Bonazzola} S.,  {Gourgoulhon} E.,  1994, Classical and Quantum Gravity, 11,
  1775

\bibitem[\protect\citeauthoryear{{Bonazzola} \& {Gourgoulhon}}{{Bonazzola} \&
  {Gourgoulhon}}{1996}]{Bonazzola1996}
{Bonazzola} S.,  {Gourgoulhon} E.,  1996, \aap, 312, 675

\bibitem[\protect\citeauthoryear{{Bonazzola}, {Gourgoulhon} \&
  {Marck}}{{Bonazzola} et~al.}{1998}]{Bonazzola1998}
{Bonazzola} S.,  {Gourgoulhon} E.,    {Marck} J.-A.,  1998, \prd, 58, 104020

\bibitem[\protect\citeauthoryear{{Bonazzola}, {Gourgoulhon}, {Salgado} \&
  {Marck}}{{Bonazzola} et~al.}{1993}]{Bonazzola1993}
{Bonazzola} S.,  {Gourgoulhon} E.,  {Salgado} M.,    {Marck} J.~A.,  1993,
  \aap, 278, 421

\bibitem[\protect\citeauthoryear{{Cardall}, {Prakash} \& {Lattimer}}{{Cardall}
  et~al.}{2001}]{Cardall2001}
{Cardall} C.~Y.,  {Prakash} M.,    {Lattimer} J.~M.,  2001, \apj, 554, 322

\bibitem[\protect\citeauthoryear{Carter}{Carter}{1970}]{Carter1970}
Carter B.,  1970, Commun. Math. Phys., 17, 233

\bibitem[\protect\citeauthoryear{Carter}{Carter}{1973}]{Carter1973}
Carter B.,  1973, in DeWitt C.,  DeWitt B.~S.,  eds, Black Holes -- Les Houches
  1972. Gordon and Breach Science Publishers, New York, p. 159

\bibitem[\protect\citeauthoryear{{Ciolfi}, {Ferrari} \& {Gualtieri}}{{Ciolfi}
  et~al.}{2010}]{Ciolfi2010}
{Ciolfi} R.,  {Ferrari} V.,    {Gualtieri} L.,  2010, \mnras, 406, 2540

\bibitem[\protect\citeauthoryear{{Ciolfi}, {Ferrari}, {Gualtieri} \&
  {Pons}}{{Ciolfi} et~al.}{2009}]{Ciolfi2009}
{Ciolfi} R.,  {Ferrari} V.,  {Gualtieri} L.,    {Pons} J.~A.,  2009, \mnras,
  397, 913

\bibitem[\protect\citeauthoryear{{Ciolfi}, {Lander}, {Manca} \&
  {Rezzolla}}{{Ciolfi} et~al.}{2011}]{Ciolfi2011}
{Ciolfi} R.,  {Lander} S.~K.,  {Manca} G.~M.,    {Rezzolla} L.,  2011, \apj,
  736, L6

\bibitem[\protect\citeauthoryear{{Ciolfi} \& {Rezzolla}}{{Ciolfi} \&
  {Rezzolla}}{2012}]{Ciolfi2012}
{Ciolfi} R.,  {Rezzolla} L.,  2012, \apj, 760, 1

\bibitem[\protect\citeauthoryear{{Colaiuda}, {Ferrari}, {Gualtieri} \&
  {Pons}}{{Colaiuda} et~al.}{2008}]{Colaiuda2008}
{Colaiuda} A.,  {Ferrari} V.,  {Gualtieri} L.,    {Pons} J.~A.,  2008, \mnras,
  385, 2080

\bibitem[\protect\citeauthoryear{{Cutler}}{{Cutler}}{2002}]{Cutler2002}
{Cutler} C.,  2002, \prd, 66, 084025

\bibitem[\protect\citeauthoryear{{Das} \& {Tandon}}{{Das} \&
  {Tandon}}{1977}]{Das1977}
{Das} M.~K.,  {Tandon} J.~N.,  1977, \apss, 49, 277

\bibitem[\protect\citeauthoryear{{Douchin} \& {Haensel}}{{Douchin} \&
  {Haensel}}{2001}]{Douchin2001}
{Douchin} F.,  {Haensel} P.,  2001, \aap, 380, 151

\bibitem[\protect\citeauthoryear{{Duncan} \& {Thompson}}{{Duncan} \&
  {Thompson}}{1992}]{Duncan1992}
{Duncan} R.~C.,  {Thompson} C.,  1992, \apj, 392, L9

\bibitem[\protect\citeauthoryear{{Easson} \& {Pethick}}{{Easson} \&
  {Pethick}}{1977}]{Easson1977}
{Easson} I.,  {Pethick} C.~J.,  1977, \prd, 16, 275

\bibitem[\protect\citeauthoryear{{{Frieben}, J. and {Rezzolla},
  L.}}{{{Frieben} \& {Rezzolla}}}{2007}]{Frieben2007}
{{Frieben}, J. and {Rezzolla}, L.}, 2007, {Rotating neutron star models with a
  toroidal mag\-ne\-tic field}, Talk at SFB/TR7 Video Seminar,
  \url{http://wwwsfb.tpi.uni-jena.de/VideoSeminar/Files/20070618-frieben.pdf}

\bibitem[\protect\citeauthoryear{{Fujisawa}, {Yoshida} \&
  {Eriguchi}}{{Fujisawa} et~al.}{2012}]{Fujisawa2012}
{Fujisawa} K.,  {Yoshida} S.,    {Eriguchi} Y.,  2012, \mnras, 422, 434

\bibitem[\protect\citeauthoryear{{Glendenning}}{{Glendenning}}{1985}]{Glendenn%
ing1985}
{Glendenning} N.~K.,  1985, \apj, 293, 470

\bibitem[\protect\citeauthoryear{{Gourgoulhon}}{{Gourgoulhon}}{2012}]{Gourgoul%
hon2012a}
{Gourgoulhon} E.,  2012, {3+1 Formalism in General Relativity}.
Vol.~846 of Lecture Notes in Physics, Springer, Berlin Heidelberg

\bibitem[\protect\citeauthoryear{{Gourgoulhon} \& {Bonazzola}}{{Gourgoulhon} \&
  {Bonazzola}}{1993}]{Gourgoulhon1993}
{Gourgoulhon} E.,  {Bonazzola} S.,  1993, \prd, 48, 2635

\bibitem[\protect\citeauthoryear{{Gourgoulhon} \& {Bonazzola}}{{Gourgoulhon} \&
  {Bonazzola}}{1994}]{Gourgoulhon1994}
{Gourgoulhon} E.,  {Bonazzola} S.,  1994, Classical and Quantum Gravity, 11,
  443

\bibitem[\protect\citeauthoryear{{Gourgoulhon}, {Haensel}, {Livine}, {Paluch},
  {Bonazzola} \& {Marck}}{{Gourgoulhon} et~al.}{1999}]{Gourgoulhon1999}
{Gourgoulhon} E.,  {Haensel} P.,  {Livine} R.,  {Paluch} E.,  {Bonazzola} S.,
   {Marck} J.-A.,  1999, \aap, 349, 851

\bibitem[\protect\citeauthoryear{{Gourgoulhon}, {Markakis}, {Ury{\= u}} \&
  {Eriguchi}}{{Gourgoulhon} et~al.}{2011}]{Gourgoulhon2012}
{Gourgoulhon} E.,  {Markakis} C.,  {Ury{\= u}} K.,    {Eriguchi} Y.,  2011,
  \prd, 83, 104007

\bibitem[\protect\citeauthoryear{{Gualtieri}, {Ciolfi} \&
  {Ferrari}}{{Gualtieri} et~al.}{2011}]{Gualtieri2011}
{Gualtieri} L.,  {Ciolfi} R.,    {Ferrari} V.,  2011, Classical Quantum
  Gravity, 28, 114014

\bibitem[\protect\citeauthoryear{{Haskell}, {Samuelsson}, {Glampedakis} \&
  {Andersson}}{{Haskell} et~al.}{2008}]{Haskell2008}
{Haskell} B.,  {Samuelsson} L.,  {Glampedakis} K.,    {Andersson} N.,  2008,
  \mnras, 385, 531

\bibitem[\protect\citeauthoryear{{Haskell}, {Samuelsson}, {Glampedakis} \&
  {Andersson}}{{Haskell} et~al.}{2009}]{Haskell2009}
{Haskell} B.,  {Samuelsson} L.,  {Glampedakis} K.,    {Andersson} N.,  2009,
  \mnras, 394, 1711

\bibitem[\protect\citeauthoryear{{Ioka} \& {Sasaki}}{{Ioka} \&
  {Sasaki}}{2003}]{Ioka2003}
{Ioka} K.,  {Sasaki} M.,  2003, \prd, 67, 124026

\bibitem[\protect\citeauthoryear{{Ioka} \& {Sasaki}}{{Ioka} \&
  {Sasaki}}{2004}]{Ioka2004}
{Ioka} K.,  {Sasaki} M.,  2004, \apj, 600, 296

\bibitem[\protect\citeauthoryear{{Jones}}{{Jones}}{1975}]{Jones1975}
{Jones} P.~B.,  1975, \apss, 33, 215

\bibitem[\protect\citeauthoryear{{Kiuchi}, {Kotake} \& {Yoshida}}{{Kiuchi}
  et~al.}{2009}]{Kiuchi2009}
{Kiuchi} K.,  {Kotake} K.,    {Yoshida} S.,  2009, \apj, 698, 541

\bibitem[\protect\citeauthoryear{{Kiuchi} \& {Yoshida}}{{Kiuchi} \&
  {Yoshida}}{2008}]{Kiuchi2008}
{Kiuchi} K.,  {Yoshida} S.,  2008, \prd, 78, 044045

\bibitem[\protect\citeauthoryear{{Kiuchi}, {Yoshida} \& {Shibata}}{{Kiuchi}
  et~al.}{2011}]{Kiuchi2011}
{Kiuchi} K.,  {Yoshida} S.,    {Shibata} M.,  2011, \aap, 532, A30

\bibitem[\protect\citeauthoryear{{Lai}, {Rasio} \& {Shapiro}}{{Lai}
  et~al.}{1993}]{Lai1993}
{Lai} D.,  {Rasio} F.~A.,    {Shapiro} S.~L.,  1993, \apjs, 88, 205

\bibitem[\protect\citeauthoryear{{Lander} \& {Jones}}{{Lander} \&
  {Jones}}{2009}]{Lander2009}
{Lander} S.~K.,  {Jones} D.~I.,  2009, \mnras, 395, 2162

\bibitem[\protect\citeauthoryear{{Lasky}, {Zink} \& {Kokkotas}}{{Lasky}
  et~al.}{2012}]{Lasky2012}
  {Lasky} P.~D.,  {Zink} B.,    {Kokkotas} K.~D.,  2012,
  preprint (arXiv:1203.3590)

\bibitem[\protect\citeauthoryear{{Lasky}, {Zink}, {Kokkotas} \&
  {Glampedakis}}{{Lasky} et~al.}{2011}]{Lasky2011}
{Lasky} P.~D.,  {Zink} B.,  {Kokkotas} K.~D.,    {Glampedakis} K.,  2011, \apj,
  735, L20

\bibitem[\protect\citeauthoryear{{Miketinac}}{{Miketinac}}{1973}]{Miketinac197%
3}
{Miketinac} M.~J.,  1973, \apss, 22, 413

\bibitem[\protect\citeauthoryear{{Naso}, {Rezzolla}, {Bonanno} \&
  {Patern{\`o}}}{{Naso} et~al.}{2008}]{Naso2008}
{Naso} L.,  {Rezzolla} L.,  {Bonanno} A.,    {Patern{\`o}} L.,  2008, \aap,
  479, 167

\bibitem[\protect\citeauthoryear{{Nozawa}, {Stergioulas}, {Gourgoulhon} \&
  {Eriguchi}}{{Nozawa} et~al.}{1998}]{Nozawa1998}
{Nozawa} T.,  {Stergioulas} N.,  {Gourgoulhon} E.,    {Eriguchi} Y.,  1998,
  \aaps, 132, 431

\bibitem[\protect\citeauthoryear{{Oron}}{{Oron}}{2002}]{Oron2002}
{Oron} A.,  2002, \prd, 66, 023006

\bibitem[\protect\citeauthoryear{{Ostriker} \& {Gunn}}{{Ostriker} \&
  {Gunn}}{1969}]{Ostriker1969}
{Ostriker} J.~P.,  {Gunn} J.~E.,  1969, \apj, 157, 1395

\bibitem[\protect\citeauthoryear{Pandharipande \& Ravenhall}{Pandharipande \&
  Ravenhall}{1989}]{Pandharipande1989}
Pandharipande V.~R.,  Ravenhall D.~G.,  1989, in Soyeur M.,  Flocard H.,
  Tamain B.,   Porneuf M.,  eds, {Nuclear Matter and Heavy Ion Collisions},
  Vol.~205 of NATO ASI Ser. B.
  Plenum Press, New York, p.~103

\bibitem[\protect\citeauthoryear{{Salgado}, {Bonazzola}, {Gourgoulhon} \&
  {Haensel}}{{Salgado} et~al.}{1994}]{Salgado1994}
{Salgado} M.,  {Bonazzola} S.,  {Gourgoulhon} E.,    {Haensel} P.,  1994, \aap,
  291, 155

\bibitem[\protect\citeauthoryear{{Sinha}}{{Sinha}}{1968}]{Sinha1968}
{Sinha} N.~K.,  1968, Aust. J. Physics, 21, 283

\bibitem[\protect\citeauthoryear{{Smarr} \& {York} Jr.}{{Smarr} \&
  {York}}{1978}]{Smarr1978}
{Smarr} L.,  {York} Jr. J.~W.,  1978, \prd, 17, 2529

\bibitem[\protect\citeauthoryear{{Sood} \& {Trehan}}{{Sood} \&
  {Trehan}}{1972}]{Sood1972}
{Sood} N.~K.,  {Trehan} S.~K.,  1972, \apss, 16, 451

\bibitem[\protect\citeauthoryear{{Stella}, {Dall'Osso}, {Israel} \&
  {Vecchio}}{{Stella} et~al.}{2005}]{Stella2005}
{Stella} L.,  {Dall'Osso} S.,  {Israel} G.~L.,    {Vecchio} A.,  2005, \apj,
  634, L165

\bibitem[\protect\citeauthoryear{{Swesty}}{{Swesty}}{1996}]{Swesty1996}
{Swesty} F.,  1996, J. Comput. Phys., 127, 118

\bibitem[\protect\citeauthoryear{{Thompson} \& {Duncan}}{{Thompson} \&
  {Duncan}}{1996}]{Thompson1996}
{Thompson} C.,  {Duncan} R.~C.,  1996, \apj, 473, 322

\bibitem[\protect\citeauthoryear{{Thorne}}{{Thorne}}{1980}]{Thorne1980}
{Thorne} K.~S.,  1980, \rmp, 52, 299

\bibitem[\protect\citeauthoryear{{Tomimura} \& {Eriguchi}}{{Tomimura} \&
  {Eriguchi}}{2005}]{Tomimura2005}
{Tomimura} Y.,  {Eriguchi} Y.,  2005, \mnras, 359, 1117

\bibitem[\protect\citeauthoryear{{Villain}, {Pons}, {Cerd{\'a}-Dur{\'a}n} \&
  {Gourgoulhon}}{{Villain} et~al.}{2004}]{Villain2004}
{Villain} L.,  {Pons} J.~A.,  {Cerd{\'a}-Dur{\'a}n} P.,    {Gourgoulhon} E.,
  2004, \aap, 418, 283

\bibitem[\protect\citeauthoryear{{Wentzel}}{{Wentzel}}{1960}]{Wentzel1960}
{Wentzel} D.~G.,  1960, \apjs, 5, 187

\bibitem[\protect\citeauthoryear{{Yasutake}, {Kiuchi} \& {Kotake}}{{Yasutake}
  et~al.}{2010}]{Yasutake2010a}
{Yasutake} N.,  {Kiuchi} K.,    {Kotake} K.,  2010, \mnras, 401, 2101

\bibitem[\protect\citeauthoryear{{Yasutake}, {Maruyama} \&
  {Tatsumi}}{{Yasutake} et~al.}{2011}]{Yasutake2011}
{Yasutake} N.,  {Maruyama} T.,    {Tatsumi} T.,  2011, J. Phys. Conf. Ser.,
  312, 042027

\bibitem[\protect\citeauthoryear{{Yoshida}, {Kiuchi} \& {Shibata}}{{Yoshida}
  et~al.}{2012}]{Yoshida2012}
{Yoshida} S.,  {Kiuchi} K.,    {Shibata} M.,  2012, \prd, 86, 044012

\bibitem[\protect\citeauthoryear{{Yoshida}, {Yoshida} \& {Eriguchi}}{{Yoshida}
  et~al.}{2006}]{Yoshida2006}
{Yoshida} S.,  {Yoshida} S.,    {Eriguchi} Y.,  2006, \apj, 651, 462

\bibitem[\protect\citeauthoryear{{Zink}, {Lasky} \& {Kokkotas}}{{Zink}
  et~al.}{2012}]{Zink2012}
{Zink} B.,  {Lasky} P.~D.,    {Kokkotas} K.~D.,  2012, \prd, 85, 024030

\end{thebibliography}

\appendix

\section{Assessment of results by Cutler (2002)}
\label{ap:newtonian}

As anticipated in Section~\ref{se:distcoef}, earlier Newtonian studies
have suggested to express the quadrupole distortion $\epsilon_{\rm
  Newt}$ induced in a self-gravitating incompressible fluid by a
toroidal magnetic field and by rotation, respectively, as a function
of the total magnetic energy $\mathscr{M}$, of the kinetic energy
$T$, and of the potential energy $W$, namely through
equation~(\ref{e:repslin1})~\citep{Wentzel1960,Ostriker1969}. For a
spherical star built with a polytropic EOS with $\gamma=2$ it is easy
to estimate that
\begin{equation} 
\label{e:twnewt} 
T = \frac{1}{5} \kappa_{1} M R^{2} \mathit{\Omega}^{2} \,,
   \quad W = - \frac{3}{4} \frac{M^{2}}{R} \,,
\end{equation}
where $\kappa_{1}=0.65345$ is a constant derived by~\citet{Lai1993}.
As a result, the ratio of the kinetic to binding energy will scale
as $T/|W| \propto R^{3}$. On the other hand, the total magnetic energy
$\mathscr{M}$ of the same body reads
\begin{equation} 
\label{e:mwnewt} 
{\mathscr{M}} = \frac{1}{8\upi} \int_{V}
    B^{2} \, \rmn{d}V \ =
    \frac{4\upi}{3} R^{3} \langle B^{2} \rangle \,,
\end{equation}
where $\langle B^{2} \rangle$ denotes the mean square average of the
magnetic field strength $B$ inside the star. It then follows that
$\mathscr{M}/|W| \propto R^{4}$. Eventually, the total distortion
$\epsilon$ can be expressed in terms of mean magnetic field strength
$\langle B^{2} \rangle^{1/2}$ and angular velocity $\mathit{\Omega}$,
namely (\cf equation~\ref{e:distcoef})
\begin{equation} 
\label{e:epslin2} 
\epsilon_{\rm Newt} = \epsilon_{B} + \epsilon_{\mathit{\Omega}} =
    - c_{B}\, \langle B^{2}_{15} \rangle \,
    + c_{\mathit{\Omega}}\, \mathit{\Omega}^{2} \,.
\end{equation}
For a Newtonian model with $M = 1.400 \, \mathrm{M}_{\sun}$ and a radius of $R
\! = \! 10.00 \, \rmn{km}$, \citet{Cutler2002} has computed $c_{B}$
according to equations~(\ref{e:twnewt}) and (\ref{e:mwnewt}) and reported
the distortion coefficients as
\begin{equation} 
\label{e:epscoef-cutler} 
c_{B}=1.600 \times 10^{-6} \,, \quad c_{\mathit{\Omega}}=7.600 \times 10^{-9} \,.
\end{equation}
However, if we use equations~(\ref{e:twnewt}) and (\ref{e:mwnewt}) and
adopt an identical model with $M = 1.400 \, \mathrm{M}_{\sun}$ and $R \!  = \!
10.00 \, \rmn{km}$, we obtain
\begin{equation} 
\label{e:epscoef0} 
     c_{B}=1.610 \times 10^{-6} \,, \quad c_{\mathit{\Omega}}=3.516 \times 10^{-9} \,,
\end{equation}
where $c_{\mathit{\Omega}}=6.725 \times 10^{-9}$ for an incompressible
fluid. Although the estimates for $c_{B}$ agree reasonably well, our
value for $c_{\mathit{\Omega}}$ is less than half the one quoted
in~\citet{Cutler2002}, which is instead close to the corresponding
value for a homogeneous sphere. To validate our estimates, we have
additionally computed the distortion coefficients with our numerical
code in the Newtonian limit and obtained
\begin{equation} 
\label{e:epscoef1} 
     c_{B}=1.511 \times 10^{-6} \,, \quad c_{\mathit{\Omega}}=3.569 \times
     10^{-9} \,,
\end{equation}
which agree well with the estimates from equation~(\ref{e:epscoef0}). The
remaining difference is due to the fact that $a_{B} = a_{\mathit{\Omega}} =
3.750$ only in the incompressible case and need to be corrected in the
compressible one. After taking into account this EOS dependence, full
agreement is achieved, as can be verified in Table~\ref{t:distnewt},
which provides a compilation of the different values for the
coefficients $c_{B}$ and $c_{\mathit{\Omega}}$ and where the coefficients
$a_{B}$ and $a_{\mathit{\Omega}}$ have been added when available.

\begin{table}
\caption{\label{t:distnewt} Distortion coefficients for a Newtonian star
with $M = 1.400 \, \mathrm{M}_{\sun}$ and $R = 10.00 \, \rmn{km}$ for a
polytropic EOS with $\gamma = 2$. For comparison, values for an
incompressible model with the same properties are included.}
\begin{tabular}{ccccc}
\hline
 $\gamma$ & $a_{B}$ & $a_{\mathit{\Omega}}$ & $c_{B}$ & $c_{\mathit{\Omega}}$ \\
\hline
$\infty$\footnotemark[1] & $3.750$ & $3.750$ & $2.013 \times 10^{-6}$ & $6.725 \times 10^{-9}$ \\
$2$\footnotemark[2] & $3.750$ & $3.750$ & $1.610 \times 10^{-6}$ & $3.516 \times 10^{-9}$ \\
$2$\footnotemark[3] & $3.518$ & $3.804$ & $1.511 \times 10^{-6}$ & $3.569 \times 10^{-9}$ \\
$2$\footnotemark[4] & $-$ & $-$ & $1.511 \times 10^{-6}$ & $3.567 \times 10^{-9}$ \\
$2$\footnotemark[5] & $3.750$ & $-$ & $1.600 \times 10^{-6}$ & $7.600 \times 10^{-9}$ \\
\hline
\end{tabular} \\
1. Distortion coefficients $c_{B}$ and $c_{\mathit{\Omega}}$ according to
equation~(\protect\ref{e:repslin1}). In the incompressible case, $T = I
\mathit{\Omega}^{2} / 2$ with $I=(2/5) M R^2$, and $W = -(3/5) M^{2}/R$. \\
2. As before but using equations~(\ref{e:twnewt}) and (\ref{e:mwnewt}). \\
3. Distortion coefficients derived from the linear perturbation of a
numerical Newtonian model. \\
4. Distortion coefficients from~\citet{Haskell2008} revised
by~\citet{Haskell2009}. \\
5. Distortion coefficients from~\citet{Cutler2002}.
\end{table}
\label{lastpage}
\end{document}